\documentclass[twocolumn]{aastex62}

\usepackage{apjfonts} 

\graphicspath{ {./figures_jj/} }

\shorttitle{\HST\/ Observations of SPIRITS Transients \& Variables}
\shortauthors{Bond et al.}

\newcommand{\HST}{{\it HST}}

\newcommand{\JWST}{{\it JWST}}
\newcommand{\Spitzer}{{\it Spitzer}}

\newcommand{\lsun}{L_\sun}

\begin{document}

\title{{\em Hubble Space Telescope\/} Imaging of Luminous Extragalactic Infrared Transients and Variables from the SPIRITS Survey\footnote{Based in part on observations and archival data from the NASA/ESA {\it Hubble
Space Telescope\/} obtained at Space Telescope Science Institute, operated by Association of Universities for Research in Astronomy, Inc., under NASA contract
NAS5-26555. Also in part on observations and archival data made with the {\it Spitzer Space Telescope}, operated by the Jet Propulsion Laboratory, California Institute of Technology, under a contract with NASA.} 
}

\author[0000-0003-1377-7145]{Howard E. Bond}
\affil{Department of Astronomy \& Astrophysics, Pennsylvania State University, University Park, PA 16802, USA}
\affil{Space Telescope Science Institute, 
3700 San Martin Dr.,
Baltimore, MD 21218, USA}

\author[0000-0001-5754-4007]{Jacob E. Jencson}
%\affil{Cahill Center for Astronomy \& Astrophysics, 
%California Institute of Technology, 
%Pasadena, CA 91125, USA}
%\affil{Present address: Steward Observatory, University of Arizona, 933 N Cherry Ave., Tucson, AZ %85721-0065, USA}
\affil{Steward Observatory, University of Arizona, 933 N Cherry Ave., Tucson, AZ 85721-0065, USA}

\author[0000-0002-4678-4432]{Patricia A. Whitelock}
\affil{South African Astronomical Observatory, P.O. Box 9, 7935 Observatory, South Africa}
\affil{Department of Astronomy, University of Cape Town, Private Bag X3, Rondebosch 7701, South Africa}

\author[0000-0001-5855-5939]{Scott M. Adams}
\affil{Cahill Center for Astronomy \& Astrophysics, 
California Institute of Technology, 
Pasadena, CA 91125, USA}
\affil{Present address: Orbital Insight, 1201 N Wilson Blvd., Suite 2100, Arlington, VA 22209, USA}

\author[0000-0001-8135-6612]{John Bally}
\affil{Center for Astrophysics \& Space Astronomy,
       Astrophysical \& Planetary Sciences Department,
       University of Colorado, UCB 389, Boulder, CO 80309, USA}

\author[0000-0002-3656-6706]{Ann Marie Cody}
\affil{SETI Institute,
339 Bernardo Ave., Suite 200,
Mountain View, CA 94043, USA
}

\author[0000-0003-1319-4089]{Robert D. Gehrz}
\affil{Minnesota Institute for Astrophysics, School of Physics \& Astronomy, University of Minnesota, 116 Church St.\ SE, Minneapolis, MN 55455, USA
}

\author[0000-0002-5619-4938]{Mansi M. Kasliwal}
\affil{Cahill Center for Astronomy \& Astrophysics, 
California Institute of Technology, 
Pasadena, CA 91125, USA}

\author[0000-0002-8532-9395]{Frank J. Masci}
\affil{IPAC, California Institute of Technology, 
1200 E California Blvd., 
Pasadena, CA 91125, USA
}

\correspondingauthor{Howard E. Bond}
\email{heb11@psu.edu}




\begin{abstract}

%Until recently, relatively little was known about variable and transient stellar phenomena that occur primarily at infrared (IR) wavelengths, arising from luminous but intrinsically cool and/or heavily dust-obscured objects. The 

SPIRITS---the SPitzer InfraRed Intensive Transients Survey---searched for luminous infrared (IR) transients and variables in nearly 200 nearby galaxies from 2014 to 2019, using the warm \Spitzer\/ telescope at 3.6 and 4.5~\micron. Among the SPIRITS variables are IR-bright objects that are undetected in ground-based optical surveys. We classify them as (1)~transients, (2)~periodic variables, and (3)~irregular variables. The transients include ``SPRITE''s (eSPecially Red Intermediate-luminosity Transient Events), having maximum luminosities fainter than supernovae, red IR colors, and a wide range of outburst durations (days to years). Here we report deep optical and near-IR imaging with the {\it Hubble Space Telescope\/} (\HST\/) of~21 SPIRITS variables. They were initially considered SPRITE transients, but many eventually proved instead to be periodic or irregular variables as more data were collected. \HST\/ images show most of these cool and dusty variables are associated with star-forming regions in late-type galaxies, implying an origin in massive stars. Two SPRITEs lacked optical progenitors in deep pre-outburst \HST\/ images; however, one was detected during eruption at $J$ and $H$, indicating a dusty object with an effective temperature of $\sim$1050~K\null. One faint SPRITE turned out to be a dusty classical nova. About half the \HST\/ targets proved to be periodic variables, with pulsation periods of 670--2160~days; they are likely dusty asymptotic-giant-branch (AGB) stars with masses of $\sim$5--$10\,M_\odot$. A few of them were warm enough to be detected in deep \HST\/ frames, but most are too cool. Out of six irregular variables, two were red supergiants with optical counterparts in \HST\/ images; four were too enshrouded for \HST\/ detection.

\null\vskip 0.2in

\end{abstract}


% \keywords{blah --- blah}

% \clearpage

\section{SPIRITS: Searching for Extragalactic Infrared Transients and Variables}

Several classes of luminous variable stars and optical transients (OTs) have been known for many decades---most famously supernovae (SNe), classical novae (CNe), and luminous blue variables (LBVs). In recent years, wide-field optical imaging surveys have been finding OTs in increasingly large numbers, and the discovery rates are now becoming enormous.  However, optical surveys are relatively insensitive to objects that are intrinsically cool, dusty, or located in obscured regions. Thus our knowledge of variable and transient phenomena occurring primarily at infrared (IR) wavelengths has, until fairly recently, been limited.

In 2014, our team started a systematic search for luminous IR transients and variables in nearby galaxies, called SPIRITS (SPitzer InfraRed Intensive Transients Survey). Our survey used the Infrared Array Camera (IRAC; \citealt{Fazio2004}) on the warm {\it Spitzer
Space Telescope\/} to search for variable extragalactic objects at wavelengths of 3.6 and 4.5~\micron. From 2014--2016, our target list contained $\sim$190 galaxies, consisting
of about 37 galaxies within 5~Mpc, 116 luminous galaxies with distances of about 5 to 15~Mpc, and the 37 most luminous galaxies in the Virgo cluster at 17~Mpc. From 2017 through the end of the survey in 2019 December, we reduced our target list to a subset of the 105 galaxies of the original sample most likely
to host new transients, including the most luminous and actively star-forming galaxies, and the 58 galaxies that had previously hosted an IR transient candidate. Observing cadences ranged from a few weeks to several years, augmented with additional data available in the \Spitzer\/ archive.\footnote{\url{https://irsa.ipac.caltech.edu/Missions/spitzer.html}} The nominal $\rm S/N=5$ limiting
magnitudes for isolated objects in our exposures are $[3.6]=20.0$ and $[4.5]=19.1$ (Vega scale).
These correspond to absolute magnitudes of $-8.5$ and $-9.4$ at 5~Mpc, and $-10.9$ and $-11.8$ at 15~Mpc. However, in practice the limiting magnitudes are affected by our ability to remove the background in image subtraction, and the detection limits can be substantially brighter than the nominal values. 

Details of the SPIRITS image-processing and variable-identification pipeline are given in \citet[][hereafter K17]{Kasliwal2017}. The pipeline includes subtraction of template reference images, for which we used available archival frames, the Post-Basic Calibrated Data (PBCD)-level mosaics, including Super Mosaics\footnote{Super Mosaics are available as Spitzer Enhanced Imaging Products through the NASA/IPAC Infrared Science Archive: \url{https://irsa.ipac.caltech.edu/data/SPITZER/Enhanced/SEIP/overview.html}}, or images from the \textit{Spitzer} Survey of Stellar Structure in Galaxies (S4G; PID 61065; PI K. Sheth). Where Super Mosaics or S4G mosaics were not available, we used stacks of archival BCD-level images. For all ``saved'' sources, those vetted by human scanners and given a SPIRITS designation, we performed aperture photometry on the difference images using a 4 mosaicked-pixel ($2\farcs4$) aperture and background annulus from 4--12 pixels ($2\farcs4$--$7\farcs2$). The extracted flux is multiplied by aperture corrections of 1.215 for [3.6] and 1.233 for [4.5], as described in the IRAC Instrument Handbook,\footnote{\url{https://irsa.ipac.caltech.edu/data/SPITZER/docs/irac/iracinstrumenthandbook/}} and converted to Vega-system magnitudes using the Handbook-defined zero-magnitude fluxes for both IRAC channels. The final photometry that we employ places a grid of apertures near each individual source position to robustly estimate the uncertainties and upper limits, as described in \citet[][hereafter J20]{Jencson2020}. For non-transient, variable sources (where the difference flux measured may be negative; see our definitions below in \S\ref{sec:classes}), if there is a plausible, identifiable counterpart source in the reference images, we add the flux of the source from aperture photometry on the reference images to our difference photometry. Otherwise, we offset the difference-flux measurements of a given light curve to bring the minimum, negative value to zero before converting to magnitudes. All photometry presented in this work is provided in an electronic format as ``data behind the figures.'' Full details of the SPIRITS survey and initial discoveries are presented in K17, and updated overviews are given by \citet{Jencson2019b} and J20.

%{\bf NOTE HEB 6/2/21: sentence about about S4G is a little unclear. Updated by JEJ 7/21/21: hopefull more clear now. }

% {\bf NOTE: the actual upper-limit arrows in Fig.~1 below are significantly brighter than the nominal limiting magnitudes given above! By 2--3 magnitudes!! Jacob, check and update the wording I used above.

% Jacob: I added a bit on the subset of galaxies for the 3 year survey extension. Otherwise I think this sounds good.}

\section{SPIRITS Variables and ``SPRITE''{\footnotesize s}}\label{sec:classes}

Based on what was known about OTs at the beginning of the SPIRITS survey, we anticipated at a minimum that we would discover members of known classes of dusty transients, as well as heavily optically obscured SNe. The category of dusty OTs includes eruptive events with peak luminosities between those of SNe and CNe, which appear to fall into two main groups:

(1)~``Intermediate-luminosity red transients'' (ILRTs), typified by NGC\,300~OT2008-1 \citep{Bond2009} and SN~2008S in NGC~6946 \citep{Szczygiel2012}. The progenitors of both of these events were detected in archival \Spitzer\/ images as luminous mid-IR sources \citep{Prieto2008, Prietoetal2008}, which were heavily obscured at optical wavelengths. The outflows from ILRTs form substantial dust, and they are bright IR sources at late times after their optical light has faded \citep[e.g.,][and references therein]{Kochanek2011}. 
ILRTs are strongly associated with spiral arms, indicating that they arise from fairly massive stars. The origin of ILRT outbursts is debated: proposed mechanisms include electron-capture SNe, low-mass core-collapse SNe, events related to LBV eruptions, and binary interactions \citep[see discussion and references in][]{Adams2016, Cai2018, Cai2019, Cai2021}. 
A recently discovered likely member of this class, M51~OT2019-1 (AT\,2019abn), had a massive, self-obscured progenitor similar to the two class prototypes, and it was shown to be variable in the 12 years of available pre-outburst archival imaging with \Spitzer/IRAC \citep{Jencson2019a}. An extended phase of early circumstellar dust destruction observed during the rise of this transient disfavors a terminal-explosion scenario, strengthening the connection between ILRTs and giant eruptions of LBVs \citep{Jencson2019a,Williams2020}.

(2)~``Luminous red novae'' (LRNe), a lower-luminosity class of dust-forming transients, generally (but not always) associated with older populations. Examples include M31~RV and V4332~Sgr \citep[][and references therein]{Bond2011, Bond2018} and V838~Mon \citep{Sparks2008,Woodward2021}. These events are probably the result of common-envelope interactions and stellar mergers \citep[e.g.,][]{Pastorello2019,Howitt2020}; this was definitely the case for the LRN V1309~Sco, which was shown to have been a close binary with a rapidly decreasing orbital period before its eruption \citep{Tylenda2011}.

% {\bf NOTE: add references to our recent papers on M51 \& NGC 45 red transients.

% Jacob: I added some texts and references for M51.}

%{\bf NOTE: We may be missing a discussion of LBV giant eruptions here, which can be similar in many ways to the ILRTs.---JEJ 05/17/2021}

As reported by K17, our initial analysis of the SPIRITS data indeed revealed numerous IR transients and variables. An unexpected finding was a new class of IR transients that lack counterparts in deep optical imaging {\it during outburst}---unlike the ILRTs and LRNe described above---and have maximum IR luminosities lying between those of CNe and SNe. We refer to these objects as ``SPRITE''s (eSPecially Red Intermediate-luminosity Transient Events). These events were defined in K17 as transients with absolute magnitudes at maximum in the range $-14<M_{[4.5]}<-11$, IR colors in the range $0.3<[3.6]-[4.5]<1.6$, and having no optical counterparts detected during the outbursts in deep ground-based images.

A second surprise emerged as the SPIRITS program continued to collect data. Several luminous stars initially considered to be transients based on a small number of observations showing a rising brightness, including several of the candidate SPRITEs, were found to be periodic variables when more \Spitzer\/ data became available. In fact, it is striking how many luminous variable stars are conspicuous in late-type galaxies, when \Spitzer\/ frames taken over several years are blinked. The contrast with what is seen at optical wavelengths, where bright variables are rare, is remarkable. These objects are most likely pulsating cool and dusty stars, with very long periods, up to several years. Most of them are probably dust-obscured AGB stars, similar to, but more luminous than, those found in the Large Magellanic Cloud and other nearby galaxies, as described by \citet{Whitelock2003}, \citet{Goldman2017}, \citet{Whitelock2018}, and \citet{Menzies2019}. They may be analogs of the OH/IR stars found close to the plane in our own Galaxy \citep[e.g.,][]{Epchtein1982, Jones1982}. 

These IR-luminous pulsators have been discussed by \citet[][hereafter K19]{Karambelkar2019}, who present a catalog of over 400 periodic or suspected periodic variables found in the SPIRITS survey. Another recent study of pulsating IR variables, based in part on SPIRITS data for galaxies in the Local Group, has been presented by \citet{Goldman2019}. However, we cannot rule out that some of these apparently periodic objects may actually be massive binaries with very long orbital periods, rather than pulsators. For example, \citet{Williams2012} have described a binary system containing a carbon-rich Wolf-Rayet (WC) star, which periodically forms dust during periastron passage when its wind collides with the outflow from a companion star. Recently, we reported six extragalactic WC Wolf-Rayet binary candidates displaying such dust-formation episodes as mid-IR SPIRITS variables, including the mid-IR counterpart to the recently discovered colliding-wind WC4+O binary candidate, N604-WRXc, in M33 \citep{Lau2021}.

A few of the SPIRITS transients are so luminous that they are very likely to be obscured SNe \citep{Jencson2017, Jencson2019b}. In a few cases these objects did prove to have detectable optical counterparts. An example is SPIRITS~16tn in NGC\,3556, at a distance of only 8.8~Mpc, which reached at least $M_{[4.5]}=-16.7$ \citep{Jencson2018}. It had been missed in optical SN surveys, but following the SPIRITS discovery it was detected in deep {\it Hubble Space Telescope\/} (\HST) and ground-based images at $I$ and {\it JHK}\null. Its optical extinction is estimated to be $A_V\simeq8$--9. This event raises the possibility that some of the SPIRITS transients with relatively few observations could have been obscured SNe that happened not to be imaged around their times of maximum light.

Young stellar objects (YSOs) are another potential source of transient phenomena at IR wavelengths. { Some low- to moderate-mass YSOs experience sudden $\sim$5 to 10~magnitude increases at visual wavelengths, followed by a gradual decline lasting years to decades.  These ``FU~Orionis'' outbursts are thought to be triggered by enhanced accretion from a circumstellar disk onto the YSO \citep[e.g.,][]{HartmannKenyon1996}.  The Spitzer YSOVAR project \citep{Rebull2014} found that most YSOs are variable in the IR, and some experience luminous outbursts. In 2015, two massive YSOs underwent $\sim$$ 4 \times 10^4 ~\lsun$ IR flares  \citep{Hunter2017,CarattioGaratti2017}.  
More luminous events are thought to be associated with explosions such as the $\sim$550-yr-old BN/KL outflow in Orion OMC1 \citep{Bally2020}.  \citet{BallyZinnecker2005} proposed that mergers of the most massive stars can produce IR flares with luminosities comparable to those of SNe.  Numerical simulations show that in dense cloud cores, orbital decay can induce massive stars and binaries to migrate rapidly to the core's center to form non-hierarchical systems.  Such systems are unstable.  $N$-body interactions tend to rearrange them into hierarchical configurations,  compact binaries, and ejected stars  \citep[e.g.,][]{Reipurth2015}.  In sufficiently dense sub-groups,  $N$-body interactions may lead to YSO mergers.   In a molecular cloud, dust shifts the UV and visual light from major accretion events and stellar mergers into the IR or sub-mm.  The resulting luminous IR flares are expected to be similar to some SPIRITS IR transients.  
}

\section{{\em HST\/} Follow-up of SPIRITS Variables}\label{sec:hstobservations}

In the present paper we report results of follow-up optical and near-IR imaging observations of SPRITEs and other SPIRITS transients and variables, made with \HST\/ and its Wide Field Camera~3 (WFC3). We received allocations of two \HST\/ orbits for an initial Director's Discretionary program in Cycle~21 (GO/DD-13935, PI H.E.B.), and eight orbits for a regular Cycle~23 program (GO-14258, PI H.E.B.). Two observing programs from another team (GO-14463 in Cycle~23, GO-14892 in Cycle~24, both with PI B.~McCollum) were devoted to imaging of two further SPIRITS transients; we have obtained these data from the public \HST\/ archive\footnote{\url{http://archive.stsci.edu/}} and include analyses of them in this paper. In addition to our newly obtained WFC3 images from these programs, we downloaded \HST\/ frames of the sites from the archive; these had been secured with the Wide Field Planetary Camera~2 (WFPC2) and the Advanced Camera for Surveys (ACS), as well as WFC3. 

The main goals of our \HST\/ imaging were: (1)~deep searches for optical\slash near-IR counterparts of SPRITEs {\it while the events were underway\/};
(2)~characterization of their stellar environments; and (3)~identification of (or limits on) progenitor objects by comparison of our new \HST\/ images with pre-outburst archival images.

Our primary considerations for \HST\/ target selection were: (1)~the outburst event appeared to satisfy the photometric criteria for SPRITEs outlined above, and was expected to be underway during the \HST\/ observation\footnote{Our observations were ``non-disruptive targets of opportunity,'' meaning that the \HST\/ observations were to be obtained no less than 3~weeks after activation. As discussed below, in a few cases of fast transients, the IR outbursts had already ended by the time of the \HST\/ observation.}; (2)~there was no detected optical counterpart in concomitant optical ground-based imaging with moderate-aperture telescopes (to rule out ordinary CNe, SNe, LBVs, and other known types of OTs); and (3)~the site of the SPRITE had been imaged previously by \HST\/ at the F814W band (or longward in a few cases). As noted by K17, SPRITEs appear to have a wide duration of outburst timescales, ranging from fast events to long-duration eruptions. We attempted to sample transients spanning this range.

Table~\ref{table:observinglog} lists the SPIRITS targets that were observed in the four \HST\/ programs. Columns~1 and 2 list the object designations and host galaxies, and column~3 gives the date of the \HST\/ observation. Column~4 contains the \HST\/ program ID number. Our WFC3 observations were made in the UVIS-channel F814W bandpass, and the IR-channel F110W and F160W filters (although in one case we omitted the F110W image). Hereafter in this paper we denote these filters as $I$, $J$, and $H$, respectively, but we emphasize that these should not be confused with the ground-based bandpasses with the same designations.

\begin{deluxetable*}{llcccccl}[!]
\tablewidth{0 pt}
\tablecaption{\HST\/ Wide Field Camera 3 Observing Log\label{table:observinglog}}
% \tabletypesize{\small}
% \tabletypesize{\footnotesize}
% \tabletypesize{\scriptsize}
\tablehead{
\colhead{SPIRITS} &
\colhead{Host} &
\colhead{Observation} &
\colhead{Program} &
\multispan3{\hfil Total Exposure Time [s] \hfil} &
\colhead{Other SPIRITS} \\
\colhead{Designations} &
\colhead{Galaxy} &
\colhead{Date} &
\colhead{ID\tablenotemark{a}} &
\colhead{F814W} &
\colhead{F110W} &
\colhead{F160W} &
\colhead{Variables in \HST\/ Field} 
}
\startdata
14aje & M101 	  & 2014-09-22 & 13935 & 1575 & 738 & 177 & $\dots$ \\
14axa & M81  	  & 2014-09-26 & 13935 & 1659 & 671 & 155 & $\dots$ \\
14qb  & NGC\,4631 & 2015-11-02 & 14258 & 1575 & 738 & 177 & $\dots$ \\
15nz  & M83       & 2016-01-18 & 14463 & 584  & $\dots$ & 796 & 14akj, 14atl\\ 
%      &           &            &       &      &     &     & 15jt, 15kh, 15nu, 15zg, 15aad,\\ 
%      &           &            &       &      &     &     & 16du, 17bb, 17ck, 17kh\\ 
15qo, 15aag  & NGC\,1313 & 2016-02-11 & 14258 & 1500 & 553 & 317 & $\dots$ \\
%      &           &            &       &      &     &     & 16tl, 17gj\\
15ahg, 14al, 14dd & NGC\,2403 & 2016-03-07 & 14258 & 1500 & 553 & 317 & 15ahp\\ 
% 15afp & NGC\,6946 & 2016-03-19 & 14258 & 1320 & 484 & 317 & 15xt \\
15afp & NGC\,6946 & 2016-03-19 & 14258 & 1320 & 484 & 317 & $\dots$ \\
15wt, 14bbc  & NGC\,7793 & 2016-04-18 & 14258 & 1290 & 484 & 317 & 14th, 17fe\\
16aj  & NGC\,2903 & 2016-05-23 & 14258 & 1425 & 738 & 177 & $\dots$ \\
15mr, 15mt  & NGC\,4605 & 2016-06-14 & 14258 & 1575 & $\dots$ & 684 & $\dots$ \\ 
16tn\tablenotemark{b}  & NGC\,3556 & 2016-09-25 & 14258 & 1425 & 738 & 403 & $\dots$ \\
16ea  & NGC\,4214 & 2017-05-17 & 14892 & $\dots$ & 2385 & 2385 & $\dots$ \\
\enddata
\tablenotetext{a}{PI for 13935 and 14258 was H.E.B\null. PI for 
programs 14463 and 14892 was B.~McCollum. In the 14463 program there were also
exposures in F625W, F606W, F125W, and F140W; in 14892 there were also exposures
in F105W.}
\tablenotetext{b}{SPIRITS 16tn has been analyzed in detail
by \citet{Jencson2018}, and interpreted as a heavily obscured core-collapse SN\null. Tabulated here for completeness, but not discussed in this paper.}
\end{deluxetable*}

For most of the observations we used small subarrays, rather than read out the entire detector, in order to obtain more exposure time during the \HST\/ visits. In four cases, we adjusted the \HST\/ pointing, and used a larger subarray, so as to include one or two additional interesting SPIRITS variables in the WFC3 field of view (FOV), as indicated by the multiple entries in column~1. All exposures were taken at three dither positions, and we downloaded the default pipeline drizzle-combined images from the \HST\/ archive. Columns 5 through 7 in Table~\ref{table:observinglog} list the total exposure times in each filter, which varied slightly due to \HST\/ visibility constraints. For nominal exposure times of 1500, 600, and 300~s, the limiting ($\rm S/N=5$) \HST\/ Vega-scale {\it IJH\/} magnitudes are 26.2, 25.2, and 23.4, respectively, based on the WFC3 exposure-time calculators (ETCs).\footnote{\url{http://etc.stsci.edu/etc/input/wfc3uvis/imaging}; the listed values are optimistic and only approximate because they neglect possible background light from the host galaxy and confusion in crowded star fields.}

% {\bf NOTE: what about putting here a statement of the IJH magnitudes expected for, say, black bodies of 500 and 1000 K that are at the limiting magnitudes of the SPIRITS survey? Or maybe at [3.6] and [4.5] values more typical of our targets, which are generally brighter than the limits?  Has anybody worked out such numbers?}

Our final observation in GO-14258 was actually made on the obscured SN SPIRITS~16tn, rather than a candidate SPRITE\null. As noted above, this target is discussed in detail by \citet{Jencson2018}; it will not be considered further in the present paper. 

% Note: I used 3-point dither patterns in both of our programs. Above limiting
% mags assumed a 1500 K BB source.
 
We also searched the SPIRITS archive for additional variable objects that happened to fall within the FOVs of our \HST\/ images, but had not been the primary targets. Five of them proved interesting enough to include in the analyses in this paper. The final column in Table~\ref{table:observinglog} lists these additional ``serendipitous'' IR variables.

%\clearpage

\section{IR Light-Curve Classification}

Table~\ref{table:hsttargets} lists information on the SPIRITS variables targeted in the \HST\/ observations discussed in this paper. For each \HST\/ pointing in the table, a header gives the host galaxy and its distance modulus, from the literature sources cited in a footnote. Column~1 gives the SPIRITS designations, and columns~2 and~3 the J2000 coordinates. The final column gives our classification of each object's light curve, as described in this section. 

%The full IR light-curve data for the 21 SPIRITS variables discussed in this paper are given in Table~xx in Appendix~A.

%{\bf 
%NOTE: Are we going to give complete
%light-curve data tables here, or say they will be published later on, or what?

%Since we have a table for the variables, maybe it makes more sense to put the peak mags and/or colors in a separate table for the transients/eruptive sources---JEJ 05/18/2021.  Yes, make it so!---HEB 5/28/21.

%}

\begin{deluxetable*}{llcl}
\tablewidth{0 pt}
\setlength{\tabcolsep}{12pt}
\tablecaption{\HST\/ Targets and Light-Curve Classifications\tablenotemark{a}\label{table:hsttargets}}
\tabletypesize{\small}
%\tabletypesize{\scriptsize}
\tablehead{
\colhead{SPIRITS} &
\colhead{R.A.} &
\colhead{Decl.} &
% \colhead{Peak} &
% \colhead{Peak} &
\colhead{Light-Curve} \\
\colhead{Designation} &
\colhead{(J2000)} &
\colhead{(J2000)} &
% \colhead{$m_{[4.5]}$} &
% \colhead{$M_{[4.5]}$} &
\colhead{Classification} 
}
\startdata
\noalign{\smallskip}
\multispan4{\hfil Field of 14aje in M101, $m-M=29.16\pm0.02$ \hfil} \\
14aje & 14:02:55.51 & +54:23:18.5 & SPRITE (fast) \\ 	  	 
\noalign{\smallskip}
\multispan4{\hfil Field of 14axa in M81, $m-M=27.79\pm0.06$ \hfil} \\
14axa & 09:56:01.52 & +69:03:12.4 & Transient\tablenotemark{b} \\ 	  	 
\noalign{\smallskip}
\multispan4{\hfil Field of 14qb in NGC\,4631, $m-M=29.33\pm0.10$ \hfil} \\
14qb  & 12:41:57.50 & +32:32:06.7 & Irregular \\ 	  	  
\noalign{\smallskip}
\multispan4{\hfil Field of 15nz in M83, $m-M=28.34\pm0.07$ \hfil} \\
15nz  & 13:37:08.37 & $-29$:50:19.7 & Periodic \\ 	  	 
14akj & 13:37:03.59 & $-29$:50:57.4 & Irregular \\ 	  	  
%14akk & 13:37:06.33 & $-29$:50:18.6 &  &  & Variable \\ 	  	  
14atl & 13:37:07.96 & $-29$:50:41.3 & Irregular \\ 	  	  
%14bpa & 13:37:05.60 & $-29$:51:17.6 &  &  & Variable \\ 	  	  
%15jt  & 13:37:08.35 & $-29$:50:54.4 &  &  & Variable \\ 	  	  
%15kh  & 13:37:11.97 & $-29$:50:07.5 &  &  & Variable \\ 	  	  
%15nu  & 13:37:09.71 & $-29$:50:09.0 &  &  & Variable \\ 	  	  
%15zg  & 13:37:04.77 & $-29$:50:35.6 &  &  & Variable \\ 	  	  
%15aad & 13:37:10.62 & $-29$:49:29.9 &  &  & Variable? \\ 	  	  
%16du  & 13:37:05.11 & $-29$:49:36.2 &  &  & Variable \\ 	  	  
%17bb  & 13:37:11.89 & $-29$:50:40.3 &  &  & Variable? \\ 	  	  
%17ck  & 13:37:12.37 & $-29$:50:34.9 &  &  & Variable \\ 	  	  
%17kh  & 13:37:11.59 & $-29$:49:42.1 &  &  & Variable? \\ 	  	  
\noalign{\smallskip}
\multispan4{\hfil Field of 15qo in NGC\,1313, $m-M=28.14\pm0.08$ \hfil} \\
15qo  & 03:18:15.26 & $-66$:30:03.4 & Periodic \\ 	  	 
%15qb  & 03:18:12.55 & $-66$:30:00.1 &  &  & SPRITE (slow)? \\ 	 
%15qw  & 03:18:14.78 & $-66$:30:28.9 &  &  &  \\ 	  	  
%15ry  & 03:18:20.31 & $-66$:30:21.9 &  &  & Variable \\ 	  	  
%15aae & 03:18:27.49 & $-66$:30:15.4 &  &  &  \\ 	  	  
15aag & 03:18:23.63 & $-66$:30:24.2 & Periodic \\ 	  	 
%16tj  & 03:18:21.22 & $-66$:29:51.1 &  &  &  \\ 	  	  
%16tl  & 03:18:18.23 & $-66$:29:47.5 &  &  & Variable \\ 	  	  
%17gj  & 03:18:15.85 & $-66$:30:10.4 &  &  & Variable \\ 	  	  
\noalign{\smallskip}
\multispan4{\hfil Field of 15ahg in NGC\,2403, $m-M=27.51\pm0.06$ \hfil} \\
15ahg & 07:36:37.40 & +65:38:02.6 & Periodic \\ 	  	 
14al  & 07:36:32.38 & +65:37:26.1 & Periodic? \\ 	  	 
14dd  & 07:36:30.04 & +65:37:57.8 & Periodic \\ 	  	 
%14de  & 07:36:30.55 & +65:37:14.7 &  &  & Variable \\ 	  	  
15ahp & 07:36:35.68 & +65:37:47.2 & Irregular  \\
%
%\multispan4{\hfil \bf Jacob \& Howard classified light curves down to here 5/10/19\hfil}\\
%I finished updating the classifications --- JEJ 05/17/2021
%
%17mg  & 07:36:38.64 & +65:37:53.1 &  &  & Variable \\ 	  	  
%17mi  & 07:36:35.94 & +65:37:26.5 &  &  & Variable? \\ 	  	  
\noalign{\smallskip}
\multispan4{\hfil Field of 15afp in NGC\,6946, $m-M=28.27\pm0.07$ \hfil} \\
15afp & 20:34:59.65 & +60:11:18.1 & Periodic? \\ 	  	  
% 15xt  & 20:34:54.51 & +60:10:54.8 &  &  &  \\ 	  	  
\noalign{\smallskip}
\multispan4{\hfil Field of 15wt in NGC\,7793, $m-M=27.77\pm0.07$ \hfil} \\
15wt  & 23:57:43.38 & $-32$:35:03.1 & Periodic \\		    
14th  & 23:57:46.32 & $-32$:34:41.3 & Irregular \\		    
14bbc & 23:57:46.28 & $-32$:35:20.6 & Periodic \\		    
%14bbd & 23:57:48.02 & $-32$:35:34.7 &  &  &  & NOTE: throw out? \\ 	 
%15uc  & 23:57:41.96 & $-32$:35:21.7 &  &  & Variable \\ 	  	  
%15wu  & 23:57:46.00 & $-32$:35:14.2 &  &  & Variable \\ 	  	  
17fe  & 23:57:44.77 & $-32$:34:58.4 & SPRITE (slow) \\ 	  	 
\noalign{\smallskip}
\multispan4{\hfil Field of 16aj in NGC\,2903, $m-M=29.82\pm0.45$ \hfil} \\
16aj  & 09:32:11.64 & +21:30:03.0 & Irregular \\ 	  	 
\noalign{\smallskip}
\multispan4{\hfil Field of 15mr in NGC\,4605, $m-M=28.72\pm0.10$ \hfil} \\
15mr  & 12:39:54.87 & +61:36:46.3 & Periodic \\ 	  	  
%14adg & 12:39:56.47 & +61:36:48.7 &  &  & Variable \\ 	  	  
%14adh & 12:39:53.68 & +61:36:52.0 &  &  & Variable \\ 	  	  
%14adi & 12:39:53.93 & +61:36:53.8 &  &  & Variable \\ 	  	  
15mt  & 12:40:06.94 & +61:36:22.3 & Periodic? \\ 	  	 
%15os  & 12:39:55.11 & +61:36:42.8 &  &  & Variable \\ 	  	  
\noalign{\smallskip}
\multispan4{\hfil Field of 16ea in NGC\,4214, $m-M=27.34\pm0.08$ \hfil} \\
16ea  & 12:15:38.61 & +36:19:46.9 & Periodic? \\ 	  	  
%14ts  & 12:15:37.14 & +36:19:33.4 &  &  & Variable \\ 	  	  
\enddata
\tablenotetext{a}{The distance moduli adopted for the host galaxies are from these sources: M101: \citet{Jang2017}; NGC\,6946: \citet{Pejcha2015}; NGC\,2903: \citet{Tully2016}; all others: \citet{Tully2013}.}
\tablenotetext{b}{Probably a classical nova; see \S\ref{sec:14axa}.}
\end{deluxetable*}

At the time the prime targets in each field were selected for \HST\/ imaging, they were all candidate SPRITE transients. However, with the accumulation of substantial additional \Spitzer\/ data since the dates of the \HST\/ observations, we are able to refine our variability information beyond what was available at the time of selection. In particular (see \S2), it became apparent that many of the candidate SPRITEs are actually not transients, but pulsators with very long periods---likely mass-losing intermediate-mass stars approaching the end of their AGB evolution.
For the purposes of this paper, we have used the following classification scheme for transients, periodic variables, and irregular variables:

(1)~{\it Transients and SPRITEs}. These are sources that were undetected, brightened once, and then declined below detectability, and for which there are sufficient data before and/or after the outburst to rule out the categories below. Specifically, we use the criteria outlined in \S1.3 of J20 to select bona fide transients. Transients falling within the absolute-magnitude and color ranges given in \S2, which were not detected from the ground at optical wavelengths, are called SPRITEs. We subdivide them into ``fast'' SPRITEs (having outburst durations of less than one typical \textit{Spitzer} visibility window of $\lesssim$6~months) and ``slow'' SPRITEs (having outburst durations extending over more than one visibility window).

%{\bf NOTE: considering that 2 of the 3 SPRITES below in Fig.~1 were actually very fast, should we reduce the fast/slow timescale to something considerably shorter than one year??

%Jacob: In my thesis and forthcoming paper, I ended up using the visibility windows as a discriminator between fast and slow. If the transient is detected in observations in more than one visibility window, it is fast. This corresponds to a timescale of ~6 months.}

%I added a reference to my thesis for the transient selection criteria and changed the slow/fast designation---JEJ 05/18/2021}

%{\bf NOTE: actually, having gone thru all of the targets, the only remaining ``transients'' are ``SPRITEs'', and one classical nova, so maybe we should just call them SPRITEs.

%\bf Jacob: the likely nova is still a ``SPRITE'' by our definition of M<-11. I think it's fine to call it one, since the SPRITE classification is based only on observational properties, not on physics. It just turns out many of them are not so mysterious, in fact MOST of the ``SPRITEs'' we found are consistent with bright, dust-forming novae.}

(2)~{\it Periodic and likely periodic variables}. As discussed above, K19 identified a set of SPIRITS variables with sufficient data to show that they were varying periodically, most likely due to long-period pulsations. We call the targets in the present study ``periodic variables'' if we have seen at least one and a half cycles of variation. If there is less time coverage, but the available light-curve shape is similar to those of the known periodic variables, we call the object ``periodic?''.

% {\bf NOTE HEB 5/24/21: however, following Jacob's recent edits of Table 2, ALL of the periodic variables are called ``likely periodic''. Maybe we should revert back to periodic, as defined above (1.5 cycles), and only say ``likely'' for more dubious cases? Also, caption of Table 3 still says ``periodic'' and ``quasi-periodic.''}

%None of our \HST\/ targets belong clearly to this class. The variables we call quasi-periodic have been seen to go through at least one cycle of variation, consistent with a periodic modulation, but without sufficient confirmatory repetitions to date. 

%Patricia's version of the above: Quasi-periodic variables: K19 identified numerous SPIRITS sources with sufficient data to indicate that they are, or might be, periodic variables. Some of our HST targets clearly belong in the same class.

(3)~{\it Irregular variables}. These objects vary irregularly, in a fashion inconsistent with either of the above classes. Specifically, they do not show obvious periodicity, nor an outbursting transient behavior as described above.

%{\bf NOTE (Jacob): Besides my comments on this above on SPRITEs, I think the classification scheme is clear and reasonable. Happy to go with this as is.}

%(4)~{\it Uncertain variable type}. These variables do not clearly fall into any of the above categories, generally because of insufficient data.

% \clearpage

\section{Astrometric Registration}

In order to search for optical counterparts of SPIRITS variables and transients in \HST\/ images, it is necessary to carry out a precise astrometric registration of the \Spitzer\/ and \HST\/ frames. This task is complicated by the fact that a large fraction of stars and background galaxies that are prominent at 3.6
and 4.5~\micron\ are faint or invisible at optical wavelengths---and vice versa. In addition, sources that are isolated at \HST\/ resolution (the WFC3 plate
scales are $0\farcs0396$ and $0\farcs128\,\rm pixel^{-1}$ in the UVIS and IR channels, respectively) are often blended at \Spitzer\/ IRAC resolution ($0\farcs6\,\rm pixel^{-1}$). These considerations make it necessary to blink the IRAC and WFC3 frames visually in order to select a sample of isolated objects common to both images---many of which are either foreground stars or compact, IR-luminous background galaxies.

In many cases, the SPIRITS variables were seen to lie in crowded locations in their host galaxies, including clusters, associations, and \ion{H}{2} regions. In these instances, we first subtracted a reference \Spitzer\/ image, in which the variable was faint or absent, from a frame showing the variable in a bright phase. We then used these ``difference'' images to determine the positions of the variables in the same astrometric framework as the reference objects in the direct frames.

We employed standard tasks in IRAF and STSDAS\footnote{The Image Reduction and Analysis Facility (IRAF) was distributed by the National Optical Astronomy Observatory, which was operated by the Association of Universities for Research in Astronomy (AURA) under a cooperative agreement with the National Science Foundation. The Space Telescope Science Data Analysis System (STSDAS) was distributed by the Space Telescope Science Institute, which is operated by AURA for NASA.} to determine centroid locations for the reference stars and galaxies in the frames. Then we used the {\tt geomap} task to map the coordinate system of the \Spitzer\/ frame to the \HST\/ image, followed by the {\tt geoxytran} task to determine the $(x,y)$ position of the \Spitzer\/ variable in the \HST\/ frame. The precision of the registrations varied depending on the number and quality of the reference objects, but generally ranged from an rms of about 0.1 to 0.3 \Spitzer/IRAC pixels ($0\farcs06$ to $0\farcs18$), or about 1.5 to 4.5 WFC3/UVIS pixels in the $x$ and $y$ directions. Depending on the quality of the target's image in the \Spitzer\/ frames, the uncertainty of its position could be somewhat larger than for the reference-frame stars.

\goodbreak

\section{Infrared Transients and SPRITE{\small s}}

\begin{figure*}[ht]
\centering
\includegraphics[scale=.375]{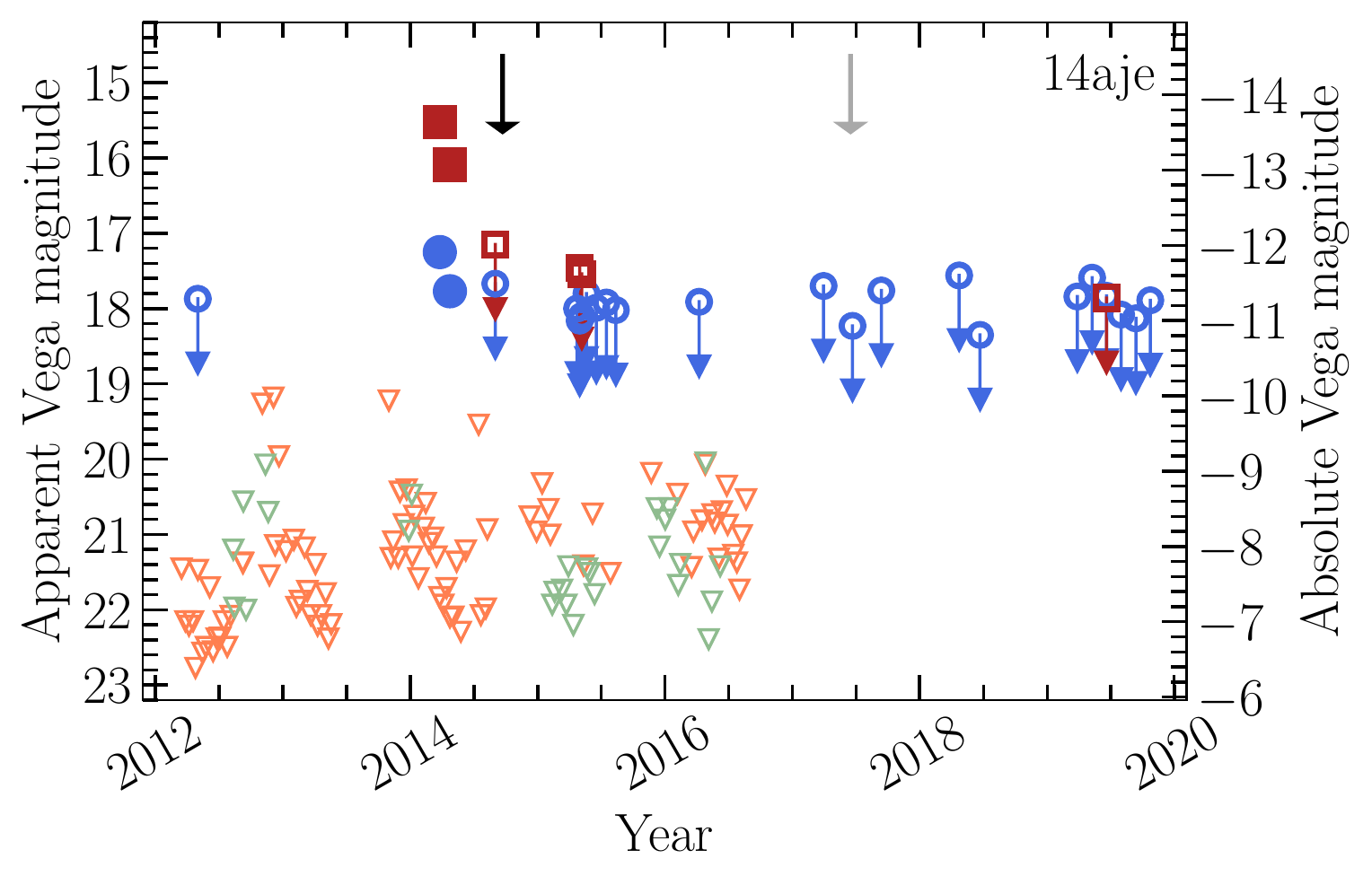}\hfill
\includegraphics[scale=.375]{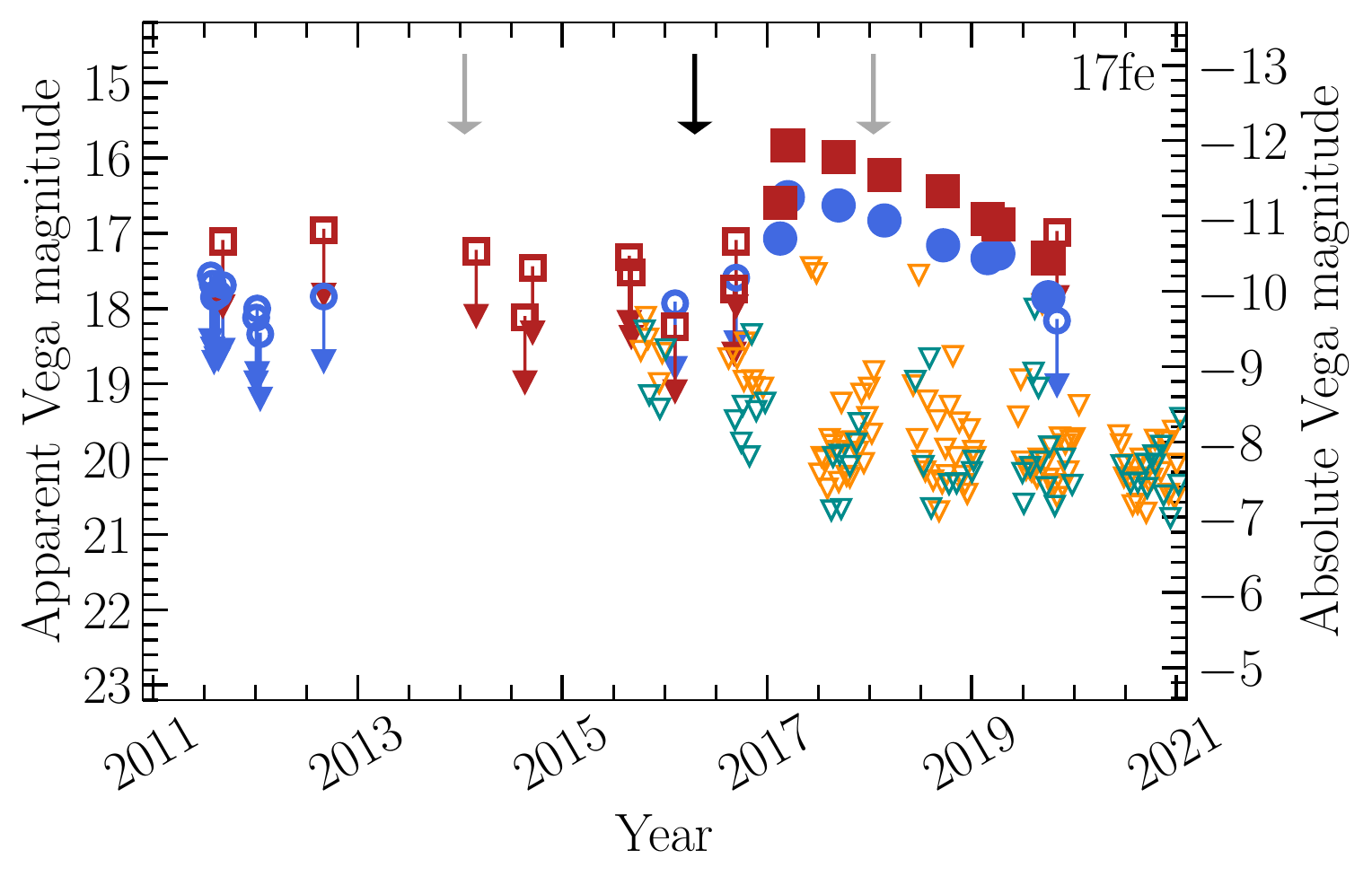}\hfill
\includegraphics[scale=.375]{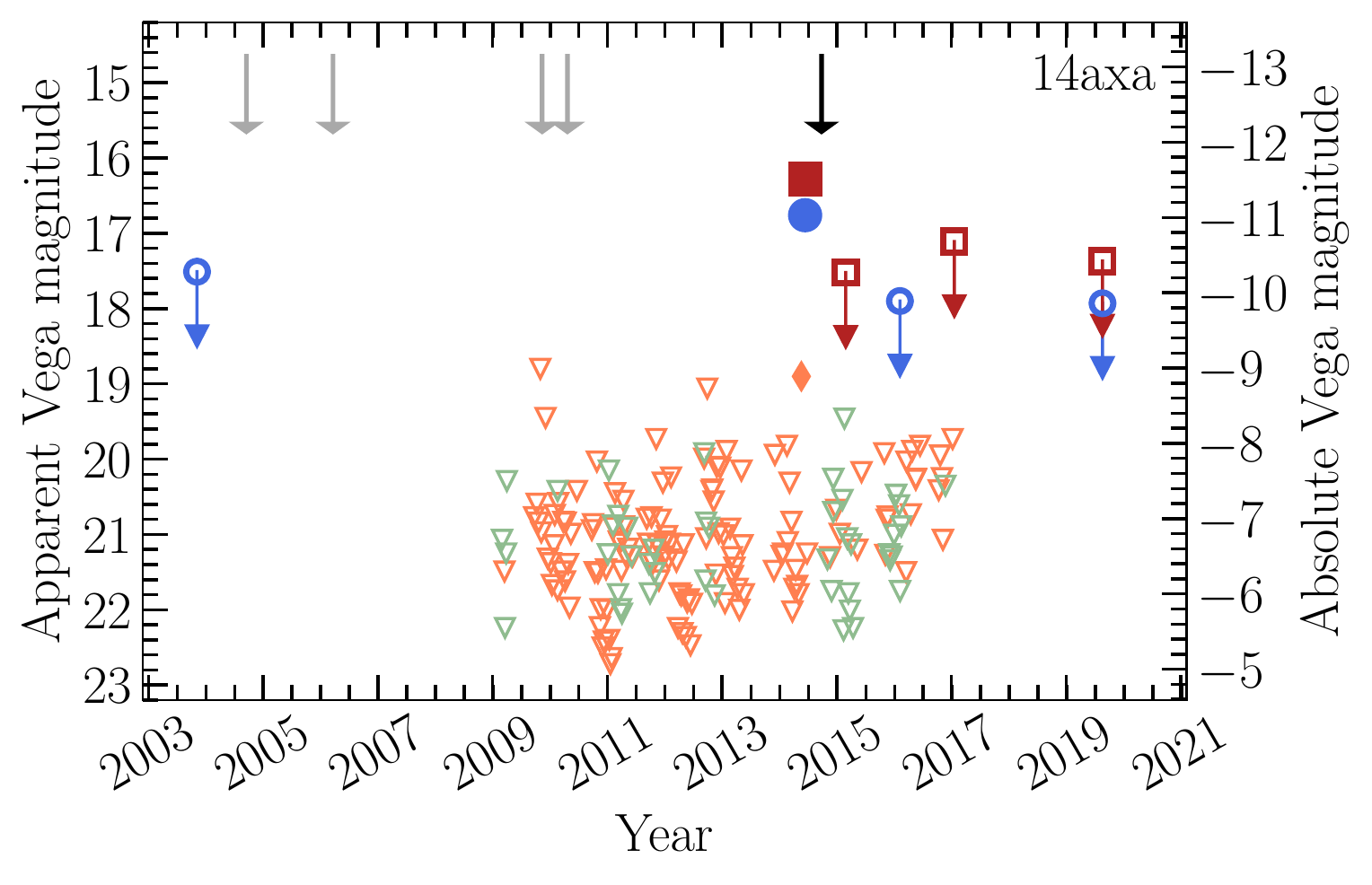}
\caption{
\Spitzer\/ IRAC light curves of two SPRITEs and a transient. In all of the light-curve figures in this paper, Vega-scale magnitudes are plotted for IRAC measurements, with {\it red squares\/} indicating [4.5] magnitudes, and {\it  blue circles\/} [3.6] magnitudes. Error bars are shown when they are larger than the plotting symbols. {\it Open symbols with down arrows\/} indicate upper luminosity limits for non-detections. Open downward triangles represent optical limits (AB magnitudes) in ATLAS-$c$ (turquoise) and -$o$ (light orange) for SPIRITS\,17fe and iptf-$g$ (green) and -$R$ (dark orange) for SPIRITS\,14aje and SPIRITS\,14axa. The detection of an optical counterpart of SPIRITS\,14axa reported by \citet{Hornoch2014} at $R = 18.9$\,mag (Vega) is also shown as the orange diamond. The epochs of our triggered \HST\/ observations are indicated at the top of each frame by {\it black arrows}, and the epochs of available archival \HST\/ broad-band images ($I$ band or longward) by {\it gray arrows}. {\it Left frame:} 14aje in M101, a luminous, very fast, and very red SPRITE, detected in only two observations; {\it middle frame:} 17fe in NGC\,7793, a slow SPRITE, which brightened suddenly and then declined over at least the subsequent 2.7~years; {\it right frame:} 14axa in M81, a likely classical nova caught in a dusty phase. 
\label{fig:transients}
}
\end{figure*}

With the accumulation of \Spitzer\/ observations and other information, only three of our targets chosen for \HST\/ follow-up have remained classified as transients. We discuss them in this section. Their IR light curves are shown in the three panels of Figure~\ref{fig:transients}. In this figure, and in subsequent light-curve plots in this paper, the epochs of \HST\/ observations are marked with vertical black or gray arrows, marking the dates of our triggered observation or of the archival observations, respectively. Since our targets are faint and extremely red, we generally only consider archival \HST\/ images taken in broad-band filters at $I$ (F814W) or longer wavelengths.

For the transients and SPRITEs, we also include optical constraints from wide-field, untargeted surveys, namely the intermediate Palomar Transient Factory (iPTF; \citealp{Cao2016}) and the Asteroid Terrestrial-impact Last Alert System (ATLAS; \citealp{Tonry2018,Smith2020}). The iPTF constraints, originally reported in Appendix~2 of J20, consist of forced-photometry measurements on the $g$- and Mould $R$-band difference images at the locations of our transients using the PTF IPAC/iPTF Discovery Engine (PTFIDE) tool \citep{Masci2017}, stacked in 10-day windows to provide deeper limits. ATLAS constraints were obtained from forced photometry\footnote{ATLAS forced-photometry server: \url{https://fallingstar-data.com/forcedphot/}} on the available ATLAS-c (``cyan'') and ATLAS-o (``orange'') difference images, again stacked in 10-day windows.

{ Table~\ref{table:transients} summarizes several properties of the light curves of the three transients, including information on the dates they were detectable, dates of maximum light, their apparent and absolute magnitudes at peak, and the $[3.6]-[4.5]$ color at peak.}

\begin{deluxetable*}{lccccccc}
\tablewidth{0 pt}
\tablecaption{Properties of Transients\label{table:transients}}
\tablehead{
\colhead{SPIRITS} &
\colhead{$t_{0}$\tablenotemark{a}} &
\colhead{Max.\ Age\tablenotemark{b}} &
\colhead{$\Delta t_{\mathrm{LC}}$\tablenotemark{c}} &
\colhead{$t_{\mathrm{peak}}$} &
\colhead{$[4.5]_{\mathrm{peak}}$} &
\colhead{$M_{[4.5],\mathrm{peak}}$} &
\colhead{$[3.6] - [4.5]$\tablenotemark{d}} \\
\colhead{Designation} &
\colhead{[MJD]} &
\colhead{[days]} &
\colhead{[days]} &
\colhead{[MJD]} &
\colhead{[mag]} &
\colhead{[mag]} &
\colhead{[mag]} 
}
\startdata
14aje & 56742.84 & 694.45 & 28.98 & 56742.84 & $15.52 \pm 0.03$ & $-13.6$ & $1.7 \pm 0.1$ \\
14axa & 56821.90 & 122.43 & $<$258.75 & 56821.90 & $16.28 \pm 0.08$ & $-11.5$ & $0.5 \pm 0.1$ \\
17fe  & 57800.71 & 127.04 & 957.17 & 57828.22 & $15.83 \pm 0.04$ & $-10.9$ & $0.69 \pm 0.06$ \\
\enddata
\tablenotetext{a}{Time of the first \textit{Spitzer}/IRAC detection.}
\tablenotetext{b}{Maximum age of transient at the time of discovery, i.e., time between first detection and the previous non-detection}
\tablenotetext{c}{Time between first and last detections.}
\tablenotetext{d}{Measured at the time of the light-curve peak, $t_{\mathrm{peak}}$.}
\end{deluxetable*}

\subsection{14aje: Fast SPRITE}

%\nobreak

%NOTE: what about an expanded light-curve plot of 14aje, to give a better idea of the timescale of the outburst? Or, maybe the words in the next paragraph are enough...
 
 The \Spitzer\/ light curves of SPIRITS 14aje in M101 ($d\simeq6.8$~Mpc) are shown in the left-hand panel of Figure~\ref{fig:transients}. This transient was first detected on 2014 March~26, at an apparent magnitude $[4.5]=15.52 \pm 0.03$. This corresponds to an absolute magnitude of $M_{[4.5]}\simeq-13.6$ at the distance of M101, well above the brightest IR luminosities observed for CNe (see \S\ref{sec:14axa}). Its color was extremely red, $[4.5]-[3.6] = 1.73$. The transient had been undetected in an archival \Spitzer\/ observation at $4.5\,\mu$m in 2012 and is not present in available Super Mosaic images (2004--2007 stack; see below for more details). 14aje faded quickly, dropping by $\sim$0.6\,mag on 2014 April~24. It was below detection at our next visit on 2014 September~2, and at all of our \Spitzer\/ observations since then. As presented in Appendix~2 of J20, limits from iPTF constrain any transient optical emission during the IR outburst to $r \gtrsim 21$\,mag. We classify 14aje as a fast SPRITE, confirming the initial classification by~K17. 
 
 We triggered our first \HST\/ SPIRITS follow-up observations on this transient, obtaining WFC3 frames on 2014 September~22. In addition to our \HST\/ data, there was subsequent archival imaging in an unrelated program in 2017 (GO-14678, PI B.\ Shappee); both dates are marked with arrows in the light-curve figure. An earlier archival observation was obtained in 2003 (GO-9492, PI F.~Bresolin), outside the plotted time interval. Our \HST\/ observations were discussed briefly by K17, but are updated here. 

\begin{figure*}[ht]
\centering
\includegraphics[width=4.5in]{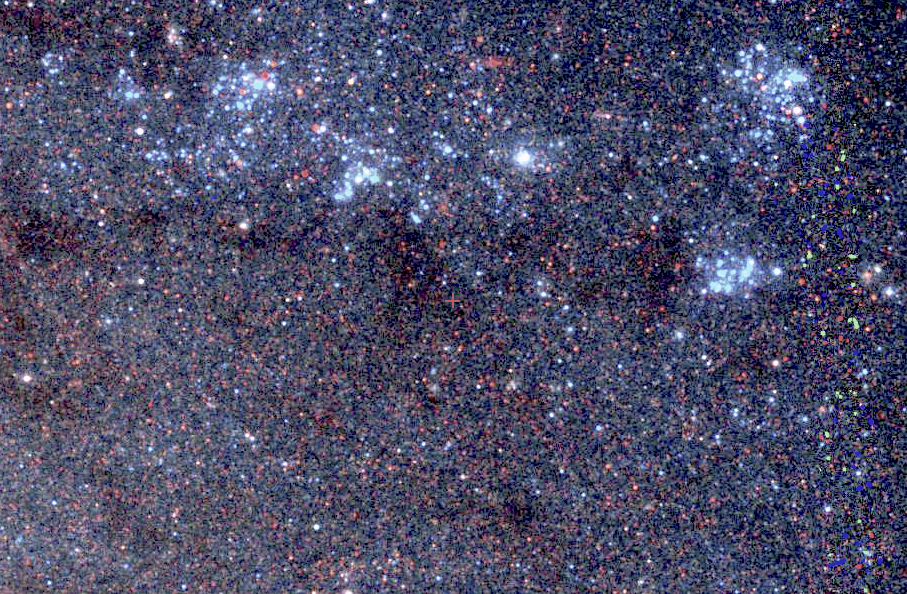}
\includegraphics[width=4.5in]{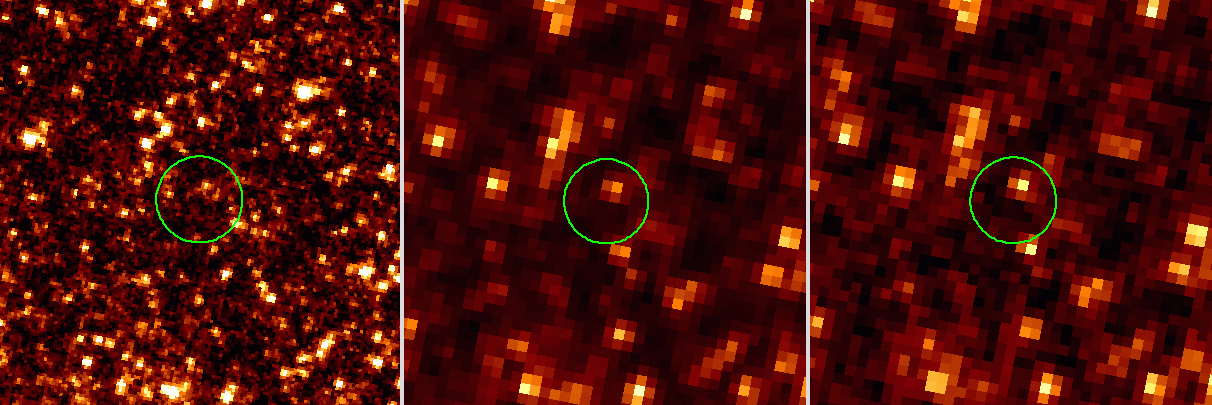}
\caption{
\HST\/ images of the site of the fast SPRITE SPIRITS~14aje in M101. {\it Top:} color rendition of the site, from $B$, $V$, and $I$ images in the Hubble Legacy Archive. The frame is $29''$ high ($\sim$950~pc at the distance of M101). The site of 14aje, marked with a red cross, lies in a dust lane in a spiral arm of M101, with several rich young associations in the vicinity. {\it Bottom row:} zooms in on \HST\/ frames we obtained several months after the outburst, in $I$, $J$, and $H$. The site of 14aje from astrometric registration of the \Spitzer\/ and \HST\/ frames is marked with green 3$\sigma$ error circles. None of the optical stars detected within the error circles in these frames appear to have varied in archival \HST\/ images taken at an epoch before, and an epoch after, the outburst (see text). Each frame is $5''$ high. All \HST\/ images presented in this paper have north at the top, east on the left.
\label{fig:14aje}
}
\end{figure*}

As the light curve shows, the 14aje event was so fast that the IR outburst unfortunately appears to have ended at an uncertain date before the epoch of our triggered \HST\/ observations. We registered \Spitzer\/ frames showing 14aje at maximum with the \HST\/ ACS images taken in 2003 and 2017, in order to determine its location in the \HST\/ frames. (We chose the ACS frames because of their large FOVs, providing more astrometric reference sources than our own WFC3 images obtained with smaller subarrays.) The top panel in Figure~\ref{fig:14aje} shows a color rendition of the SPRITE's environment, taken from the Hubble Legacy Archive\footnote{\url{http://hla.stsci.edu/hlaview.html}} (HLA)\null. The site lies in a spiral arm of M101, with several young associations nearby containing blue supergiants and a few red supergiants. The transient is located in a dark dust lane, but does not appear to lie within a rich association. The bottom three panels in Figure~\ref{fig:14aje} zoom in on the site in the WFC3 images we obtained several months after the outburst in, from left to right, $I$, $J$, and $H$\null. The green circles show the 3$\sigma$ error locations from the astrometric registration. There are several faint stars within the error circle in the $I$ frame, the brightest of which is more conspicuous in the $J$ and $H$ images. The apparent magnitudes (Vega scale) of this star are $I=25.8$, $J=23.3$, and $H=22.2$, according to photometry from the Hubble Source Catalog (HSC),\footnote{\url{http://archive.stsci.edu/hst/hsc} . In this paper we frequently give photometry for sources detected in \HST\/ images; in many cases these values are quoted from the HSC \citep[see][for an overview]{Whitmore2016}. HSC magnitudes are on the AB scale, and are determined using small photometric apertures. Throughout this paper we have corrected the HSC magnitudes to infinite apertures, using the values at \url{https://archive.stsci.edu/hst/hsc/help/HSC_faq/ci_ap_cor_table_2016.txt}. Where appropriate, we then converted the AB magnitudes to the Vega scale, using the zero-points for the WFC3 camera available at \url{http://www.stsci.edu/hst/instrumentation/wfc3/data-analysis/photometric-calibration} , and for the ACS camera from \citet{Sirianni2005}.}  available from the HLA display of the field. However, none of the stars within the error circle varied significantly in the $I$-band frames taken in 2003, 2014, and 2017. Thus we conclude that we did not detect the transient with \HST\/ a few months after its IR eruption had ended. Moreover, aside from the possibility that the object was able to return to essentially the same quiescent level it had before outburst, we have no compelling identification of an optical progenitor. These observations, along with the rapidity of the IR transient, make this SPRITE event qualitatively different from ILRTs like NGC\,300~OT2008-1, SN\,2008S, and M51~OT2019-1, which would have been detected easily by \HST\/ a few months after their outbursts at $I$, $J$, and~$H$.

%\textbf{NOTE JEJ 7/21/21: the J, H mags, I think, are consistent with an AGB. $M_J = -5.6$. This is maybe consistent with it being a less rich (older) association, but if the extinction is really large it could be much more luminous. And we can't confirm if it's the counterpart anyways.}

%\textbf{added by Jacob}: 

There is also no pre-eruption IR counterpart detected in the available archival \Spitzer/IRAC imaging. We examined the location of the event in the SHA Super Mosaics in all four IRAC channels, which consist of stacks of images taken between 2004--2007, and derived $5\sigma$ limiting magnitudes based on the faintest detected sources within a $40''$ radius in PSF-photometry catalogs constructed for M101 by K19. Our limiting (and absolute) magnitudes of $[3.6] > 19.7$ ($-9.5$), $[4.5] > 19.2$ ($-10.0$), $[5.8] > 16.9$ ($-12.3$), and $[8.0] > 15.9$ ($-13.3$) are sufficient to rule out an obscured progenitor as luminous and massive as those observed for the ILRTs mentioned in the previous paragraph. All of them had $M_{[4.5]} < -10$ before their outbursts. 

One possibility is that SPIRITS\,14aje was a heavily obscured core-collapse SN, similar to those presented in \citet{Jencson2017,Jencson2018,Jencson2019b}, but for which the luminous IR peak was missed during the gap in \textit{Spitzer}/IRAC coverage between 2012 and the start of the SPIRITS survey in 2014. The red $[3.6] - [4.5]$ color would not be unusual for a late-phase core-collapse SN (see, e.g., \citealp{Tinyanont2016,Szalai2019,Jencson2019b}). The deep optical limits from iPTF shown in Figure~\ref{fig:transients} to $R \gtrsim 21$\,mag would then imply many magnitudes of extinction ($A_V \gtrsim 9$\,mag for an SN peaking at $M_R = -16$\,mag). Such high obscuration, perhaps by a dense molecular cloud, would explain the lack of a conspicuous progenitor star in the archival {\it HST} imaging. Still, given our relatively weak constraints on the timescale and peak brightness of 14aje, we are unable to confirm this scenario, and its definite nature thus remains elusive.

\subsection{17fe: Archetypal Slow SPRITE}

%{\bf NOTE (2/4/20 HEB) This subsection updated because Jacob found a counterpart in 2018 HST images!}

The \Spitzer\/ IR light curves of SPIRITS\,17fe in the Sculptor-group galaxy NGC~7793 ($d\simeq3.6$~Mpc) are shown in the middle panel of Figure~\ref{fig:transients}. We caught it rising in brightness on 2017 February 16, and it had brightened another 0.8\,mag on 2017 March~16. At that date, marking the maximum brightness seen in our data, it had apparent magnitudes of $[3.6]=16.52$ and $[4.5]=15.83$, corresponding to absolute magnitudes of $-11.3$ and $-11.9$, respectively. Subsequently, 17fe slowly faded until going below our detection limit in our final \Spitzer\/ observations on 2019 November~1. Thus the outburst duration was at least 988~days. There were no detections of this object prior to the 2017 outburst throughout the available \Spitzer\/ imaging at 3.6 and 4.5~$\mu$m, as shown by the upper limits in the middle panel of Figure~\ref{fig:transients}. Also shown are the optical limits derived from the ATLAS forced-photometry light curves in the ATLAS-c and -o bands, which constrain the presence of an optical counterpart to $\gtrsim$19--20 AB mag in both bands for nearly the entire duration of the IR transient. We thus classify 17fe as a prototypical slow SPRITE---luminous in the IR, undetected in ground-based optical data.

%{\bf NOTE: what can we add about ground-based observations or limits, to verify that it is a SPRITE??}

\begin{figure*}[ht]
\centering
\includegraphics[width=5in]{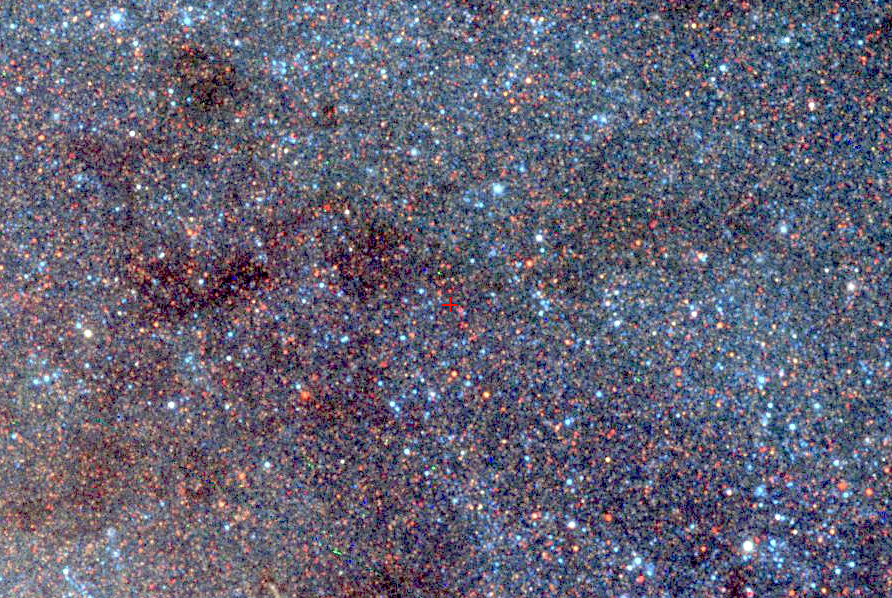}
\includegraphics[width=5in]{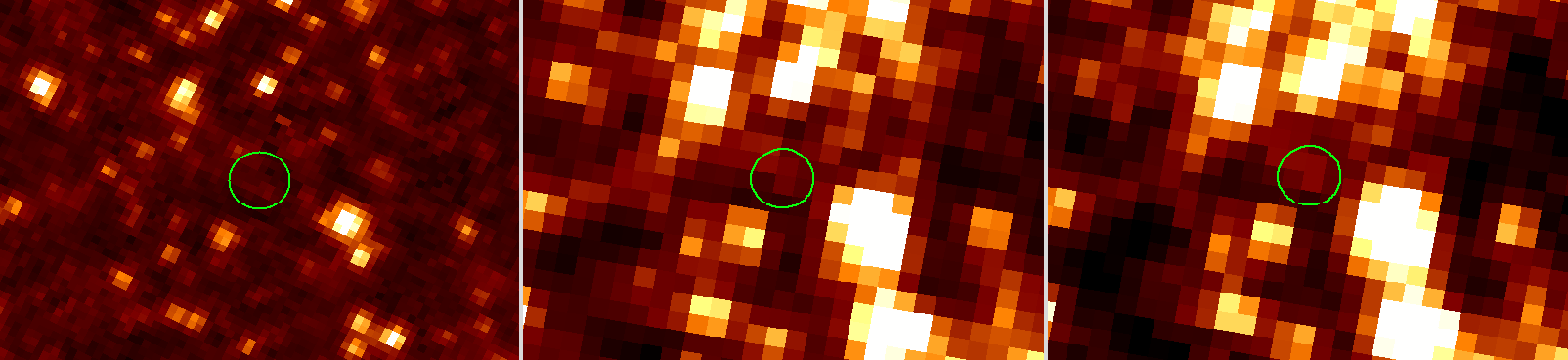}
\includegraphics[width=3.33in]{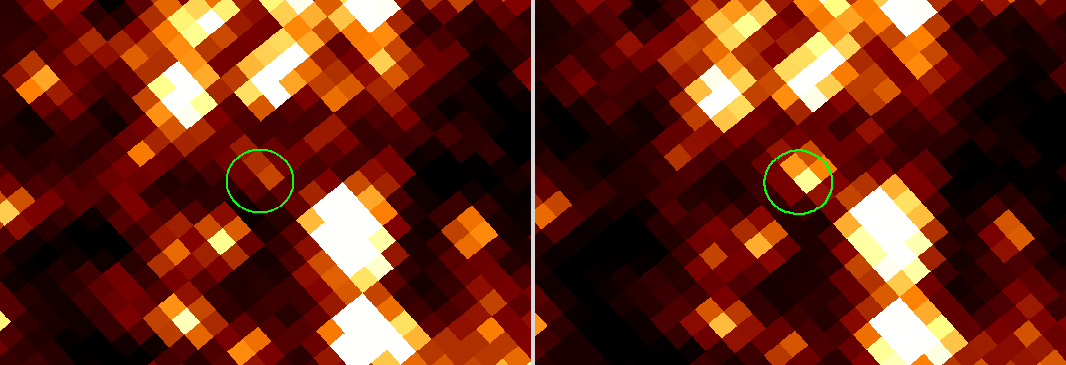}
\caption{
\HST\/ images of the site of the slow SPRITE SPIRITS~17fe in NGC\,7793. {\it Top:} color rendition of the environment, created in the Hubble Legacy Archive from $B$, $V$, and $I$ frames. Frame is $30''$ high ($\sim$525~pc at the distance of NGC\,7793). The site of 17fe, marked with a red cross, lies in a spiral arm, with dust lanes and numerous young blue stars and red supergiants in the vicinity.
{\it Middle row:} zooms in on \HST\/ frames taken before the outburst, in $I$, $J$, and $H$. The $I$ frame was taken 13.2~yr before the eruption, and $J$ and $H$ 0.8~yr before. The site of 17fe from astrometric registration of the \Spitzer\/ and \HST\/ frames is marked with green 3$\sigma$ error circles. Each frame is $2\farcs4$ high.
{\it Bottom row:} \HST\/ frames taken in $J$ and $H$ 0.8~yr after the maximum of the IR outburst, while the eruption was still underway. A near-IR counterpart is detected at $J$ and is bright at $H$.
\label{fig:17fe}
} 
\end{figure*}

The site of 17fe serendipitously lies within the \HST\/ field that we imaged for SPIRITS~15wt (discussed below in \S\ref{sec:15wt}), for which our triggered observations were obtained on 2016 April~18. This was 304~days {\it before\/} our first \Spitzer\/ detection of the 17fe event. We astrometrically registered a \Spitzer\/ 4.5~$\mu$m difference-image frame, taken at 17fe's maximum light, with an archival \HST\/ ACS $I$-band frame obtained in 2003 (GO-9774, PI S.~Larsen). (The 2003 frame, rather than our 2016 image obtained with a WFC3 subarray, was chosen for the registration because of its larger FOV\null.) The top picture in Figure~\ref{fig:17fe} shows a color rendition of the site from the HLA\null. Like the fast SPRITE 14aje, the 17fe event occurred in a spiral arm of its host galaxy, with numerous young blue stars, red supergiants, and dust lanes in its vicinity. The frames in the middle row of Figure~\ref{fig:17fe} zoom in on \HST\/ images of the site, with green circles marking the 3$\sigma$ location from the astrometric registration. From left to right, these frames show $I$ in 2003, and $J$ and $H$ from our own pre-outburst observation in 2016. In addition to these frames, there are archival \HST\/ images in the $I$ band obtained in 2001 (two epochs: GO-8599, PI T.~Boeker; and GO-9042, PI S.~Smartt) and in 2014 (GO-13364, PI D.~Calzetti). There were no changes in brightness of any objects inside the error circle in all of these pre-outburst images. Outside the error circles on the southwest side is a bright red star, which is a high-amplitude variable in the \HST\/ frames; however, it is too far outside the circle to be related to 17fe. 

%{\bf NOTE HEB 5/30/21: Jacob, I verified that the HSC magnitudes below include the aperture corrections; in case you want to try an SED fit.}

In addition to the pre-outburst \HST\/ images, there are fortuitous archival WFC3/IR frames obtained in the $J$ and $H$ bandpasses on 2018 January~16 (GO-15330, PI D.~Calzetti), 306~days after the date of the IR maximum. The IR outburst was still underway at this epoch. Cutouts from these frames are shown in the bottom row of Figure~\ref{fig:17fe}. A faint, very red object has appeared near the center of the astrometric error circle at both wavelengths, making it a very likely near-IR counterpart of the SPRITE in outburst. Based on aperture photometry relative to HSC stars in the nearby field, we find Vega-scale magnitudes for this star of $J=24.2$ and $H=21.4$. There is a hint that this object is present in our 2016 pre-outburst $J$ and $H$ frames, but the field is crowded with overlapping faint stars. There is no convincing progenitor in the pre-outburst $I$ frames; there is a partially resolved star inside the error circle in the 2003 $I$-band image just southwest of the center, with a Vega-scale magnitude of $I\simeq27.1$. However, this star does not coincide with the object that appeared in the 2018 frames. 

\begin{figure}
\centering
\includegraphics[width=0.45\textwidth]{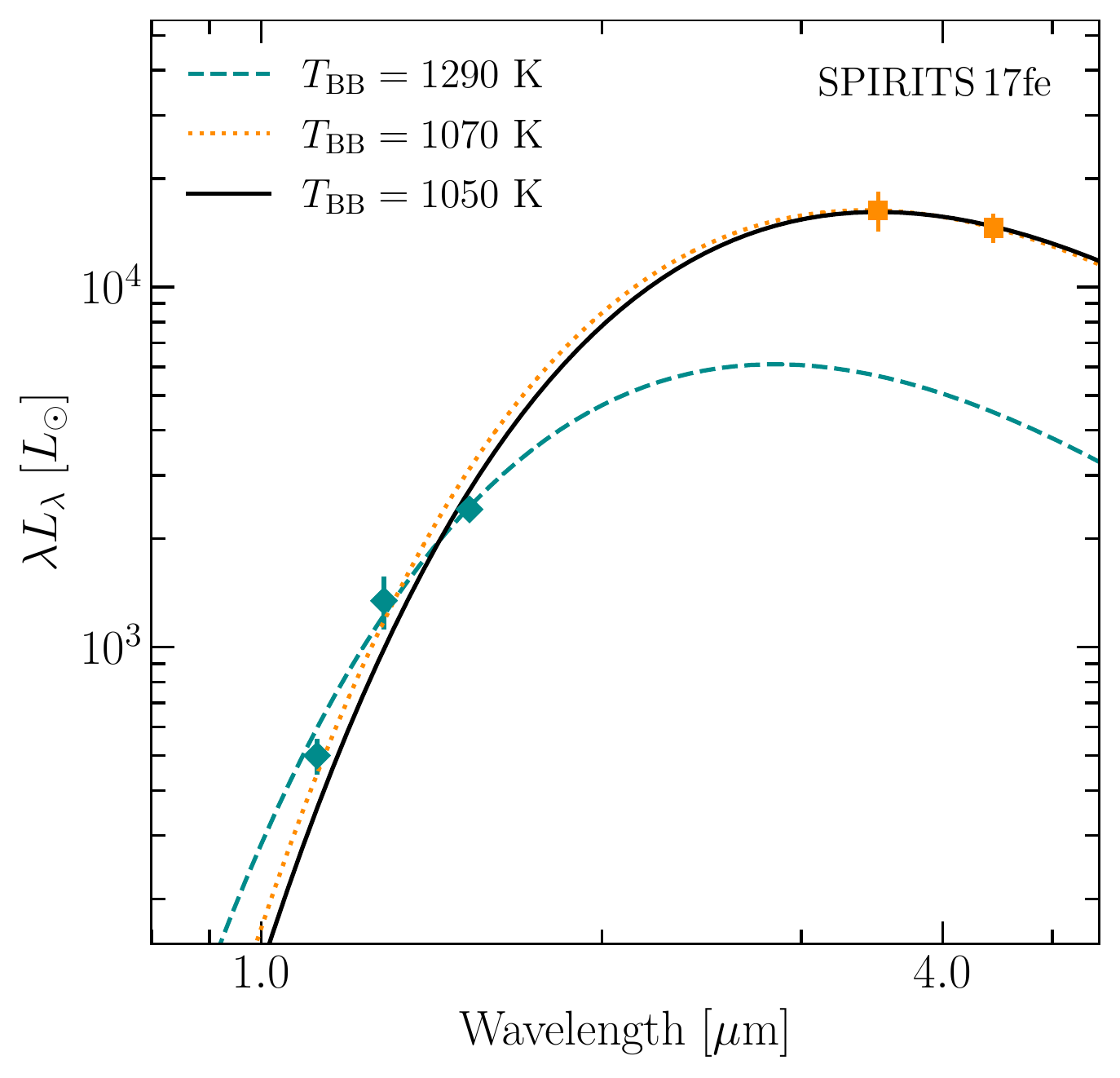}
\caption{
\label{fig:17fe_SED}
SED of the slow SPRITE, SPIRITS\,17fe, constructed from photometry of the {\it HST\/} WFC3/IR F110W, F128N, and F160W images taken on 2018 January 16.2 (blue diamonds), and from the {\it Spitzer}/IRAC [3.6] and [4.5] light curves interpolated to the same epoch (orange squares). Blackbody fits to only the {\it HST\/} points (blue dashed curve), only the {\it Spitzer\/} points (orange dotted curve), and all points together (black solid curve) are shown with their corresponding temperatures, $T_{\mathrm{BB}}$, given in the legend in the upper-left corner. 
} 
\end{figure}

In Figure~\ref{fig:17fe_SED}, we show the spectral-energy distribution (SED) of 17fe, constructed from the 2018 January 16
WFC3/IR detections and interpolations of the \textit{Spitzer}/IRAC [3.6] and [4.5] light curves to the same epoch, 333 days after the first detection of the event with \textit{Spitzer}. The SED is very red, appearing to peak in the IR around 3\,$\mu$m, at a band luminosity of $\lambda L_{\lambda}\sim10^4~L_{\odot}$. The near- to mid-IR SED can be approximated by a blackbody spectrum of temperature $T_{\mathrm{BB}} \simeq 1050$\,K\null. We also tried fitting the \textit{HST\/} and \textit{Spitzer\/} points with two separate blackbodies. In this case, the \HST\/ data alone indicate a slightly warmer temperature of $T_{\mathrm{BB}} \simeq 1290$\,K; however, there is not strong evidence for two components. These values are near the temperatures for dust condensation \citep[e.g.,][]{Ney1978,Gehrz1980a}, suggesting the presence of newly formed, warm dust. 

A stellar merger is a compelling scenario for the origin of LRNe and at least some slow SPRITEs. For SPRITEs like 17fe, early dust formation is required to obscure or dramatically shorten the associated OT\null. In models by \citet{Pejcha2016a,Pejcha2016b}, elaborated by \citet{Metzger2017}, the secondary light-curve peak seen in many LRNe can be explained by the shock-interaction of the dynamical merger ejecta with equatorially concentrated material ejected from the binary during the pre-dynamical in-spiral phase. The dense, rapidly cooling regions behind radiative shocks are favorable locations for dust formation. \citet{Metzger2017} find that, for certain binary configurations, namely those involving giant stars and having long phases of pre-merger mass loss, dust may form early enough to completely obscure the associated shock-powered transient at optical wavelengths.

For a similar, but more luminous, slowly evolving IR transient, SPIRITS\,19fi, an associated faint, short-duration ($\approx$10\,day), red OT was detected in stacked observations from the Zwicky Transient Facility (see J20). For two other slow SPRITES, SPIRITS\,17ar and SPIRITS\,18nu, their near-IR outburst spectra show strong molecular absorption features akin to those of a late M giant (Jencson et al., in prep.). These features are also seen in late-time spectra of several optically bright LRNe \citep[e.g.,][]{Kaminski2015,Blagorodnova2017,Blagorodnova2020}. These observations, together with the late-time SED of SPIRITS\,17fe suggestive of warm dust, lend credence to a stellar merger accompanied by early dust formation as a viable origin of many slow SPRITEs. 

%The brightest object inside the circle, just below center, has an $I$ magnitude of about 27.1 (Vega scale), based on a comparison with nearby objects in the field that have magnitudes given in the HSC\null. This object is not variable in the available \HST\/ material, and is not very red, since it is barely detected, if at all, in the $J$ and $H$ frames. Whether it is related to 17fe could only be determined by obtaining new \HST\/ images of the site. 

%{\bf NOTE: add some discussion here of its possible nature?? Jacob sent this comment, which we should incorporate: This may be a good candidate for an LRN given the slowly evolving light curve. Metzger and Pechja (2017) suggested slow SPRITEs may be stellar mergers that are self-obscured by early dust formation. Need to say something about whether it was ever detected in the optical or not.}

%Jacob adds: There also looks to be a faint counterpart in the Spitzer reference images. I'll work on getting mag estimates for these. UPDATE: There's blended emission, but no convincing counterpart at that location in Viraj's catalogs.}

%{\bf NOTE: Howard to update this subsection to discuss 2018 counterpart found by Jacob!}

% {\bf NOTE: 17fe light curve needs gray arrow at 2018-01-16}

%SPIRITS\,14axa in M81 ($d=3.6$~Mpc) was initially classified by us (K17) as a fast SPRITE\null. We detected it with \Spitzer\/ at only one epoch, 2014 June~13, as shown by its light curve in the right-hand panel of  Figure~\ref{fig:transients}. There were \Spitzer\/ non-detections 122~days before this observation (2014 February~11), and 239 days afterward (2015 February~7). Our triggered \HST\/ ob

\subsection{14axa: Classical Nova?}\label{sec:14axa}

SPIRITS\,14axa in M81 ($d\simeq3.6$~Mpc) was initially classified by us (K17) as a fast SPRITE\null. We detected it with \Spitzer\/ at only one epoch, 2014 June~13, as shown by its light curve in the right-hand panel of  Figure~\ref{fig:transients}. There were \Spitzer\/ non-detections 122~days before this observation (2014 February~11), and 239 days afterward (2015 February~7). Our triggered \HST\/ observations on 2014 September~26 were obtained between the dates of the \Spitzer\/ detection and the subsequent non-detection, so it is unknown whether the IR event was still underway. The absolute magnitude at the single \Spitzer\/ detection was $M_{[4.5]}\simeq-11.5$. 

%\begin{figure}[ht]
%\centering
%\includegraphics[scale=.525]{Spitzer_lcs_hst_14axa.pdf}
%\caption{
%\Spitzer\/ IRAC light curves of SPIRITS 14axa in M81, a possible classical nova. Plotting symbols as in Figure~1. 
%\label{fig:14axalightcurve}
%}
%\end{figure}

We learned later that this event had also been detected at optical wavelengths and reported as a CN, designated PNV J09560160+6903126.\footnote{We are grateful to D.~Bishop for maintaining a website devoted to extragalactic novae, at \url{https://www.rochesterastronomy.org/novae.html}, which alerted us to this optical detection. The apparent coincidence of 14axa with the nova was also noted by \citet{Oskinova2018}.} The initial discovery was by \citet{Hornoch2014}, who reported an unfiltered (approximately $R$) optical magnitude of 18.9, on 2014 May~21.9. The transient had been fainter than magnitude 21.7 two nights earlier, demonstrating an extremely fast rise time. Five nights later, \citet{HornochStoev2014} observed the object with a wide-field camera on the 2.5-m Isaac Newton Telescope. A narrow-band filter confirmed strong emission at H$\alpha$, and yielded a Sloan $r'$ magnitude of 19.6. All of this information is consistent with the transient being a CN, although to our knowledge there is no direct spectroscopic confirmation of this conclusion. Our detection with \Spitzer\/ suggests that the nova was in a dust-forming post-maximum phase at the time of our observation.

\begin{figure}[ht]
\centering
\includegraphics[width=3.25in]{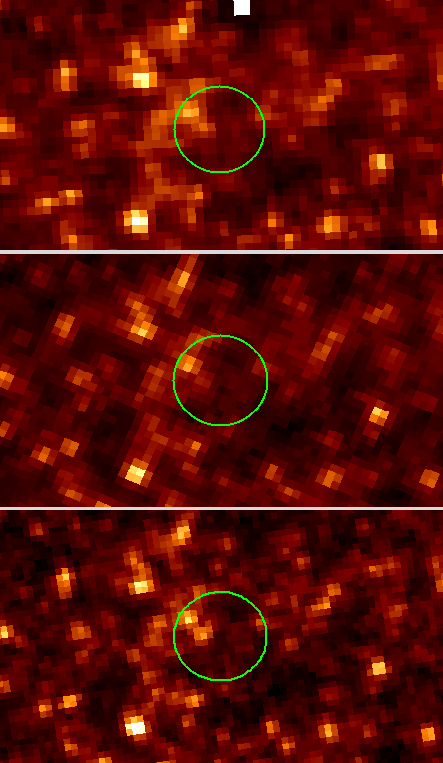}
\caption{False-color renditions of $I$-band
\HST\/ images of the site of SPIRITS~14axa in M81. {\it Top:} ACS image from 2002; {\it middle:} ACS image from 2004; {\it bottom:} WFC3 image taken by us on 2014 September~26, triggered by the \Spitzer\/ discovery 3.5~months earlier. Each frame is $1\farcs6$ high. The {\it green circles\/} show the 3$\sigma$ positional error circles, based on an astrometric registration of the 2004 image with a \Spitzer\/ image showing the transient. A faint star inside the error circle, seen in 2002 and 2004, had brightened in the 2014 image, and is likely to be the optical counterpart of this probable classical nova.
\label{fig:14axa}
}
\end{figure}

The site of 14axa lies between two spiral arms of M81, in a crowded sheet of stars which appears to lack a young population (in contrast to most of the SPIRITS IR transients, which strongly tend to be associated with spiral arms, young associations, and dusty environments). There are three available $I$-band \HST\/ images of the site, two obtained with ACS in 2002 and 2004 (GO-9353, PI S.~Smartt, and GO-10250, PI J.~Huchra), and our triggered WFC3 observation in 2014. We performed an astrometric registration of the \Spitzer\/ 4.5~$\mu$m image showing 14axa with the 2004 ACS image, in order to locate the site in the \HST\/ frames. Figure~\ref{fig:14axa} shows renditions of the three \HST\/ $I$-band frames, with green circles marking the 3$\sigma$ error positional locations. Not far from the centers of the circles is a faint star that noticeably brightened in the 2014 image. Approximate $I$ magnitudes (AB scale), determined from aperture photometry relative to nearby stars with HSC magnitudes, are 25.2 in 2002, 25.8 in 2004, and 24.7 in 2014. (In our 2014 $J$ and $H$ frames, the object is badly blended with the neighboring star to the northeast.) The $I$-band luminosity at the 2014 observation is consistent with a CN about four months past maximum; see, for example, the $I$-band and near-IR light curves of T~Pyx presented by \citet{Walter2012} and \citet{Evans2012}. However, the pre-outburst object is unusually luminous compared to the progenitors of typical CNe. Unless it is a chance superposition, the pre-outburst detection suggests that the binary system has a red-giant donor star, similar to Galactic (recurrent) novae such as T~Pyx and RS~Oph \citep[e.g., see the reviews of][]{Schaefer2010, Mukai2015}. The object was below detection in archival pre-outburst \HST\/ images at the $V$ band, showing that it was indeed red.

% 10/16/21: I updated/corrected the I mags given above. They are now AB mags, corrected to infinite aperture--HEB

As discussed in J20, the majority of SPIRITS transients fainter than $M_{[4.5]}
\simeq -12.5$ at peak are similar to 14axa: that is, they were detected
only within a single \Spitzer\/ visibility window---implying an evolutionary
timescale $\lesssim$6~months. At their peak they have red IR colors of $[3.6] -
[4.5]$ between 0.5 and 1.2.  Two events, SPIRITS\,15bb and SPIRITS\,15bh,
were also preceded by short ($\lesssim$2\,months), faint ($M_g \approx -8$)
optical outbursts recovered in stacked, archival iPTF images. All 15 of these
fast and faint IR transients in SPIRITS were found in just seven galaxies within
$\approx$5~Mpc, and nearly half of them were in M81. This high rate from a small
number of the nearest galaxies suggests they are common events. CNe are thus an
attractive scenario for their origin; the rate in large galaxies like the Milky
Way is estimated to be $\approx$40--80\,yr$^{-1}$
\citep[e.g.,][]{Shafter2017,De2021}. As with 14axa, the optical precursors
detected for 15bb and 15bh were in the luminosity range typical of novae, further
supporting this hypothesis. Some novae, namely those of the DQ~Her class, such as NQ~Vul \citep{Ney1978}, LW~Ser \citep{Gehrz1980a}, and V705~Cas \citep{Gehrz1995a,Evans1997,Evans2005}, form optically thick dust shells, while others that form optically thin shells still produce strong IR emission, peaking on a timescale of $\approx$50--80~days at
$M_\mathrm{[4.5]} \approx -11$ to $-12.5$ \citep[e.g., V1668~Cyg;][]{Gehrz1980b}. These properties are generally consistent with the fast
and faint IR transients discovered by SPIRITS (J20). Further study on the
implications of this population for the rate of strongly dust-forming novae and
their impact on the chemical enrichment and dust budget of galaxies
\citep[e.g.,][]{Gehrz1988,Gehrz1999,Evans2012} is thus warranted. A less likely alternative is that this was the coronal-emission phase of an ONe nova such as QU~Vul \citep{Greenhouse1988,Greenhouse1990}.   These novae appear to be several absolute magnitudes fainter at $4.5\,\micron$ during maximum light than dusty novae (see, e.g., \citealt{GehrzJones1995b}).

%\clearpage

\section{Luminous Periodic Infrared Variables}\label{sec:PerVar}

As noted above (\S2), a significant fraction of the suspected transients discovered early in the SPIRITS survey proved eventually to be periodic, or likely periodic, variables when more \Spitzer\/ data, over longer time baselines, had accumulated. The periods associated with these sources are nearly all longer than 1000 days. This led to us triggering \HST\/ observations of variables that were only recognized as being periodic later on. In typical cases, the objects were in the rising phase of their light curves during the first few SPIRITS observations, leading to our initial classifications of them as candidate SPRITEs.

In view of the importance of the periodicity to understanding these sources, we first summarize our analysis of the \Spitzer\/ photometry and the consequent insight into the nature of this sample of periodic sources. We then describe the \HST\/ observations of individual objects and the additional understanding they provide.

%Similar luminous periodic IR variables discovered in nearby galaxies by SPIRITS were discussed by K19; they suggested that these objects are related to the dusty OH/IR stars found in our Galaxy and the LMC\null. 

Characteristics of the periodic and likely periodic variables that we observed with \HST\/ are summarized in Table~\ref{table:periodicvars}. 
We derived the periods given in column~2 from the \Spitzer\/ light curves using the procedure described in K19 (which in turn follows \citealt{Vanderplas2015}), allowing the [3.6] and [4.5] magnitudes to be analyzed simultaneously. Two of the variables have periods quoted as lower limits, because there is \Spitzer\/ photometry covering only about a single cycle.  Coverage of the other sources is rather better than for most of the variables discussed by K19, although few have complete coverage for as much as two cycles.  Given that Mira light curves are known not to repeat exactly from cycle to cycle, our derived periods can only be good to about 5--10\%, and are best judged by examining the individual light curves.
The mean apparent magnitudes (denoted [3.6] and [4.5]) are given in columns~3 and 5, the peak-to-peak amplitudes ($\Delta[3.6]$ and $\Delta[4.5]$) in columns~4 and 6, and the mean colors ($[3.6]-[4.5]$) in column~7. Columns~8 and 9 give the the mean absolute magnitudes at [3.6] and [4.5], calculated using the distance moduli in Table~\ref{table:hsttargets}. The absolute magnitudes cover the range $-11.07 >M_{[4.5]} >-12.31$, except for one unusually luminous variable, 15mr, at $M_{[4.5]}\simeq-13.76$.

\begin{deluxetable*}{lrccccccc}
\tablewidth{0 pt}
\tablecaption{Properties of Periodic and Suspected Periodic Variables\label{table:periodicvars}}
\tablehead{
\colhead{SPIRITS} &
\colhead{Period} &
\colhead{[3.6]\tablenotemark{a}} &
\colhead{$\Delta[3.6]$\tablenotemark{a}} &
\colhead{[4.5]\tablenotemark{a}} &
\colhead{$\Delta [4.5]$\tablenotemark{a}} &
\colhead{$[3.6]-[4.5]$} &
\colhead{$M_{[3.6]}$} &
\colhead{$M_{[4.5]}$} \\
\colhead{Designation} &
\colhead{[days]} &
\colhead{[mag]} &
\colhead{[mag]} &
\colhead{[mag]} &
\colhead{[mag]} &
\colhead{[mag]} &
\colhead{[mag]} &
\colhead{[mag]} 
}
\startdata
15nz\tablenotemark{b}  & 1614 & $\dots$ & $\dots$ & 16.66 & 1.07 & 1.5 & $\dots$  & $-$12.17 \\
15qo  		       & 1232 & 16.69	& 1.37    & 16.11 & 1.43 & 0.58 & $-$11.45 & $-$11.68 \\ 
15aag 		       & 1233 & 18.10	& 1.63    & 17.07 & 1.55 & 1.04 & $-$10.04 & $-$11.07 \\
15ahg\tablenotemark{c} & 1163 & 16.49 	& 0.93    & 16.13 & 1.16 & 0.36 & $-$11.02 & $-$11.38 \\
14al  		       & $\sim$2160 & 16.60	& 0.96    & 15.20 & 0.82 & 1.40 & $-$10.91 & $-$12.31 \\
14dd  		       & 1418 & 16.89	& 0.95    & 15.92 & 1.11 & 0.97 & $-$10.62 & $-$11.59 \\
15afp 		       & >1650& 17.63	& 0.79    & 16.87 & 1.13 & 0.76 & $-$10.64 & $-$11.40 \\
15wt\tablenotemark{c}  & 1188 & 16.55 	& 0.90    & 16.03 & 0.94 & 0.53 & $-$11.22 & $-$11.75 \\
14bbc\tablenotemark{c} & 1498 & 16.29 	& 1.12    & 15.57 & 1.13 & 0.71 & $-$11.48 & $-$12.20 \\
15mr\tablenotemark{c}  & 1113 & 15.71 	& 1.37    & 14.96 & 1.32 & 0.76 & $-$13.01 & $-$13.76 \\
15mt\tablenotemark{c}  & >1800& 16.86 	& 1.67    & 16.69 & 2.16 & 0.17 & $-$11.86 & $-$12.03 \\
16ea\tablenotemark{b,c} & $\sim$670 & $\dots$ & $\dots$ &  16.51 & 1.11 & 1.6 & $\dots$ & $-$10.83 \\
\enddata
\tablenotetext{a}{[3.6] and [4.5] denote the mean apparent magnitudes of the variables, and $\Delta[3.6]$ and $\Delta[4.5]$ denote the peak-to-peak amplitudes of their variations.} 
\tablenotetext{b}{Insufficient data at [3.6] to determine mean magnitude, amplitude, and absolute magnitude over the pulsation cycle (see text). The approximate color given in column~7 is based on an estimate of the likely mean [3.6] magnitude.} 
\tablenotetext{c}{15ahg, 15wt, 14bbc, 15mr, and 15mt had detected or suspected optical/near-IR counterparts in \HST\/ imaging (see text). 16ea is equivocal (see \S\ref{sec:16ea}). The others had no convincing counterparts at $I$, $J$, or $H$ in \HST\/ images.}
\end{deluxetable*}

\begin{figure*}[ht]
\centering
\includegraphics[width=5.25in]{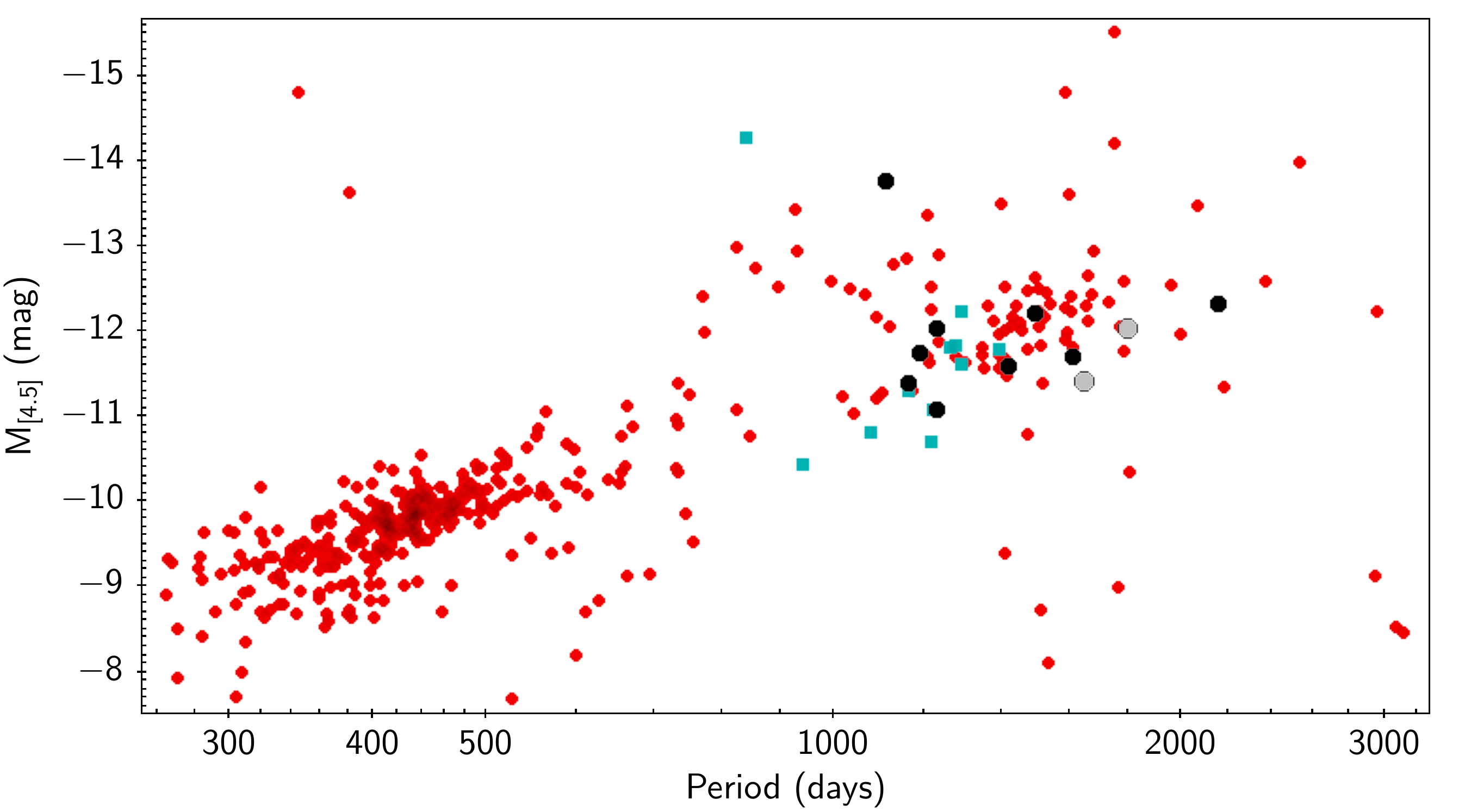}
\caption{
{Mean [4.5] absolute magnitude versus pulsation period for
the periodic and suspected periodic variables from this paper, shown as black points, or grey points if only one pulsation cycle or less has been covered. Red points mark the periodic variables from K19, and cyan squares the LMC OH/IR stars  from \citet{Goldman2017}. The isolated and unusually luminous black point is SPIRITS 15mr (see \S\ref{subsec:15mr}).}
\label{fig:plrelation}
}
\end{figure*}

A similar class of luminous periodic IR variables discovered in nearby galaxies by SPIRITS was discussed by K19; they suggested that some of these objects are related to the dusty OH/IR stars (intermediate-mass, oxygen-rich Mira variables) found in our Galaxy and the LMC\null. In Figure~\ref{fig:plrelation} we plot the period-luminosity relation for our variables, using filled grey circles for objects with one variation cycle or less, and filled black circles for those with more. Also plotted are the data from K19 for SPIRITS periodic variables in nearby galaxies (filled red circles), and for LMC OH/IR stars from \citet{Goldman2017} (filled cyan squares).  All but one of our variables fall on, or close to, the clump of variables that K19 identify with intermediate-mass AGB stars ($1000<P<2000$~days and $-11>M_{[4.5]}> -13$), and which itself falls close to the extrapolated period-luminosity relation found for Mira variables in the LMC \citep{Riebel2015, Whitelock2017}. These sources have large variation amplitudes (e.g., K19, their Figure~4), as do our sources which are in the approximate range $0.8 <\Delta [4.5]< 2.2$ mag. The range of colors,  $0.17<[3.6]-[4.5] < 1.5$, is also comparable to the values discussed by K19 (cf.\ their Figure~5). Footnotes in column~1 of Table~\ref{table:periodicvars} mark the five variables with \HST\/ detections or possible detections, as we discuss later in this section.

%It is not surprising (based on the discussion in the introduction to this section) that these four variables are among the bluest objects in the table; their mean colors lie in the range $0.17<[3.6]-[4.5]<0.76$ (see below). 

%{\bf NOTE Jacob do you  want  to say something about the accuracy of the colors, perhaps quote only one decimal? }

Most of these variables have extremely red \Spitzer\/ colors, indicating very low effective temperatures and circumstellar dust, a consequence of high mass-loss rates. For example, colors of $[3.6]-[4.5]\simeq0.6$, 1.0, and 1.5 correspond to blackbody temperatures of $\sim$1000, 800, and 500~K, respectively (cf.\ Figure~7 in K19).

At such very low effective temperatures, we generally would not expect to detect these objects with our one-orbit \HST\/ observations. In a typical case of a variable with an apparent IRAC channel~2 magnitude of $[4.5]\simeq16.5$, and the nominal exposure times we used for our \HST\/ observations (\S\ref{sec:hstobservations}), the WFC3 ETC indicates that the source would not be detected ($\rm S/N<5$) in the \HST/WFC3 $I$, $J$, or $H$ bandpasses if its blackbody temperature were below $\sim$750~K\null. Above 750~K it would be detectable only at $H$\null. For a temperature above 850~K, it would be detected also at $J$\null. For detection at $I$, a source with $[4.5]=16.5$ would have to be hotter than $\sim$1275~K\null. However, the use of blackbodies in these estimates is only a rough approximation, since OH/IR stars have strong molecular absorptions, due to H$_2$O and CO in particular, which influence the $[3.6]-[4.5]$ and other colors. Moreover, these estimates are optimistic, since they neglect background light and dust extinction.

%{\bf Note Do we want to leave the above in place? While it is a good first test, the OH/IR stars have very strong molecular absorptions, $H_2 O$ and CO in particular, which influence the [3.6]-[4.5] and other colors. The paragraph below is an alternative, but we could leave both in. HEB: the material in question has been moved to the end of section 8.}

%{\bf NOTE: the following paragraph was provided by Patricia, but I'm not sure where, or if, to fit it in; for example, it doesn't actually mention absolute magnitudes.}

It is also possible to estimate the anticipated \HST\/ magnitudes, using what is known about the LMC OH/IR stars that are illustrated in Figure~\ref{fig:plrelation}, and assuming that our variables are similar.  We use the $JHK(L)$ lightcurves for these OH/IR stars from \citet{Whitelock2003} and $I$-band light curves from \citet{Soszynski2009}; note, however, that the very reddest LMC sources were not detected in the ground-based $I$- or $H$-bands.
Unfortunately there is only single-epoch \Spitzer\/ photometry for these, so we do not have mean \Spitzer\/ magnitudes or colors. The single-epoch colors are in the range $0.3<[3.6]-[4.5]<0.8$ for the AGB stars, and $[3.6]-[4.5]=0.9$ for the single supergiant. The AGB amplitudes are in the ranges $1.5<\Delta H <1.9$ and $2.8<\Delta I <4.0$, while  the supergiant has  $\Delta H=0.4 $ and $\Delta I =1.5$. The very large amplitudes of the AGB variables, particularly at the shortest wavelengths, complicates any predictions of \HST\/ flux. Furthermore the \citet{Goldman2018} comparison of the SMC with the LMC suggests that the mass-loss rates of these O-rich stars are a function of metallicity; therefore it seems likely that the AGB stars under discussion will have thicker shells than similar stars in the LMC and thus fainter magnitudes at \HST\/ wavelengths. 
Nevertheless, these provide a useful comparison and the following ranges of colors found for the LMC sources are used to estimate the expected \HST\/ magnitudes for our sources:
 $7.6<[I-4.5]<10.7$ and $2.6<H-[4.5]<4.8$. 
However, additional dust extinction is possible and will make the \HST\/ magnitudes even fainter than these values would predict.

We also note that there are significant differences between the bandpasses of the ground-based $IJHK$ filters and the similar filters used by \HST\null. Although these differences are potentially problematic, particularly for cool stars with extended atmospheres where strong molecular absorption dominates the colors, they are not important for the comparisons made here, where the large-amplitude variations dominate.

\begin{figure*}[ht]
\centering
\includegraphics[scale=.375]{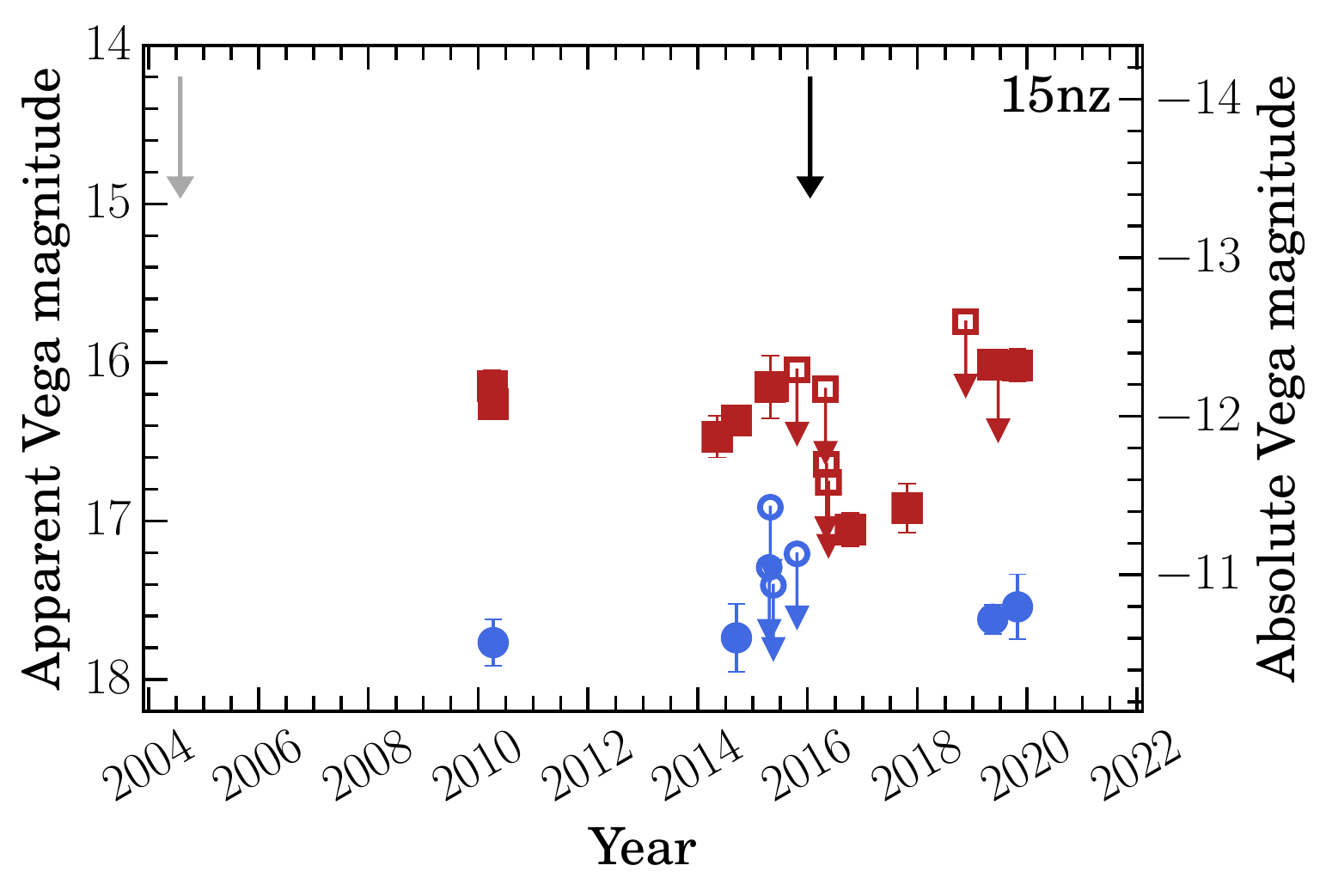}\hfill
\includegraphics[scale=.375]{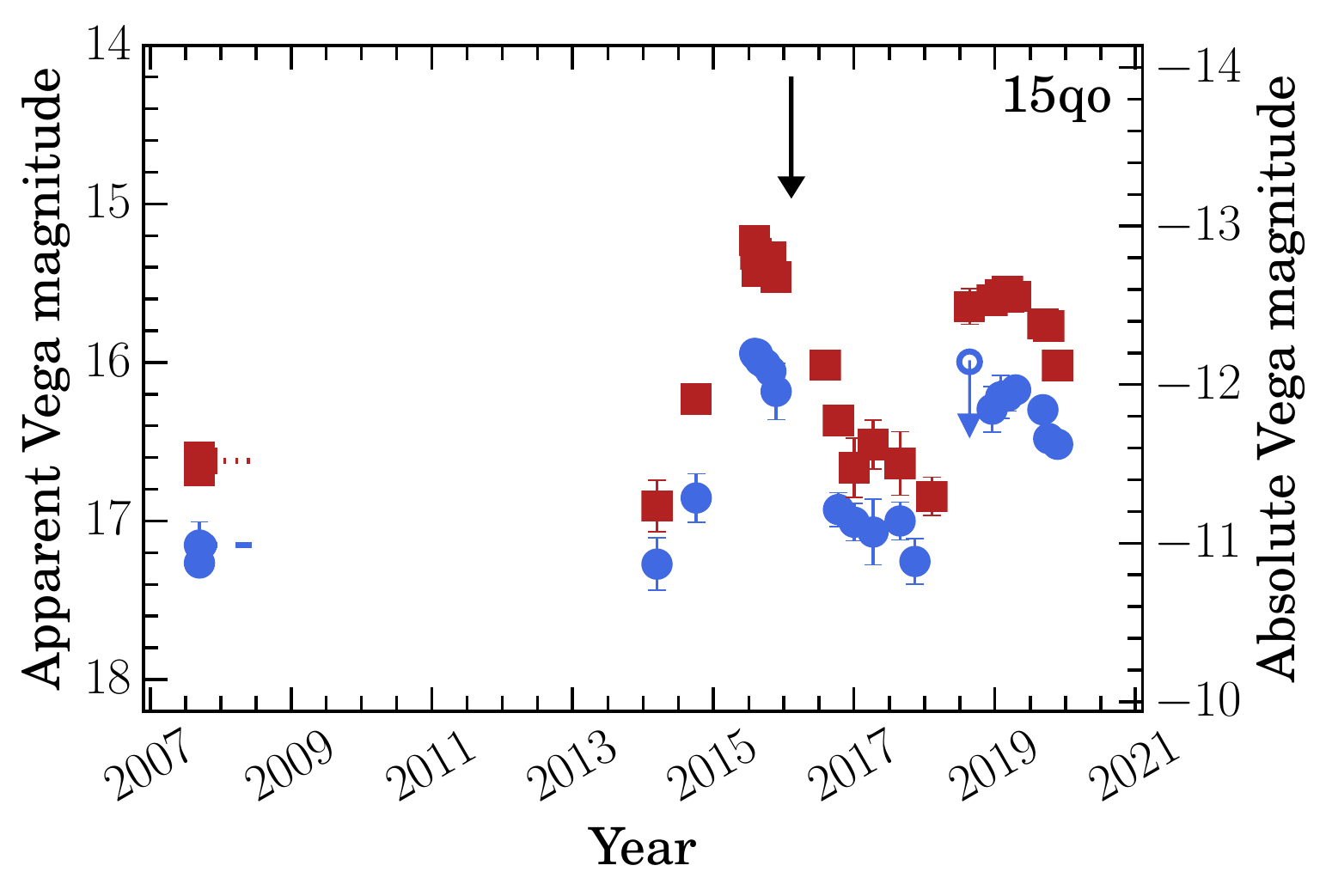}\hfill
\includegraphics[scale=.375]{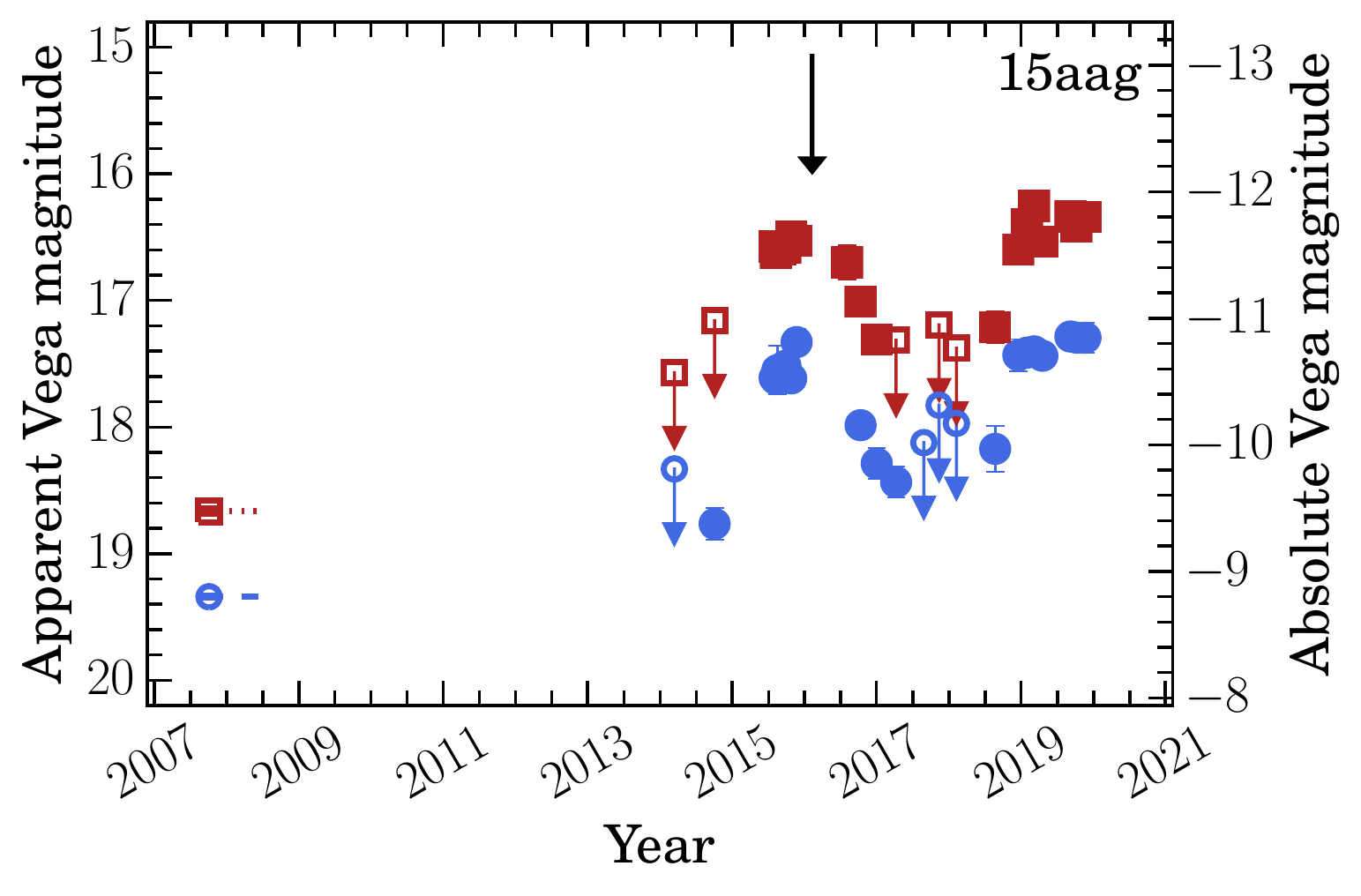}\hfill
\includegraphics[scale=.375]{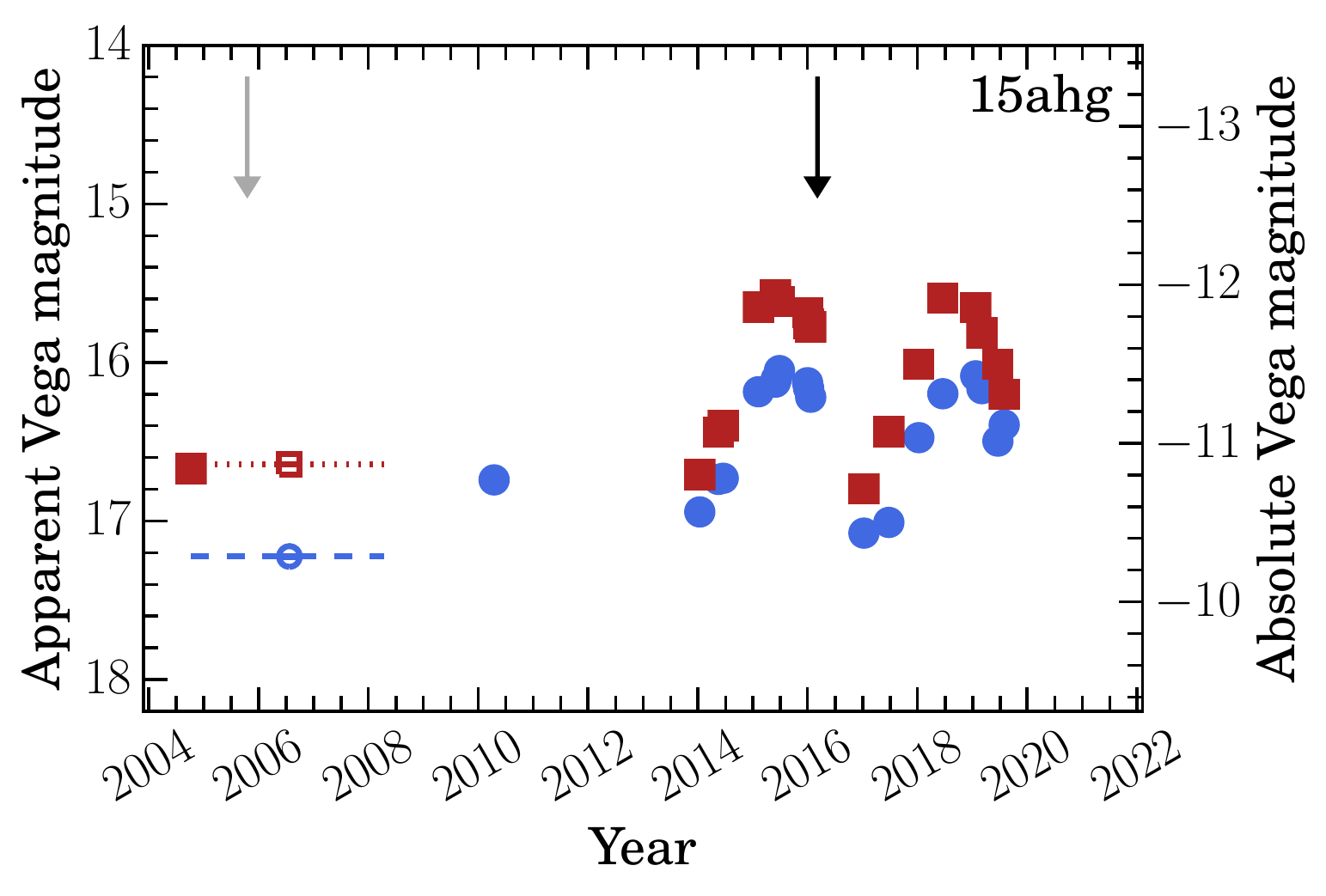}\hfill
\includegraphics[scale=.375]{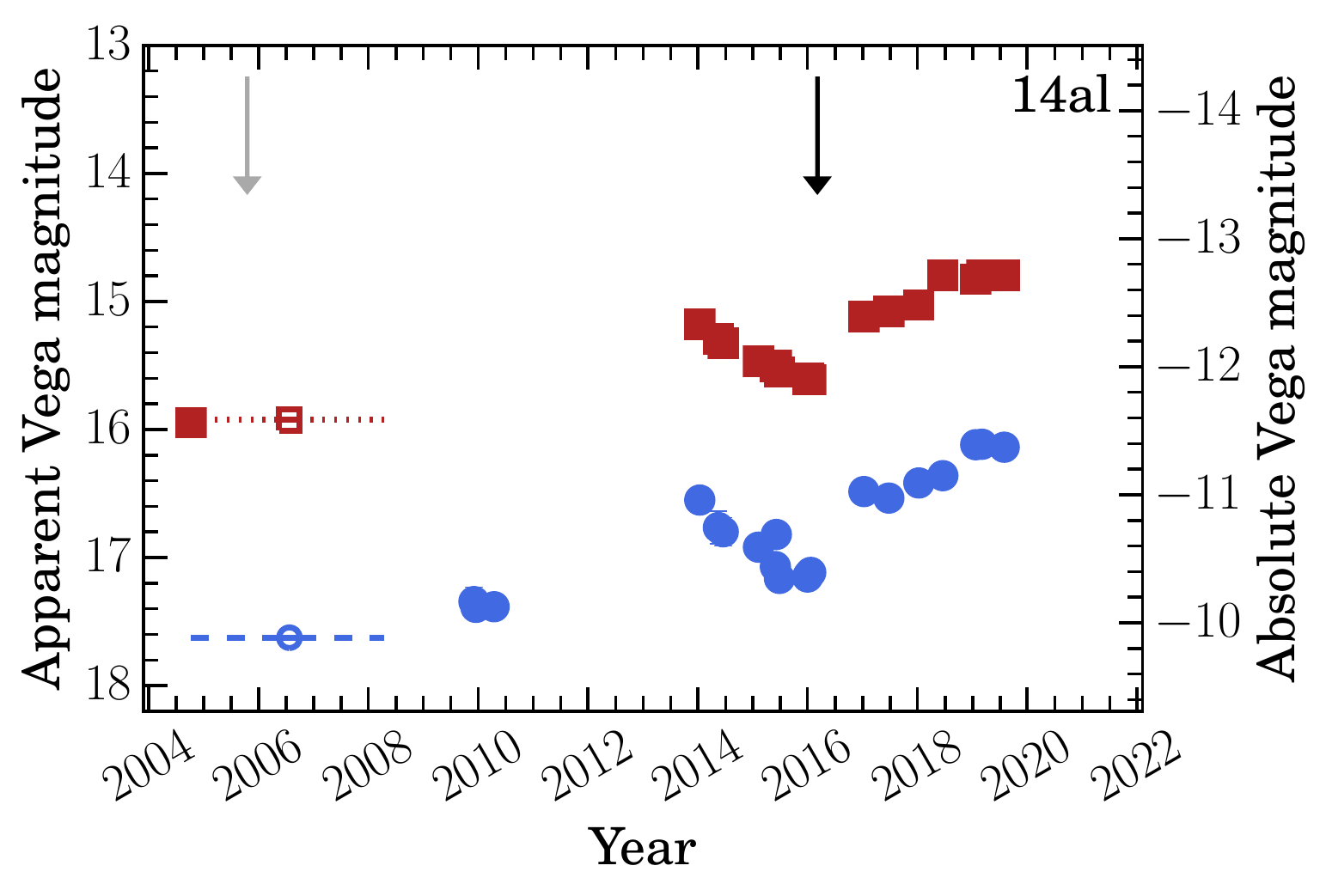}\hfill
\includegraphics[scale=.375]{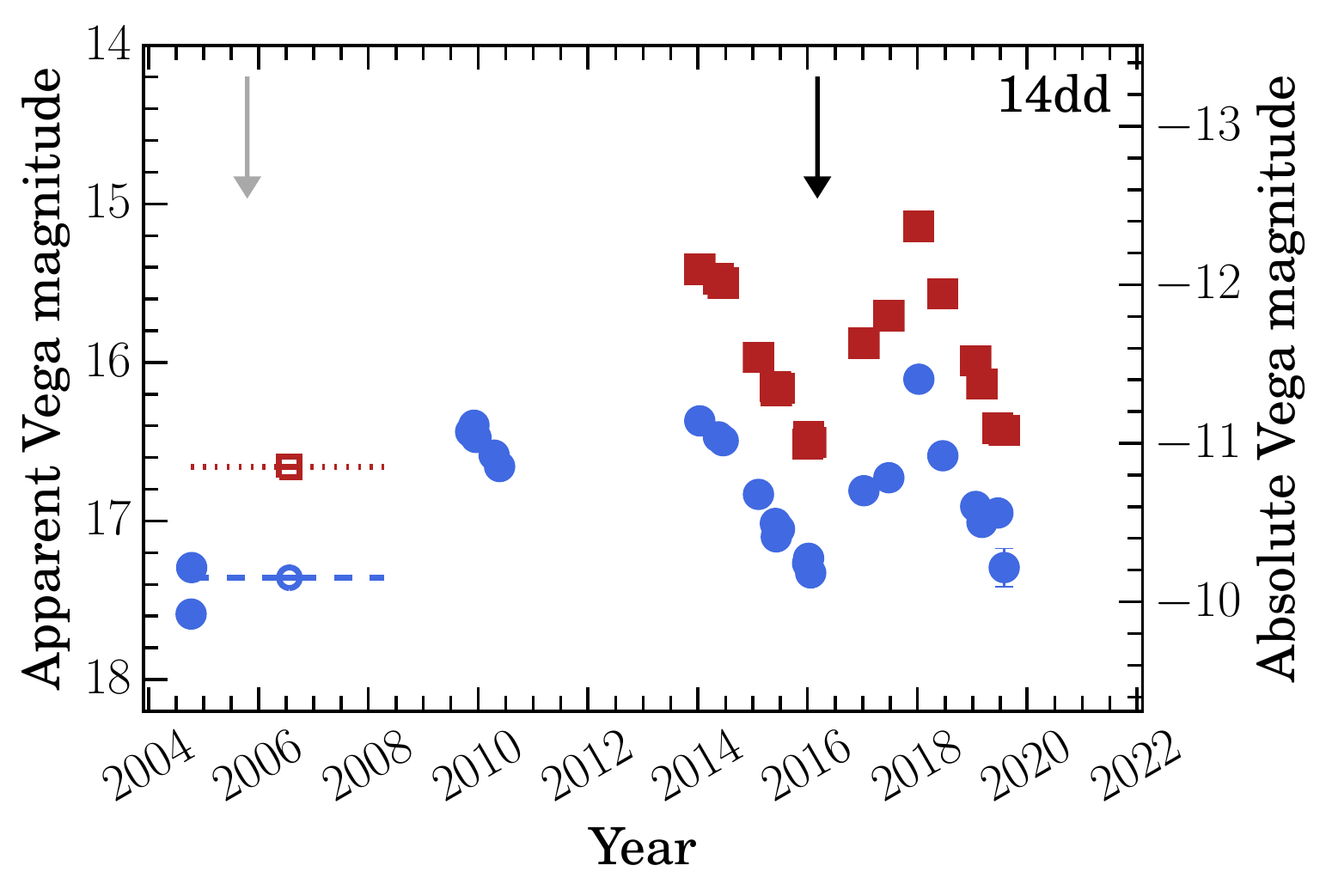}\hfill
\includegraphics[scale=.375]{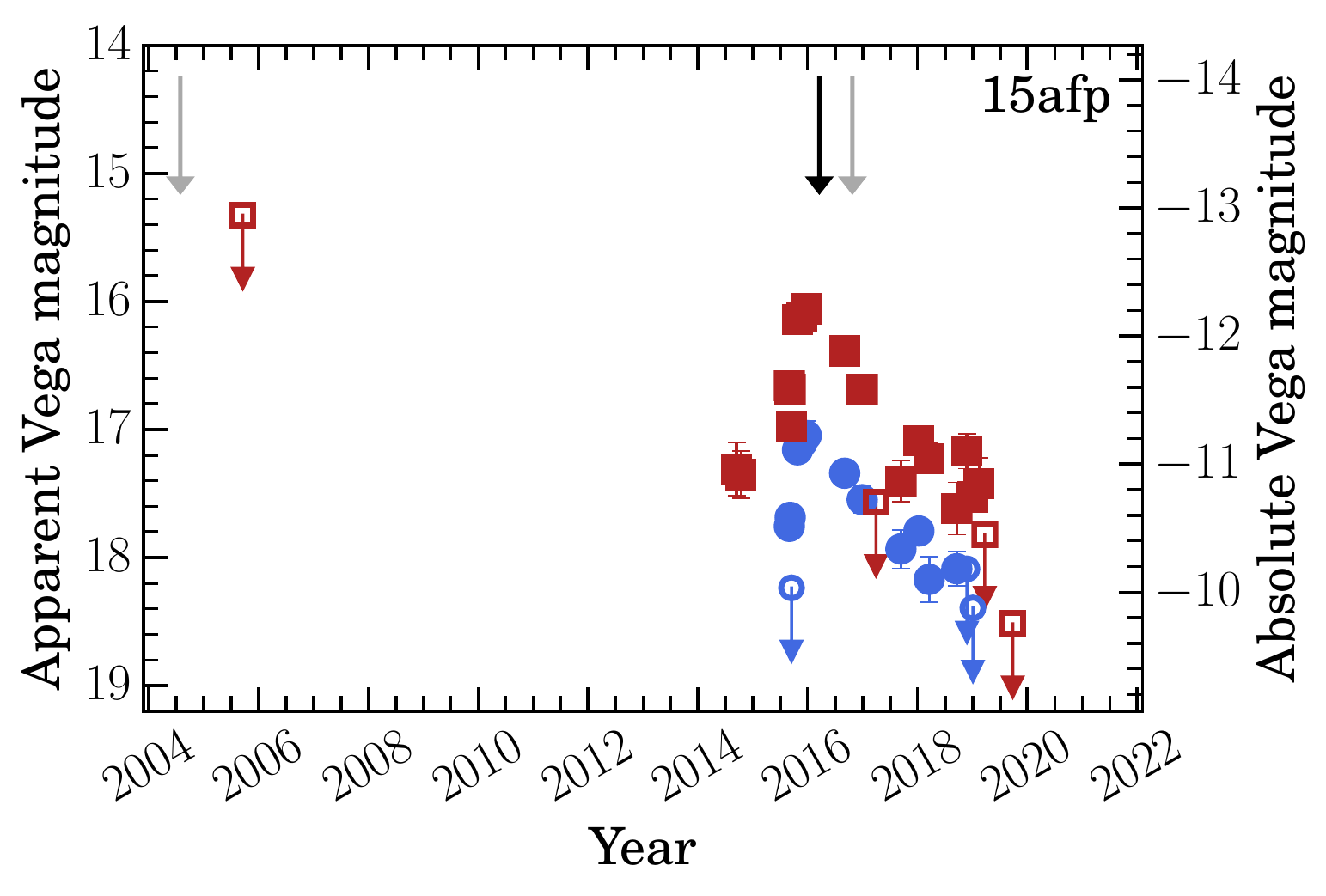}\hfill
\includegraphics[scale=.375]{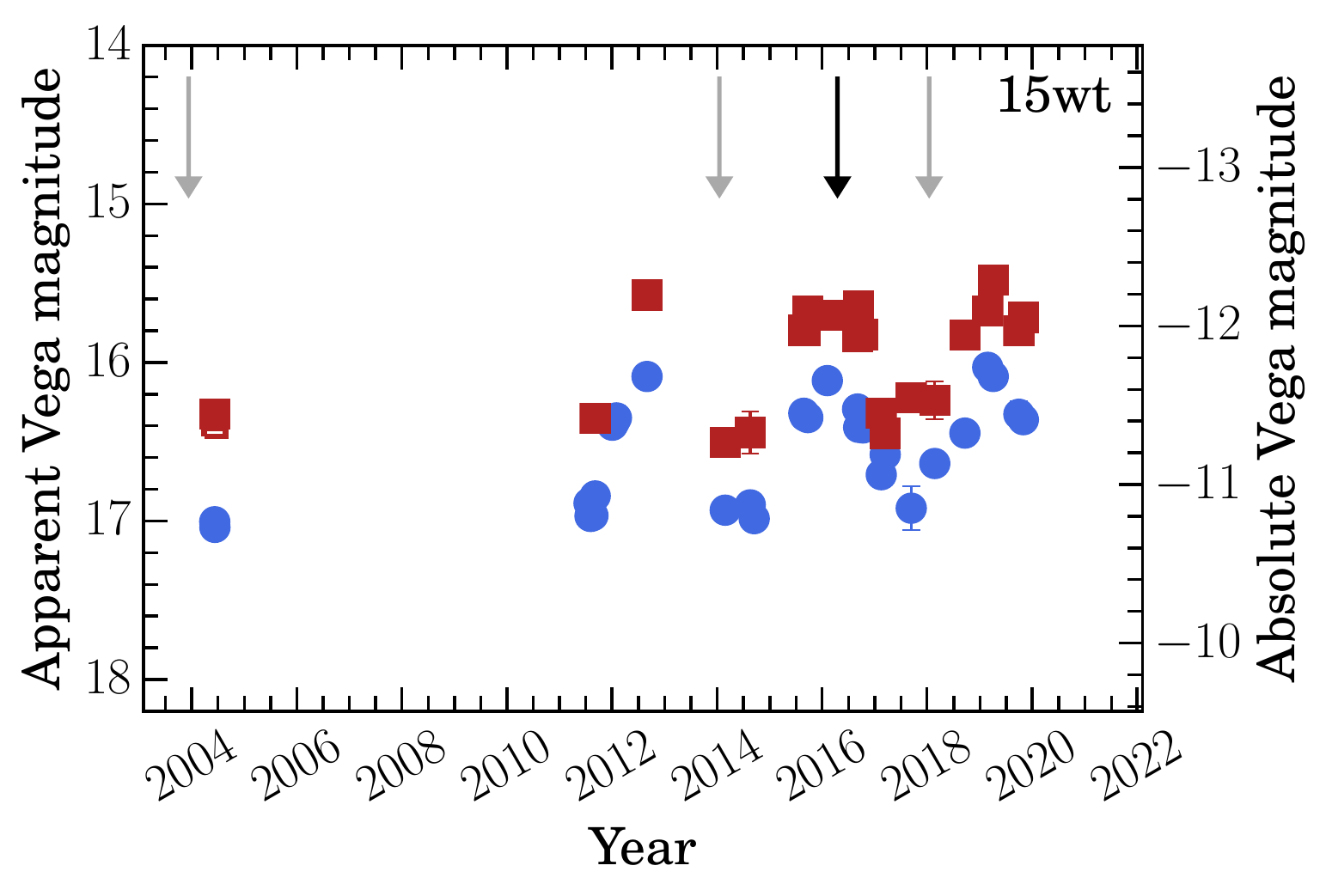}\hfill
\includegraphics[scale=.375]{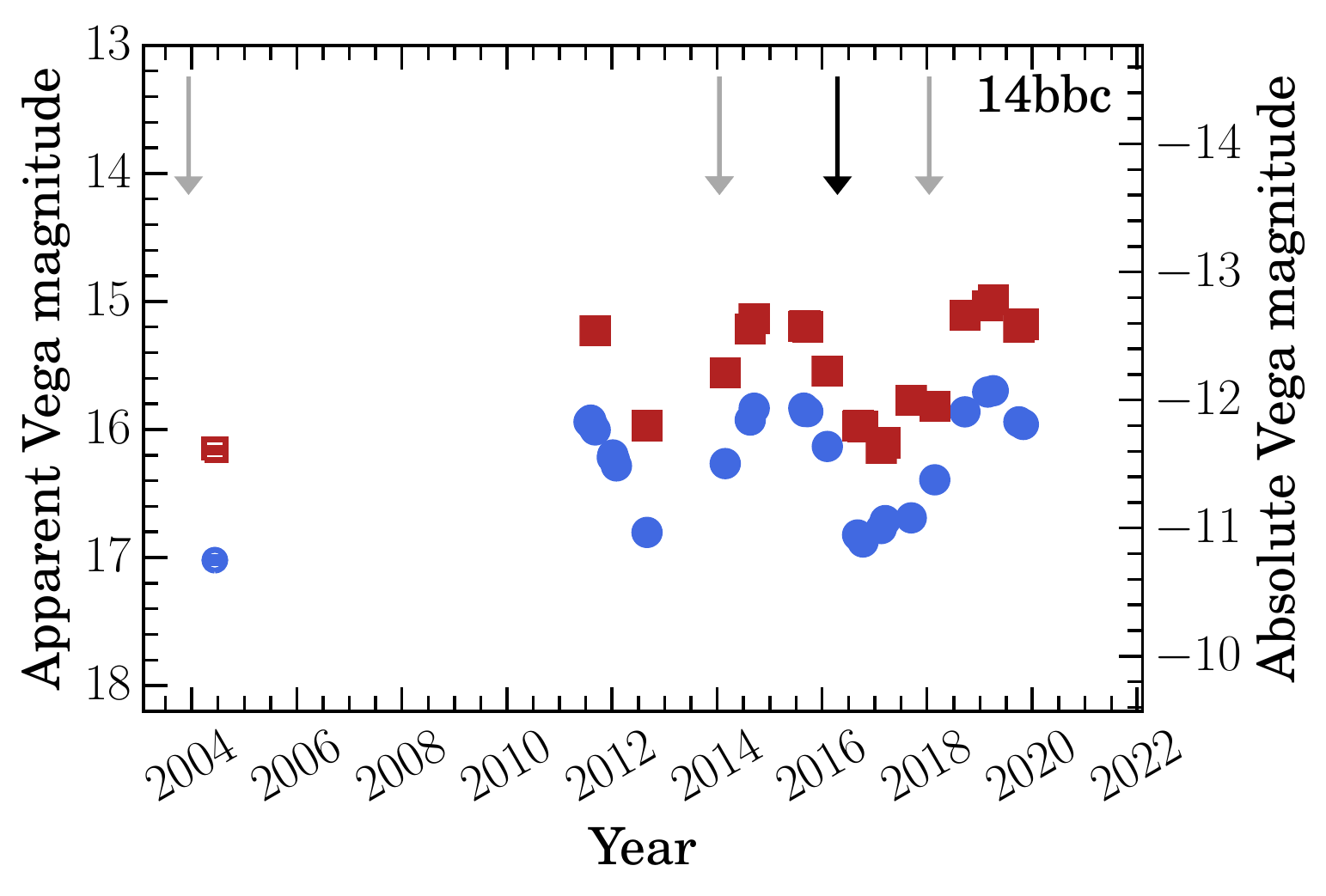}\hfill
\includegraphics[scale=.375]{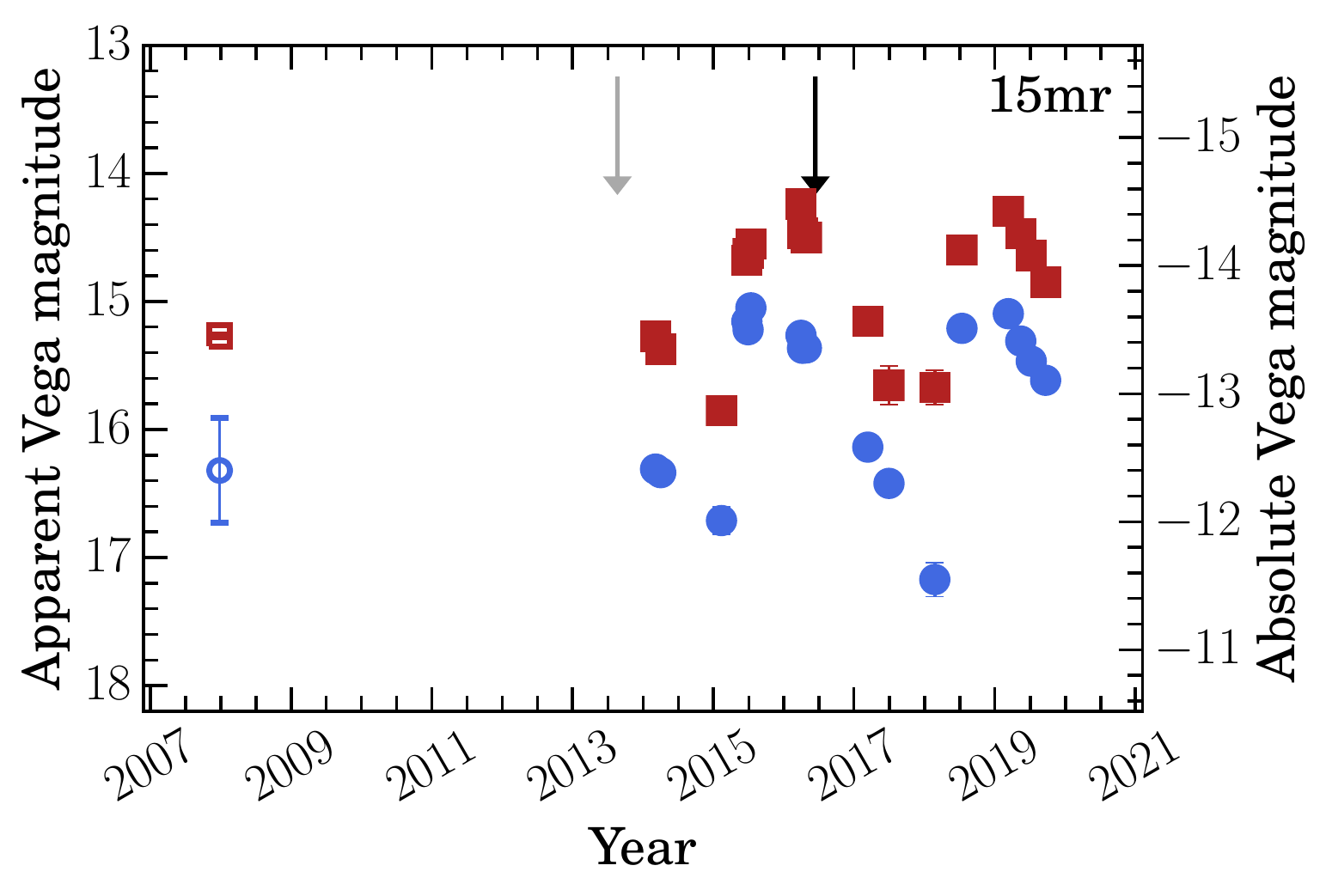}\hfill
\includegraphics[scale=.375]{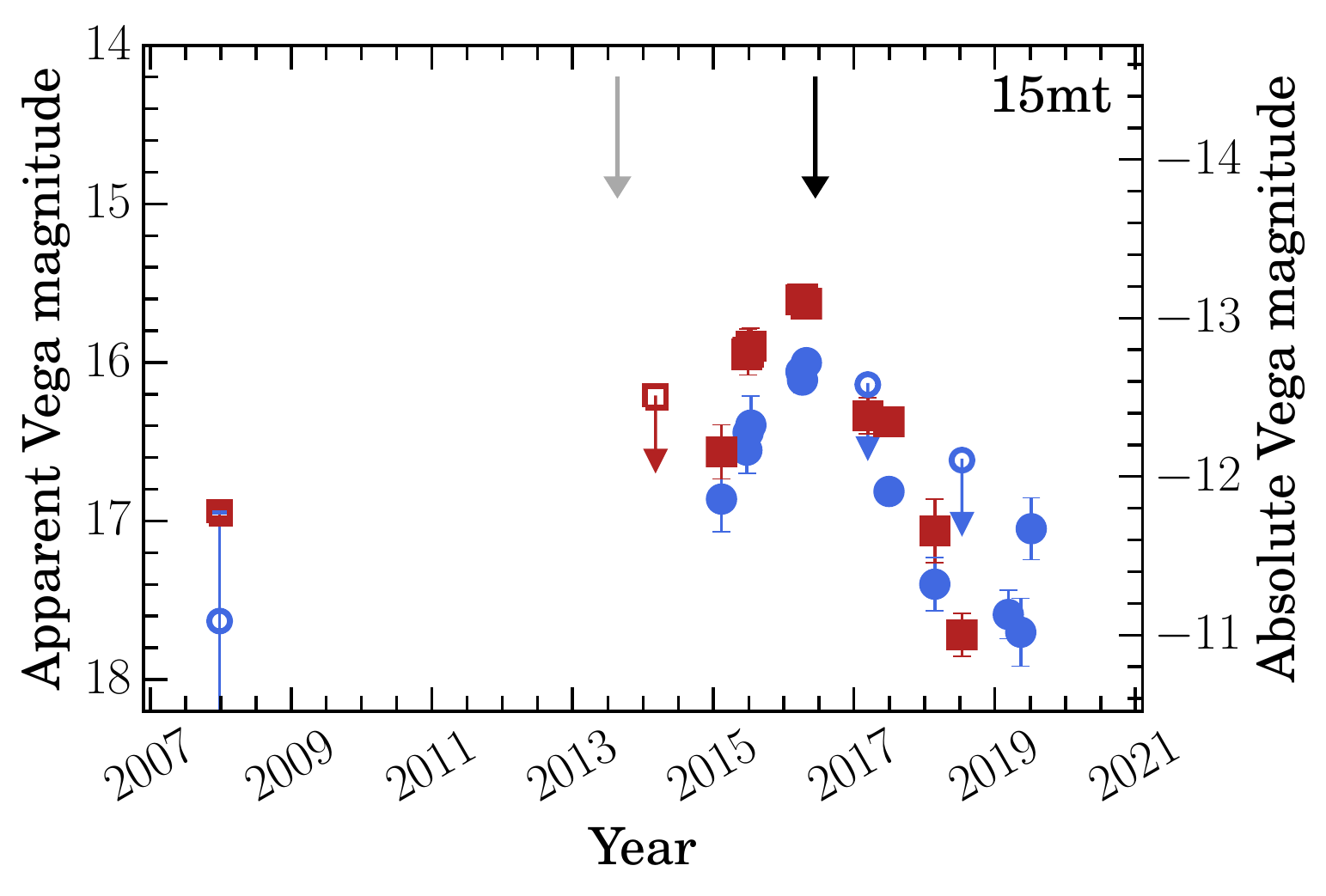}\hfill
\includegraphics[scale=.375]{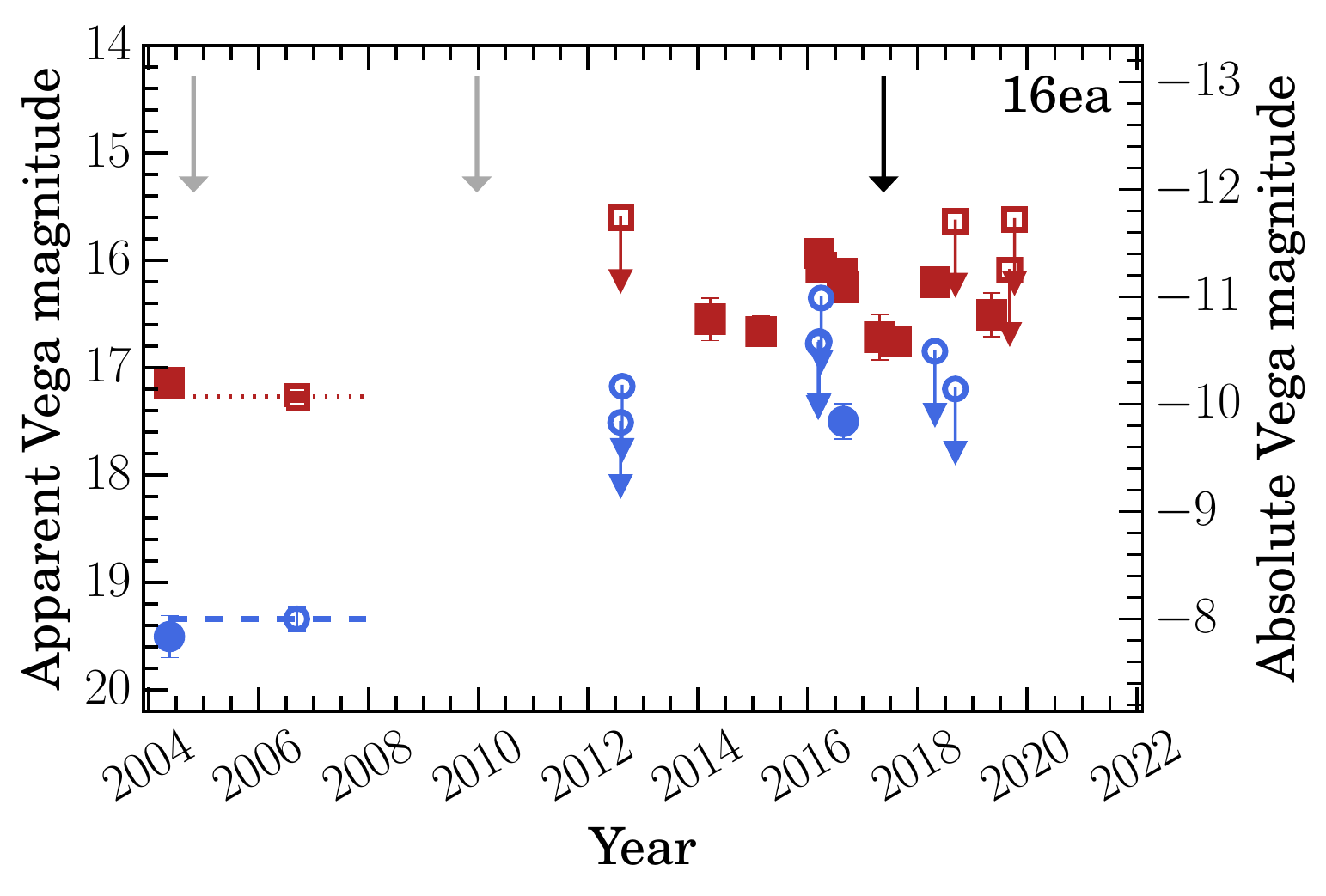}\hfill
\caption{
\Spitzer\/ IRAC light curves of periodic and suspected periodic variables from difference imaging. Plotting symbols as in Figure~1. Baseline magnitudes (aperture photometry on reference images) are shown as open symbols, where the dashed horizontal bars indicate the range of epochs that were included in the reference image stacks.}
\label{fig:periodic_array}
\end{figure*}

We now discuss the individual periodic and likely periodic variables that we imaged with \HST\null. Their \Spitzer\/ light curves are collected in Figure~\ref{fig:periodic_array}. 

%(note that here [4.5] is on the Vega magnitude system while $H$ and $I$ are on the AB system): $9.0<[4.5]-I<12.1$ and $3.7<[4.5]-H<6.2$. 

%{\bf NOTE: need to adjust above and below to the Vega scale.}

%In the following estimated $H$-band and $I$-band magnitudes are provided in the AB system for comparison with \HST\/ values. 

%In this section, we first present the results of \HST\/ imaging of each object. We then give a general discussion of the properties of this class of periodic IR variables.

%\begin{figure}[ht]
%\centering
%\includegraphics[scale=.525]{Spitzer_lcs_hst_15nz.pdf}
%\includegraphics[scale=.525]{Spitzer_lcs_hst_15qo.pdf}
%\includegraphics[scale=.525]{Spitzer_lcs_hst_15aag.pdf}
%\caption{ \Spitzer\/ IRAC light curves of quasi-periodic variables. Plotting symbols as in Figure~1.
%\label{fig:quasiperiodic}}
%\end{figure}

\subsection{15nz}\label{sec:15nz}

SPIRITS 15nz was announced by \citet{Jencson2015} as a possible IR transient in M83 ($d\simeq4.7$~Mpc), based on our initial few SPIRITS observations. These data showed a slowly rising [4.5] luminosity during 2014 and early 2015. In response to this announcement, a separate team (GO-14463, PI B.~McCollum) obtained \HST\/ imaging of the site. Since 2015 we have accumulated additional IR observations, and there are also archival \Spitzer\/ data obtained in 2010. The IR light curves of 15nz are shown in the top-left panel of Figure~\ref{fig:periodic_array}. The data at [4.5] indicate a period close to 1600~days. There are insufficient [3.6] detections to analyze for periodicity, but they can be used for an estimate of the color, which at $[3.6]-[4.5] \simeq 1.5$ makes 15nz the reddest of the periodic variables discussed in this section.

%These data reveal an apparently periodic variation, which has gone through a little over one cycle since the first SPIRITS observation in 2014. The data can be fit with a period of about 1600  days. 

%{\bf NOTE: actually, there appears to be an earlier observation, so we actually have about 2 cycles.}

%At this period, and a mean absolute magnitude of about $M_{[4.5]}=-11.8$, 15nz's location in the period-luminosity diagram agrees well with that found by K19 (their Figure~3) for a sample of over 400 luminous long-period IR variables discovered in the SPIRITS survey. The IR color of 15nz is very red, at about $[3.6]-[4.5]\simeq1.4$ (?).

As noted in the introduction to this section, detection of a source as cool as 15nz at \HST\/ wavelengths is not expected. We registered a \Spitzer\/ frame taken at maximum brightness with \HST\/ frames to find the precise location of the variable in the latter. At this site in the disk of M83 there is a dense sheet of faint stars. There are several faint objects within a 3$\sigma$ error circle, but comparing $I$-band frames taken at three epochs (2009, 2010, 2016), we see no significant variation of any of these stars. Only two epochs of \HST\/ imaging at $J$ and $H$ are available (2009, 2016), but again we see no variation of any detected stars within the error circle. It thus appears that, as was expected, 15nz has no optical or near-IR \HST\/ counterpart at $I$, $J$, and $H$\null. From a comparison with the OH/IR stars in the LMC, we would expect 15nz to have 
$I>27.4$ and $H>21.5$, or probably much fainter given how red it is.  

\null

\medbreak

\subsection{15qo and 15aag}

SPIRITS 15qo was likewise announced by \citet{Jencson2015} as an IR transient, lying in NGC~1313 ($d\simeq4.2$~Mpc). A second transient or variable in the same galaxy, SPIRITS~15aag, was announced by \citet{Jencson2016a}; it lies only $54''$ away from 15qo. We adjusted our \HST\/ pointing so as to capture both of these objects in the same image frames.

The \Spitzer\/ light curves of these two variables are shown in the top-middle and top-right frames of Figure~\ref{fig:periodic_array}. As in the case of 15nz, both objects were caught in rising phases in the first few SPIRITS observations, and were announced as transients. However, as the light curves show, they later proved to be periodic variables. Curiously, their pulsation periods are nearly the same, both being about 1230 days, and they are also nearly in phase. They are, however, definitely distinct objects.

%{\bf NOTE (PAW): I find it  "disturbing" (or very interesting) that the periods of 15qo and 15aag are identical. They are also in-phase with each other. The mags and colors are different but .... HEB: I changed the wording above a little, but there's no doubt they are different objects... although maybe Jacob wants to check this to be absolutely positive.}

%{\bf NOTE: we should make the time intervals the same in Figure 9 for the two objects in NGC 1313}

We registered a \Spitzer\/ image showing both objects near maximum light with archival \HST\/ images as well as the new images obtained in our program. At the site of 15qo there is an extremely rich star field, lying in an actively star-forming region in NGC~1313. Several faint stars fall within a 3$\sigma$ error circle, including a bluish star near the center. None of these stars appear to be significantly variable in $I$-band frames taken in 1994, 2003, 2004 (two epochs), and 2016. Our $J$ and $H$ frames show no very red star at the site. We conclude that there is no detectable optical/near-IR counterpart. 

%{\bf NOTE: the final sentence above was added by PAW, but it needs to be moved to the discussion at the end of this section. }

% and a comparison with the LMC OH/IR stars suggests we might expect $H\sim 20$.

The results are similar for 15aag, which is significantly redder than 15qo. It also lies in a rich field, not far from several \ion{H}{2} regions. There is again a faint star within the 3$\sigma$ error circle, which showed no significant variability in $I$-band frames taken in 2001, 2003, 2004 (two epochs), and 2016. This star is detected in our $J$ and $H$ observations, but the source is not extremely red; unfortunately these are the only frames in $J$ and $H$ in the \HST\/ archive. As in the case of 15qo, we find no convincing optical/near-IR counterpart.

%{\bf NOTE (HEB): do we need to show any pictures of the fields with the error circles? Since we aren't claiming an optical/NIR counterpart, such images might not be very useful. For now, I am just giving verbal descriptions.}

\subsection{15ahg, 14al, and 14dd}\label{sec:15ahg}

SPIRITS 15ahg is another object announced as an IR transient by \citet{Jencson2016a}, lying in the M81-Group spiral galaxy NGC~2403 ($d\simeq3.2$~Mpc). Two previously identified variables in the SPIRITS database, 14al and 14dd, lie close to the position of 15ahg, and we adjusted the telescope pointing so that we could include all three in the \HST\/ frames. The site of these variables lies near a spiral arm on the northwest side of NGC~2403, in an extremely rich star field. Several giant \ion{H}{2} regions are nearby.

\begin{figure*}[ht]
\centering
\includegraphics[width=4.5in]{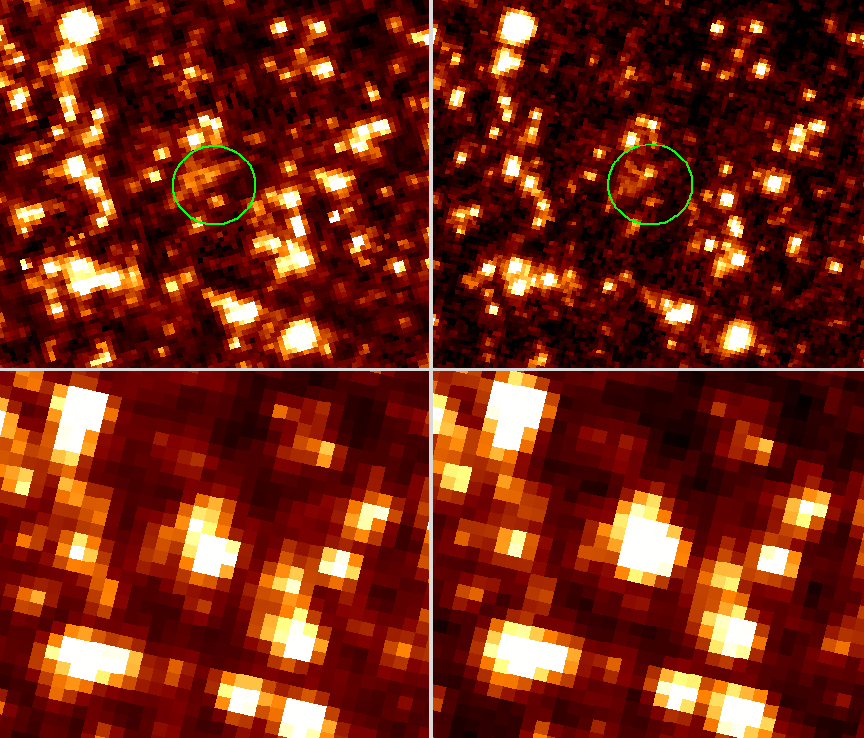}
\caption{
False-color renditions of \HST\/ images of the site of the periodic variable SPIRITS 15ahg in the nearby galaxy NGC~2403. {\it Top row:} $I$-band frames taken in 2005 (left) and 2016 (right). The green circles mark 3$\sigma$ error positions of the \Spitzer\/ variable. Just north of the center is a star that brightened significantly and is a likely optical counterpart of 15ahg. {\it Bottom row:} \HST\/ images of the site in 2016, taken in $J$ (left) and $H$ (right). The candidate counterpart is bright at $J$ and very bright at $H$\null. Height of each frame is $3\farcs6$.
\label{fig:15ahg}
}
\end{figure*}

The \Spitzer\/ light curves of 15ahg are shown in the left panel in the second row of Figure~\ref{fig:periodic_array}. Its period is about 1160 days and it is relatively blue, with a mean $[3.6]-[4.5]=0.36$
There are two $I$-band \HST\/ observations of this site in the archive, one taken in 2005 (GO-10402, PI R.~Chandar), and the other our frame obtained in 2016. We registered these images with a \Spitzer\/ IRAC channel~2 frame from the SPIRITS program, taken near maximum brightness of 15ahg. Inside the 3$\sigma$ error circle in the $I$-band frames is a faint star that brightened significantly from 2005 to 2016, consistent with the phasing of the IR variability shown in Figure~\ref{fig:periodic_array}. In the 2005 image, the variable lies in a blended clump of several faint stars, which is possibly a compact sparse cluster. In 2016 it had risen to an \HST\/ $I$ magnitude of about 24.5 (Vega scale, based on aperture photometry relative to several HSC stars in the field). This star is optically very red, and is well detected in our $J$ and $H$ frames. The HSC Vega-scale $J$ and $H$ magnitudes for the object are 20.79 and 18.94, respectively. Thus there is little doubt that this object is the optical/near-IR counterpart of the \Spitzer\/ variable. The four panels in Figure~\ref{fig:15ahg} show small postage stamps from the two $I$-band frames (top row), and from our $J$ and $H$ images (bottom row). The astrometric 3$\sigma$ error circle is shown in green in the top two frames. 

A comparison with the LMC  OH/IR stars (see introduction to this section) would predict 
$18.7 < H < 20.9$ and $1.1<(J-H)<2.0$ for 15ahg.
These are consistent with the measured $H\simeq18.9$ and $J-H\simeq1.8$ and support our suggestion that this object is a pulsating AGB variable.

The light curves of 14al and 14dd are shown in the second and third panels in the second row of Figure~\ref{fig:periodic_array}.
14al is extremely red, with a mean $[3.6]-[4.5]=1.4$, and its {approximate} period, 2160 days (5.9~years), is the longest of any of the periodic variables. Its peak-to-peak amplitudes, $\Delta [3.6] \simeq 1.0$ and $\Delta [4.5] \simeq 0.8$~mag, although slightly less than those of the other periodic variables, are still large.   14al lies in a rich star field. There are several faint stars at the site, but none varied significantly in the $I$ frames from 2005 and 2016. None of them are bright at $J$ and $H$. We conclude there is no convincing optical/near-IR counterpart.

14dd is another likely periodic variable, with a period of 1420~days and a red color of $[3.6]-[4.5]\simeq1.0$. As in the case of the nearby 14al, there are no variable objects at the site in the two available \HST\/ $I$-band frames, nor any conspicuous objects within the error circle at $J$ and $H$\null. Again, we find no credible optical/near-IR counterpart of this very cool object.

\subsection{15afp}

The IR light curves of SPIRITS\,15afp are shown in the left-hand panel in the third row of Figure~\ref{fig:periodic_array}. This object rose by over one magnitude from the first SPIRITS observation in 2014 to a peak in late 2015, leading us to announce it as a candidate transient \citep{Jencson2016a}. We triggered our \HST\/ imaging in 2016 March. 15afp lies in a spiral arm of the face-on and actively star-forming galaxy NGC\,6946 ($d\simeq4.5$~Mpc). This variable could be periodic, but our observations cover less than one cycle of a period of around 1650 days. We also note our photometry is likely contaminated by another nearby, but unrelated, variable source. The amplitudes and color are similar to the other periodic variables discussed in this section. However, we cannot rule out that the object is a slow SPRITE transient, based on our relatively limited data.

We registered a \Spitzer\/ image taken at the maximum brightness of 15afp with an ACS $I$-band frame obtained in 2016 October (GO-14786, PI B.~Williams). The site lies in a dense sheet of stars, with moderately high extinction. There are several faint stars within a 3$\sigma$ error circle. None of them appeared to vary between the ACS frame and our own WFC3 $I$-band frame taken 7~months earlier, nor in comparison with a shallow ACS frame obtained in 2004 (GO-9788, PI L.~Ho). There are no conspicuous sources at the site in our $J$ and $H$ frames. On the basis of these fairly limited data, we see no convincing evidence of an optical or near-IR counterpart.

\subsection{15wt and 14bbc}\label{sec:15wt}

The \Spitzer\/ light curves of SPIRITS\,15wt are plotted in the middle panel in the third row of Figure~\ref{fig:periodic_array}. 15wt rose by nearly 1~mag over the first two years of SPIRITS observations of the host galaxy, NGC~7793 in the nearby Sculptor Group ($d\simeq3.6$~Mpc). This slow eruption of an apparent transient prompted us to trigger our \HST\/ imaging, obtained on 2016 April~18. However, our subsequent observations, as well as archival data, clearly reveal that 15wt is actually a periodic variable of the type discussed by K19, with a well-determined period of 1190~days. It is relatively ``blue,'' with a mean color of $[3.6]-[4.5]\simeq0.5$.

We registered an archival \HST/ACS $I$-band image of NGC~7793 obtained on 2003 December~10 (GO-9774, PI S.~Larsen) with a \Spitzer\/ frame taken near maximum brightness of 15wt, in order to locate the site in the \HST\/ frames. We then registered the ACS frame with WFC3 $I$-band frames taken on 2014 January~18 (GO-13364, PI D.~Calzetti) and on 2016 April~18 in our own program. Just inside the 3$\sigma$ registration error circle is a star that was not detected in 2003 and 2014, but had brightened in 2016. This object is very bright in the $J$ and $H$ frames that we obtained at the same time as our $I$ image in 2016, and the date of the \HST\/ imaging is close to the time of maximum brightness in the \Spitzer\/ frames. Thus we conclude that the optical/near-IR object is the counterpart of the IR variable. In confirmation, there are also archival WFC3/IR frames of the site taken on 2018 January~16 in $J$ and $H$ (GO-15330, PI D.~Calzetti), when the IR variable was near minimum light, as shown in Figure~\ref{fig:periodic_array}. The counterpart is significantly fainter in these frames than in 2016.

\begin{figure*}[ht]
\centering
\includegraphics[width=4.5in]{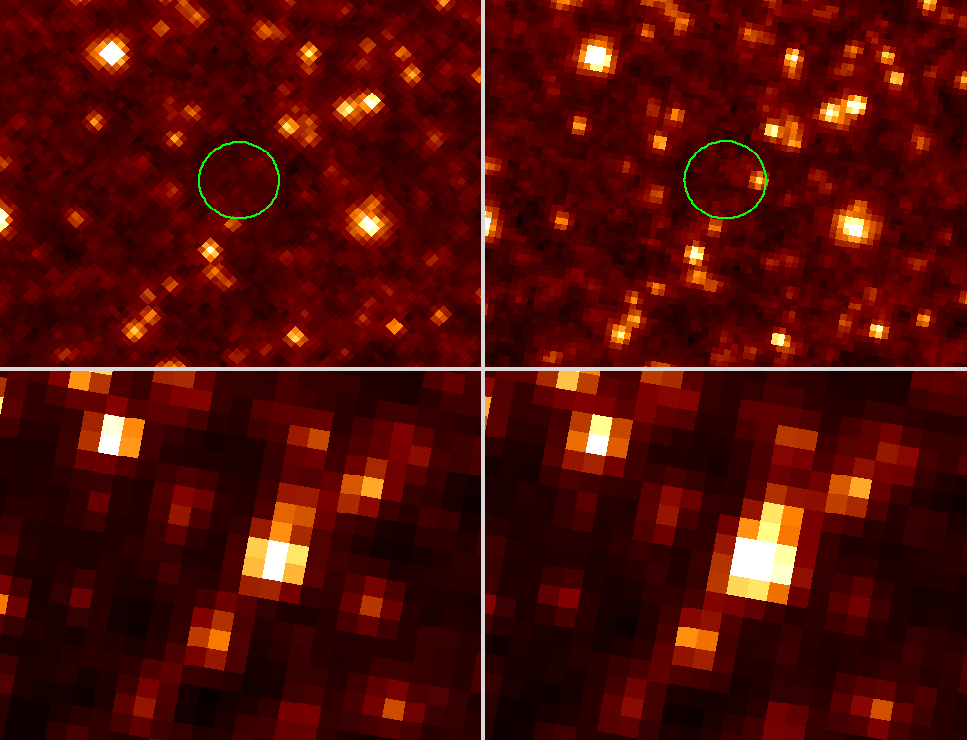}
\caption{
False-color renditions of \HST\/ WFC3 images of the site of the periodic IR variable SPIRITS 15wt in the nearby galaxy NGC~7793. {\it Top row:} $I$-band frames taken in 2014 (left) and 2016 (right). The green circles mark 3$\sigma$ error positions of the \Spitzer\/ variable. Inside the circles is a star that brightened significantly around the time of maximum IR luminosity, and is a likely optical counterpart of 15wt. {\it Bottom row:} near-IR \HST\/ images of the site in 2016, taken in $J$ (left) and $H$ (right). The counterpart is bright at $J$ and very bright at $H$\null. Height of each frame is $2\farcs4$.
\label{fig:15wt}
}
\end{figure*}

Figure~\ref{fig:15wt} presents false-color renditions of the $I$-band frames from 2014 and 2016 in the top row, revealing the star just inside the error circle that brightened in 2016. The bottom row shows that this star is very bright in $J$ and $H$, in the WFC3 frames taken during our same \HST\/ visit in 2016. The HSC gives the following magnitudes (Vega scale) for this object at the 2016 epoch: $I=24.23$, $J=20.25$, and $H=18.70$. The LMC OH/IR stars (see introduction to this section) would have $26.7>I>23.6$ and $20.8>H>18.6$ 
at this distance, consistent with our suggestion that the \Spitzer\/ variable and its \HST\/ optical/near-IR counterpart is a pulsating AGB star. 

%{\bf NOTE: HEB needs to check the 2018 frames of 15wt pointed out by Jacob.}

%{\bf NOTE: the 15wt light curve needs gray arrows at 2003-12-10 and 2014-01-18. Also needs one added at 2018-01-16 }

%{\bf NOTE: the 14bbc light curve also needs gray arrows at 2003-12-10 and 2014-01-18. But the gray arrow in 2011 should be removed. Also needs gray arrow at 2018-01-16}

%{\bf NOTE: should we present an SED, discuss this source further??}

Another IR variable, 14bbc, had been discovered earlier during the SPIRITS program. It lies close to the position of 15wt, and we adjusted the \HST\/ pointing for our triggered observation in 2016 so as to include both 15wt and 14bbc in the images. The \Spitzer\/ light curves of 14bbc are shown in the right-hand panel in the third row of Figure~\ref{fig:periodic_array}. This is another periodic variable, which has undergone two full cycles during the archival and SPIRITS \Spitzer\/ observations (plus an earlier observation in 2004). Its period is 1500~days, and its color is ($[3.6]-[4.5]\simeq 0.7$).

\begin{figure*}[ht]
\centering
\includegraphics[width=5in]{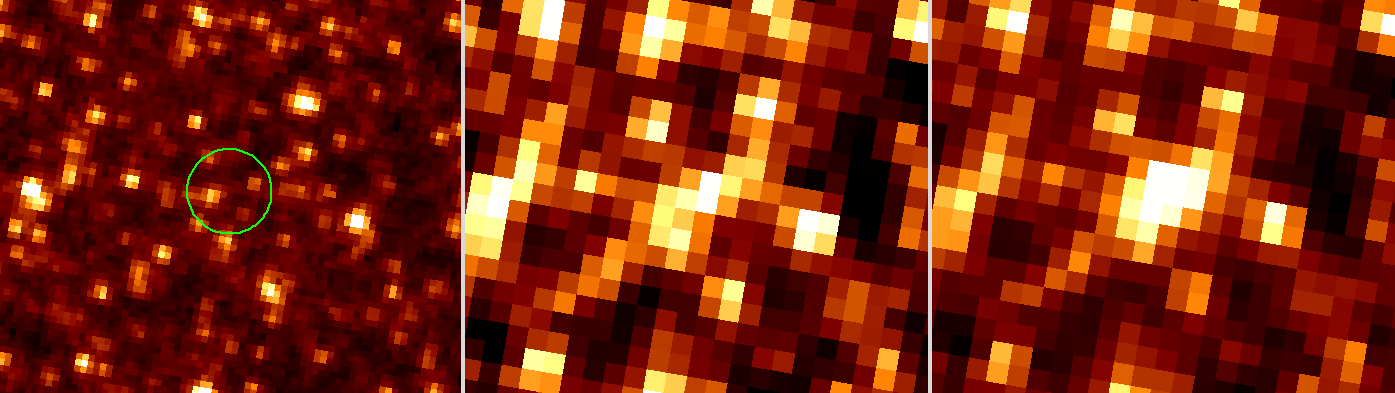}
\caption{
False-color renditions of \HST\/ WFC3 images of the site of the periodic IR variable SPIRITS 14bbc in NGC~7793, obtained on 2016 April~18. From left to right the frames were taken in $I$, $J$, and $H$\null. The green circle in the $I$ image is the 3$\sigma$ error position of the \Spitzer\/ variable. As discussed in the text, here is no obvious counterpart of 14bbc in the $I$ frame (none of the faint stars inside the circle varied across several \HST\/ epochs), but it is detected at $J$ and is bright at $H$\null. Height of each frame is $2\farcs4$.
\label{fig:14bbc}
}
\end{figure*}

The site of 14bbc lies in an extremely dense star field in a spiral arm of NGC~7793. We located the site in the available \HST\/ images through astrometric registration, as just described for 15wt. As shown in the $I$-band frame in the left panel of Figure~\ref{fig:14bbc}, there are several faint stars inside the 3$\sigma$ error circle. None of them varied between archival \HST\/ frames obtained in 2003, 2014, and our own frames from 2016, and we believe the variable was not detected in the $I$ band. However, as shown in the figure, a blended source does appear in the $J$ frame, and it is bright at $H$; the HSC gives a Vega-scale magnitude for this object of $H=19.42$. In the archival frames obtained on 2018 January~16 used for 15wt, the candidate has faded significantly at both $J$ and $H$, consistent with the phasing of 14bbc seen in Figure~\ref{fig:periodic_array}. Based on its extremely red optical color, location near the center of the error circle, and variability at $J$ and $H$, we conclude that this star is the near-IR counterpart of the \Spitzer\/ variable. Its magnitude is consistent with the $20.4>H>18.2$ anticipated for an AGB variable at the distance of NGC\,7793, using the LMC OH/IR star colors (see introduction to this section).

\subsection{15mr and 15mt \label{subsec:15mr} }

The IR variability of SPIRITS\,15mr was reported by \citet{Jencson2015}, who suggested it as a transient and possible SPRITE\null. It belongs to the star-forming barred spiral galaxy NGC\,4605 ($d\simeq5.5$~Mpc). Another SPIRITS variable, 15mt, lies nearby, and for our triggered \HST\/ observation on 2016 June~14 we adjusted the telescope pointing to include both objects. The \Spitzer\/ light curves of both variables are plotted in the bottom-left and bottom-middle panels in Figure~\ref{fig:periodic_array}. As the figure shows, both objects were rising in brightness up to the date of our 2016 \HST\/ imaging, leading at that time to our classification of both as possible transients. However, our subsequent observations show that 15mr is a periodic variable that has gone through two cycles, with a period of 1110~days. 

%{\bf NOTE: this paragraph belongs in the discussion subsection:}

The position of 15mr in the period-luminosity relation (Figure~\ref{fig:plrelation}) shows it to be about 2~mag brighter than the other sources with similar periods. However, its color, $[3.6]-[4.5]\simeq 0.8$, and peak-to-peak amplitudes, $\Delta[3.6]=1.4$ and $\Delta[4.5]=1.3$, are very similar to those of the other pulsators. As Figure~\ref{fig:plrelation} shows, a few of the variables in K19 also lie in this region of the period-luminosity relation. The nature of these luminous variables is unclear, although this is where we expect to find mass-losing red supergiants. The LMC OH/IR supergiant, IRAS\,04553$-$6825, the most luminous source from \citet{Goldman2017}, is an example of this population.

15mt is also an apparent periodic variable with a longer period ($\sim$1800~days), but the classification is less certain because it has only gone through a single cycle in the available \Spitzer\/ data. It has a very ``blue'' IR color, $[3.6]-[4.5]\simeq0.2$, and especially large peak-to-peak amplitudes,  $\Delta [3.6] \simeq 1.7$ and $\Delta [4.5] \simeq 2.2$. 

\begin{figure*}[ht]
\centering
\includegraphics[width=4.5in]{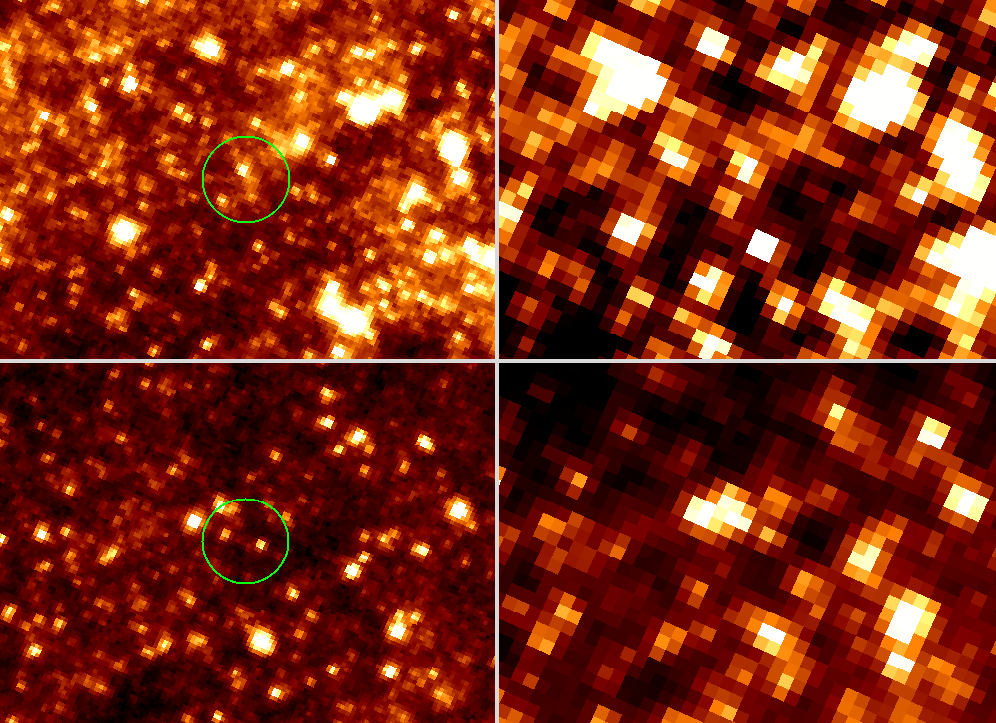}
\caption{
False-color renditions of \HST\/ WFC3 images of the site of the periodic IR variables SPIRITS 15mr and 15mt in NGC~4605. {\it Top row:} the site of 15mr in $I$- and $H$-band frames taken in 2016. The green circle in the $I$ frame marks the 3$\sigma$ error position of the \Spitzer\/ variable. The brightest star inside the circle is bright at $H$, and may be a counterpart of 15mr. {\it Bottom row:} $I$ and $H$ frames from the same two \HST\/ images, showing the site of 15mt. A bright star just outside the northeast edge of the error circle is much brighter in $H$ and might be a counterpart of the \Spitzer\/ variable. However, there is also a source just inside the circle on the northeast side that is bright at $H$ but undetected at $I$, making it another candidate counterpart. Height of each frame is $3\farcs6$.
\label{fig:15mrpix}
}
\end{figure*}

We registered a \Spitzer\/ IRAC 4.5~$\mu$m frame from 2016 April, showing both 15mr and 15mt near maximum brightness, with an archival \HST\/ WFC3 $I$-band image obtained in 2013 (GO-13364, PI D.~Calzetti), and we compared the 2013 image with our triggered frame taken in 2016. 15mr lies in a rich star field with moderate extinction, with several young associations in the vicinity. As shown in the top left image of Figure~\ref{fig:15mrpix}, inside the 3$\sigma$ registration error circle are several resolved stars and numerous, partially resolved fainter objects. The site lies near or within a rich young association to the southwest, from which it is separated by a dark dust lane. The brightest star within the error circle is also bright in $H$, as shown in the top right image in Figure~\ref{fig:15mrpix}, making it a candidate counterpart of 15mr. However, this star is not especially red at optical wavelengths, and is even detected in the $u$ band (F336W filter); the HSC gives magnitudes (Vega scale) of $u=23.39$, $B=24.12$, $V=23.70$, and $I=22.99$. Yet the object also has a near-IR excess, as shown by the HSC Vega-scale magnitude of $H=20.36$. By comparing the two available $I$-band images from 2013 and 2016, we see no significant variation of this star. Without further information, it remains unclear whether the bluish star represents a binary companion of (or a chance alignment with) the IR variable, or whether the situation is more complicated. The LMC OH/IR supergiant mentioned above, IRAS\,04553$-$6825, has $I-[4.5] \simeq 8.7$ and $H-[4.5] \simeq 3.8$, so 15mr  would be expected to have $I\simeq23.7 $ and $H\simeq 18.8$. Taking into account the limited information we have on supergiant colors and 15mr's large amplitude, the measured $H=20.36$ cannot be used to rule out the possibility that it is an OH/IR supergiant. The shorter-wavelength magnitudes may be too bright for the OH/IR-supergiant interpretation, but, as already noted, they could plausibly originate from a chance alignment in this rich star field, or conceivably a binary companion. 

%$I-[4.5] \sim 10.1$ and $H-[4.5] \sim 5.2$ (AB mags for $I$ and $H$, Vega mags for [4.5]), 

%{\bf NOTE: the 15mr and 15mt light-curve plots need a gray arrow at 2013-08-22. Why does 15mr have IRAC points in 2007 but 15mt doesn't?}

15mt also lies in a rich field, overlain with clumpy dust absorption, as shown in the $I$-band image in the bottom left frame in Figure~\ref{fig:15mrpix}. There are several resolved stars inside the 3$\sigma$ error circle, and a brighter star on the northeast side just outside the circle. The $H$-band image in the lower right panel in the figure shows two partially resolved bright stars, one corresponding to the bright $I$-band star just outside the circle. This bright star is extremely red; its Vega-scale HSC magnitudes are $V=25.18$, $I=22.47$, and $H=19.44$, making it a plausible optical/near-IR counterpart of 15mt---especially since the variable's $[3.6]-[4.5]$ color is relatively ``blue.'' However, we see no significant variation of this star in the $I$ band between 2013 and 2016. The second $H$-band star, just inside the error circle, appears to be undetected in the $I$ frame. An LMC OH/IR AGB star (see introduction to this section) would have 
$27.4>I>24.3$ and $21.5>H>19.3$,
so it seems possible that the star measured is the \Spitzer\/ source, but that it is confused at shorter wavelengths by bluer stars. Alternatively the  second $H$-band star may be the \Spitzer\/ source. 15mt is discussed further in the summary below.

%{\bf NOTE: do you measure H for this second source, outside the error  circle?}

%{\bf NOTE: need to correct final sentence above to Vega scale.}

%As with 15mr, it is difficult to reach further conclusions from the existing data.

%\begin{figure}[ht]
%\centering
%\includegraphics[scale=.525]{Spitzer_lcs_hst_15mr.pdf}
%\includegraphics[scale=.525]{Spitzer_lcs_hst_15mt.pdf}
%\includegraphics[scale=.525]{Spitzer_lcs_hst_16ea.pdf}
%%\includegraphics[scale=.525]{Spitzer_lcs_hst_14dd.pdf}
%\caption{
%\Spitzer\/ IRAC light curves of three quasi-periodic variables. Plotting symbols as in Figure~1.
%\label{fig:15mrlightcurve}
%}
%\end{figure}

\subsection{16ea}\label{sec:16ea}

%This subsection still being written...

%{\bf NOTE HEB 5/24/21: I moved 16ea from the SPRITES section to here. Need to make text consistent with being in the periodic variables section.}

%{\bf NOTE: %at least for now, I am calling 16ea a fast SPRITE, but this is open for discussion. 
%There are some issues with the light curve, e.g., upper limits on the same dates where an actual measurement is given; and no limits plotted on dates when there were observations (including what look to me like marginal detections on some of those dates when I inspect the SPIRITS website).}

%{\bf NOTE 1/4/20: based on ongoing discussion with Jacob, we may move 16ea to periodic variables, or eruptive...}

%{\bf NOTE 2/3/21 (Jacob): I am fairly convinced that 16ea belongs in the "likely periodic" category, especially when viewing the difference light curves in flux space.}

SPIRITS 16ea was announced by \citet{Jencson2016b} as a possible IR transient in the nearby starburst irregular galaxy NGC\,4214 ($d\simeq2.9$~Mpc). In response to this publication, a separate team (GO-14892, PI B.~McCollum) obtained \HST\/ imaging observations of the site of 16ea with WFC3.

The \Spitzer\/ light curves of 16ea are plotted in the bottom-right panel of Figure~\ref{fig:periodic_array}. Unfortunately, the quality of the light curves is relatively poor and difficult to interpret. This is due to the presence of many image and subtraction artifacts in the vicinity of 16ea in our difference images, leading to large uncertainties and numerous nondetections. Still, the handful of detections at [4.5] shows evidence of multiple peaks, consistent with a periodic variable. Hence we classify it as probably periodic, although the properties listed in Table~\ref{table:periodicvars} should be viewed with caution. The inferred color (from the limited [3.6] detections) is very red at $[3.6] - [4.5] = 1.6$\,mag. 
%No object was detected by \Spitzer\/ over several years at its site at 3.6 and 4.5~\micron, until 16ea first appeared at its maximum observed brightness on 2016 March~12. Our previous \Spitzer\/ observation had been made 209~days earlier, on 2015 August~16. The brightness declined over the ensuing 188 days, with 16ea last being detected on 2016 September~16. The transient was undetected at two epochs in 2017, but then reappeared at a single epoch, 2018 April~28, before again becoming undetectable in the rest of our data. 
The \HST\/ observations in GO-14892 were obtained on 2017 May~17 and~21; unfortunately this imaging was done only in the WFC3 IR channel. The status of the variable at this epoch is uncertain, but it appears to have been near a minimum in the pulsation cycle.%; the \HST\/ date fell 23 days after a previous non-detection with \Spitzer\/ observation, and 87~days before the next \Spitzer\/ visit, at which 16ea was also undetected.

%16ea qualifies as a ``transient,'' since it was undetected in all previous \Spitzer\/ observations, up until its appearance in early 2016. Following the 2016 outburst, it was again undetected, apart from a single apparent detection at [4.5] in 2018, up until our final observation near the end of the \Spitzer\/ mission. We tentatively classify 16ea as a fast SPRITE, although with an absolute magnitude at maximum of $M_{[4.5]}=-11.1$ (check this) it barely qualifies. The object, when detected, was very red, with $[3.6]-[4.5]$ typically around 1.0 (check this).

\begin{figure*}[ht]
\centering
\includegraphics[width=5in]{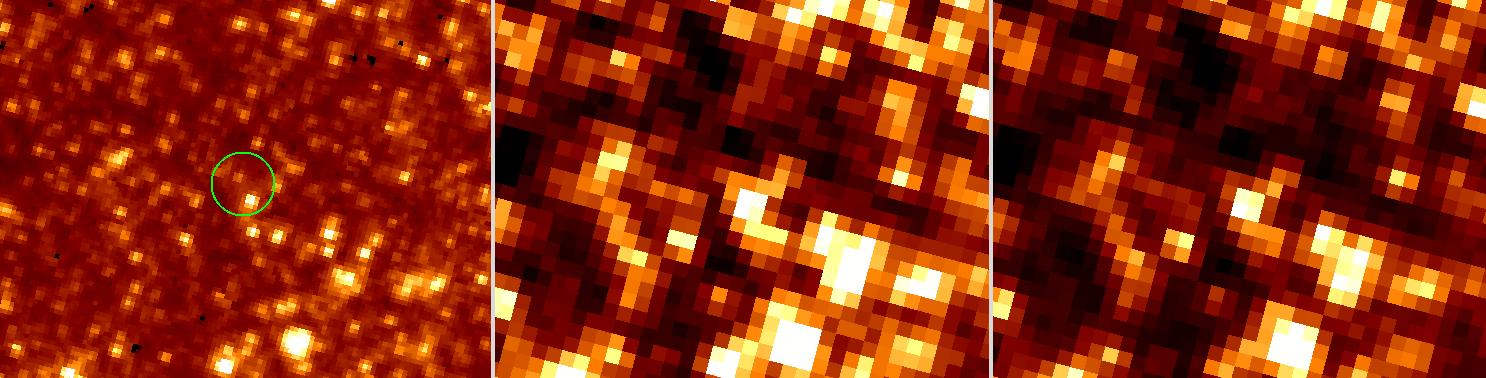}
\caption{
False-color renditions of \HST\/ WFC3 images of the site of the IR variable SPIRITS 16ea in NGC~4214, obtained in 2009. From left to right the frames were taken in $I$, $J$, and $H$\null. The green circle in the $I$ image is the 3$\sigma$ error position of the \Spitzer\/ object in the WFC3 frame. A bright, very red star inside the southwestern edge of the circle, which is also detected at $J$ and $H$, is a likely optical and near-IR counterpart of the \Spitzer\/ variable.  Height of each frame is $3\farcs3$.
\label{fig:16eapix}
}
\end{figure*}

We astrometrically registered a \Spitzer\/ 4.5~\micron\ difference image with an archival \HST\/ WFC3 $I$-band frame obtained on 2009 December~23 (GO-11360, PI R.~O'Connell). Images in the WFC3 IR channel $J$ and $H$ bandpasses were obtained at the same time in the O'Connell program. Figure~\ref{fig:16eapix} shows false-color renditions of the 16ea site from these $IJH$ frames, with the 3$\sigma$ registration error circle shown in the left panel. Inside the circle is a bright star, which is very red, making it a plausible optical and near-IR counterpart of 16ea. The HSC gives magnitudes (adjusted to Vega scale) for this star of $I=22.30$, $J=20.75$, and $H=19.88$. In comparison with an ACS $I$-band image of the site obtained on 2004 October~24 (GO-10332, PI H.~Ford), we see no significant change in brightness of this candidate. Comparing the 2009 and 2017 frames in $J$ and $H$, there is again no compelling evidence for variability. Given the fragmentary data from both \HST\/ and \Spitzer, it is difficult to reach firm conclusions about this object, beyond our identification of a probable optical/near-IR counterpart. 

%{\bf NOTE: according to the SPIRITS website, there were Spitzer observations of 16ea on 2015-08-16 but I don't seen any upper limits plotted in the light curve at that date. There were 2 observations in 2017, but only one upper limit, in 4.5 only, during 2017. Data seem to be missing in 2019 as well. Also gray arrows should be added at 2004-10-25 and 2009-12-23. Also, I'm puzzled by there being only one plotted detection at 3.6, since the images on the SPIRITS website seem to me to show the object at 3.6 at several epochs.}

%{\bf NOTE (Jacob): The missing data is probably because the SNR is too low in our photometry. I cull photometry from the final light curves that has very high errors to clean them up.}

\subsection{Periodic Variables: Discussion}

The analysis of the \Spitzer\ data for the periodic variables suggests that the majority (at least nine) of them are highly evolved AGB stars, similar to those discussed by K19 and to the OH/IR AGB variables in the Galaxy and LMC\null. Where we have \HST\/ detections they support this conclusion. These stars are important in several respects. Their progenitors must have had intermediate masses (approximately 5 to $10\,M_{\odot}$), and they represent a brief and poorly understood phase of late stellar evolution dominated by convection and mass loss. They are among the most significant dust producers in their respective galaxies, returning significant quantities of processed material to the interstellar medium. Intermediate-mass AGB variables, such as these, will be among the most distant individual stars observable with the {\it James Webb Space Telescope\/} (\JWST), and are likely to be important probes of their respective populations.

It is not surprising that the four variables that we detected at \HST\/ wavelengths, and identified as possible AGB stars, are among the bluest objects, lying in the range $0.17<[3.6]-[4.5]<0.76$  in Table \ref{table:periodicvars}. 15qo is the bluest  ($[3.6]-[4.5]=0.58$) variable that was {\it not\/} detected by \HST; however, its host galaxy is 0.4~mag more distant than that of 14bbc, the reddest object that {\it was\/} detected. Thus the detections, or lack thereof, are consistent with our conclusion that these sources are AGB variables. 

The fifth variable that was also also detected by \HST\/, but does not appear to be an AGB star, is 15mr. As discussed above, it is more luminous than the AGB variables, and is possibly a red supergiant. However, as discussed by K19 for variables in this part of the period-luminosity relation, it could alternatively be a dust-producing binary system.

%Although not bright at \HST\/ wavelengths these stars are all very luminous at infrared wavelengths and are likely to be dominant sources of emission in many JWST surveys and will also be important variables in the LSST era. {\bf this needs more though and maybe moved to conclusions  I'll return to it PAW!}

One source that lies in a position in the period-luminosity consistent with the AGB variables, but may be something else, is 15mt. Its blue \Spitzer\/ colors and very large variation amplitudes are not typical, and alternative explanations should be considered. While we might speculate that its periodicity could be due to orbital motion in a binary, we would expect an IR-luminous binary to be dusty and thus have red colors. We observed only a single cycle of variation, so it remains possible that 15mt is actually a transient. Unfortunately we have insufficient information to be confident one way or the other.

14al is a better candidate for an AGB variable, but its period, estimated to be longer than 2160~days, is poorly defined because of insufficient time coverage. There are only a few pulsating stars known with primary periods above 2000~days. If confirmed as a pulsating AGB star, 14al is potentially very interesting. It may be a candidate super-AGB star, i.e., a star with a progenitor mass of about $10\,M_{\odot}$ that could become an electron-capture SN \citep{Siess2007, Doherty2015, Doherty2017}. It is well worth a more detailed study with \JWST.

\section{Luminous Irregular Infrared Variables}

\begin{figure*}[ht]
\centering
\includegraphics[scale=.375]{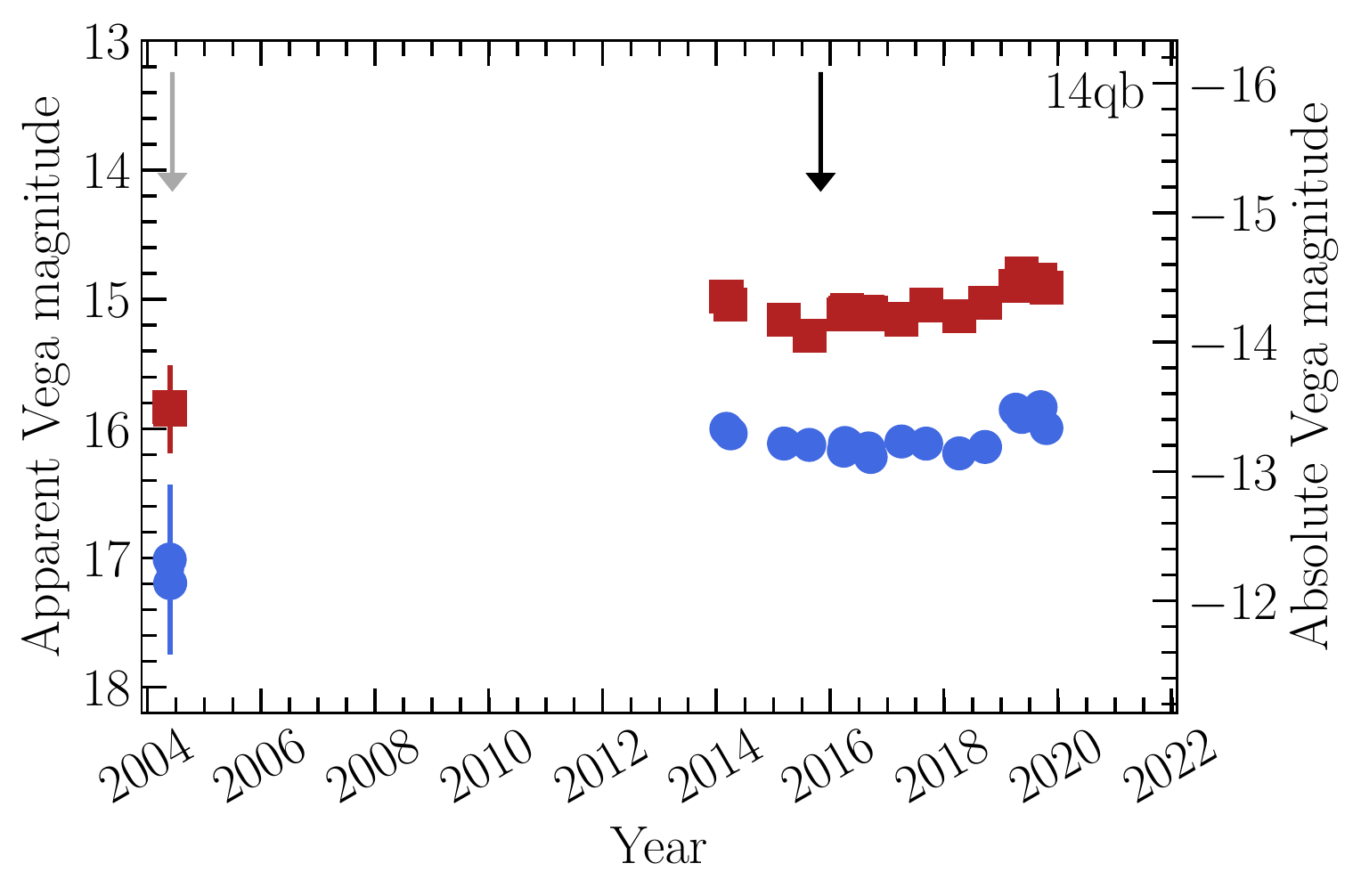}\hfill
\includegraphics[scale=.375]{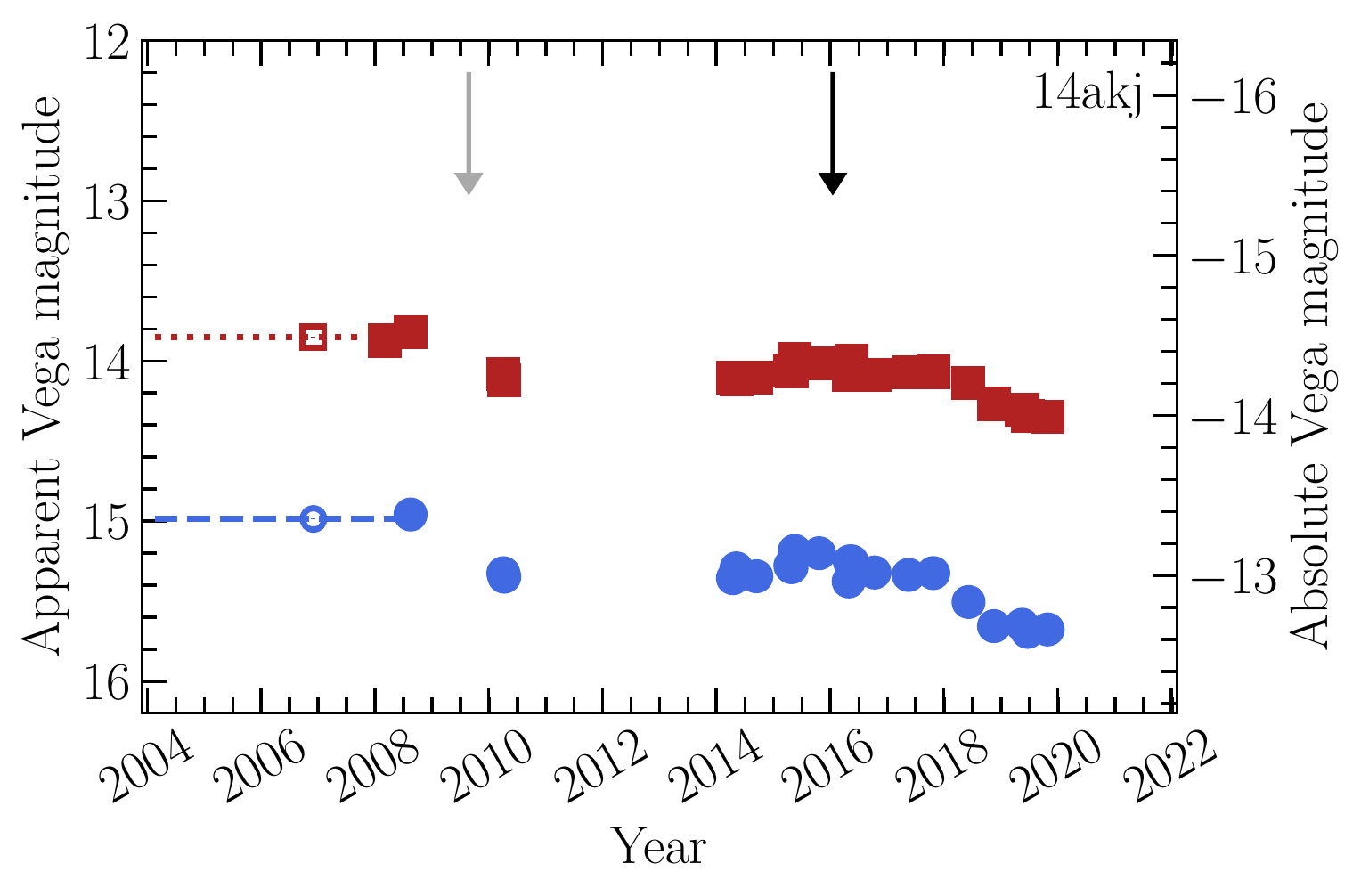}\hfill
\includegraphics[scale=.375]{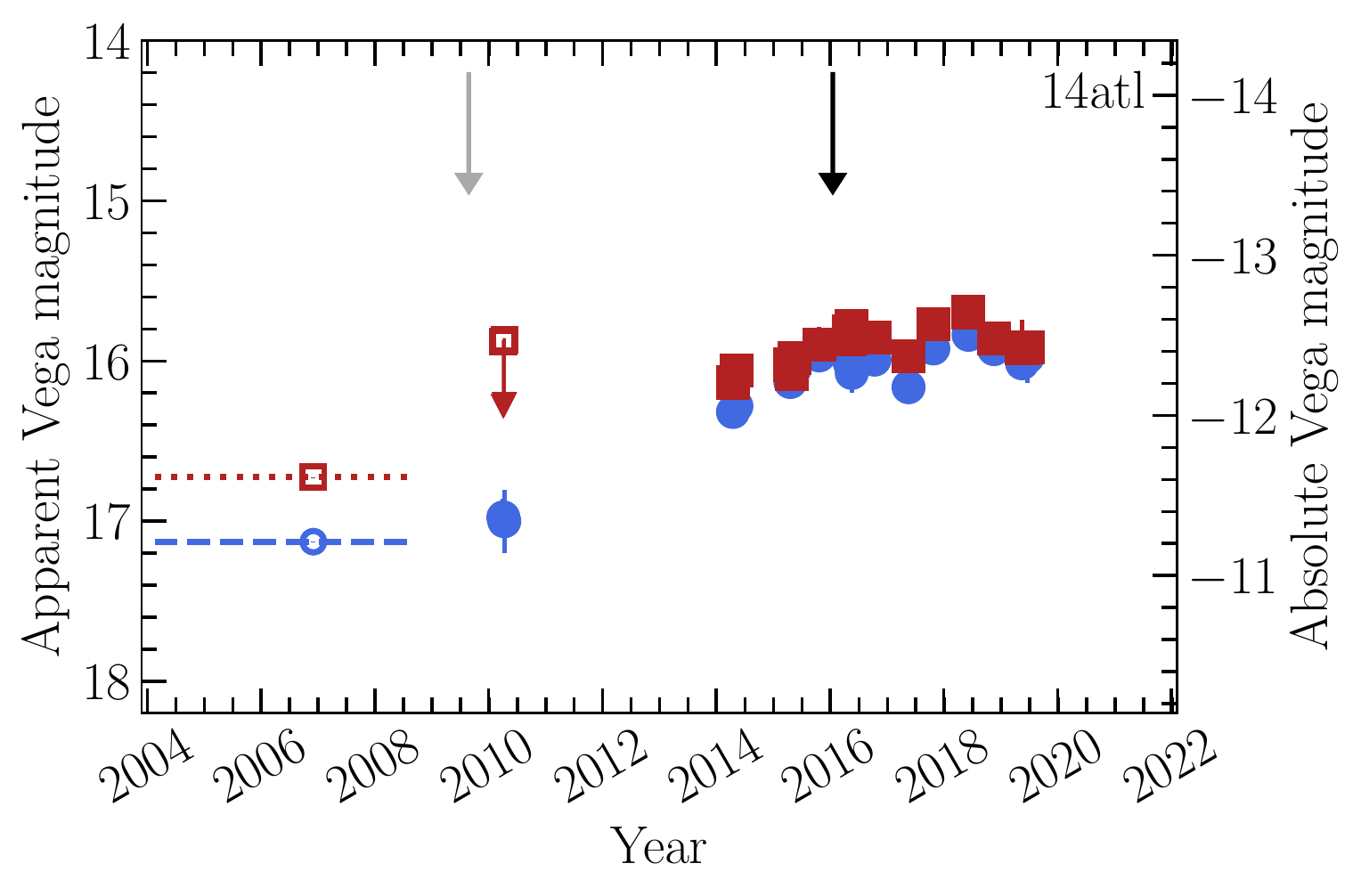}
\includegraphics[scale=.375]{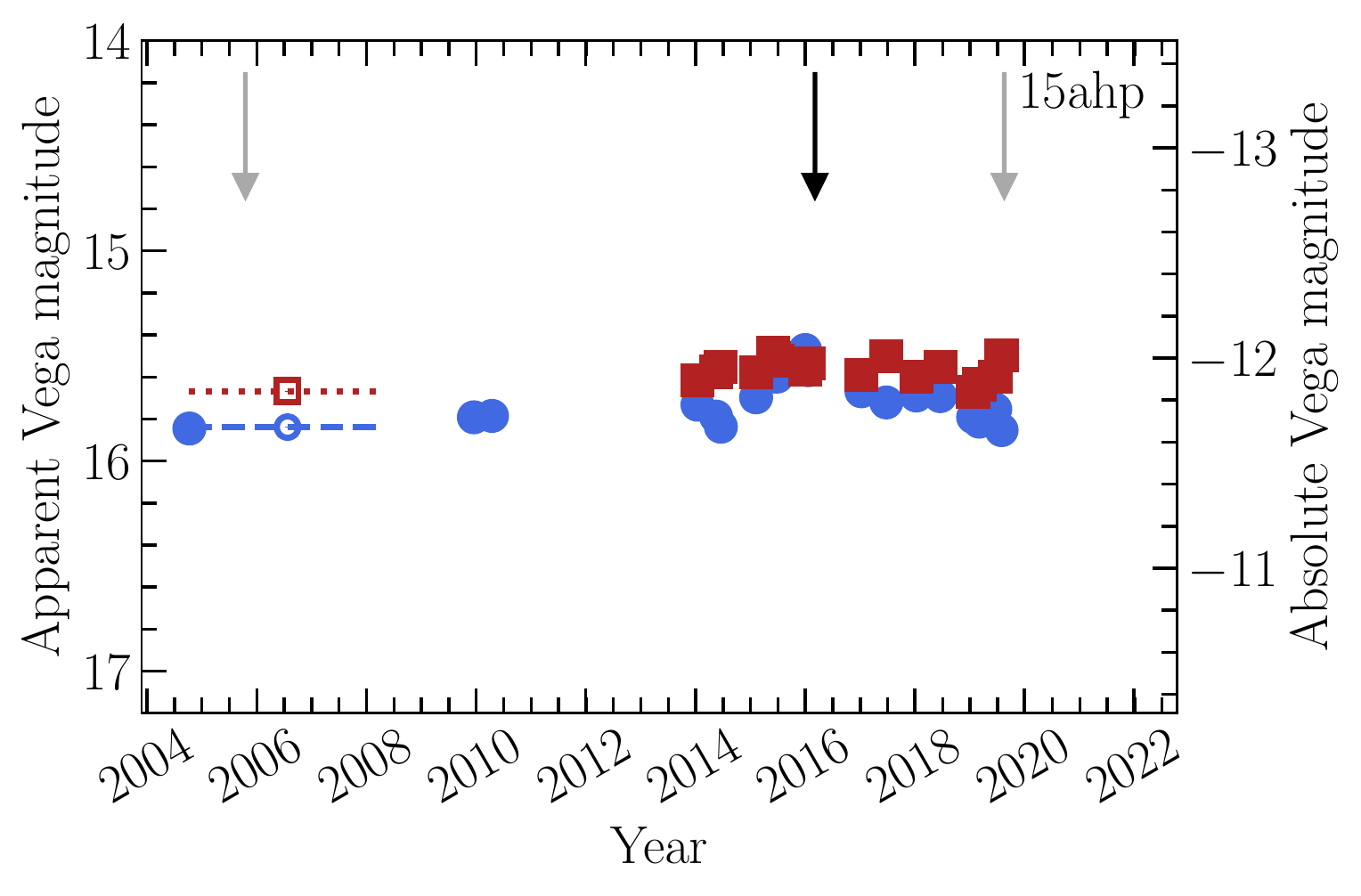}\hfill
\includegraphics[scale=.375]{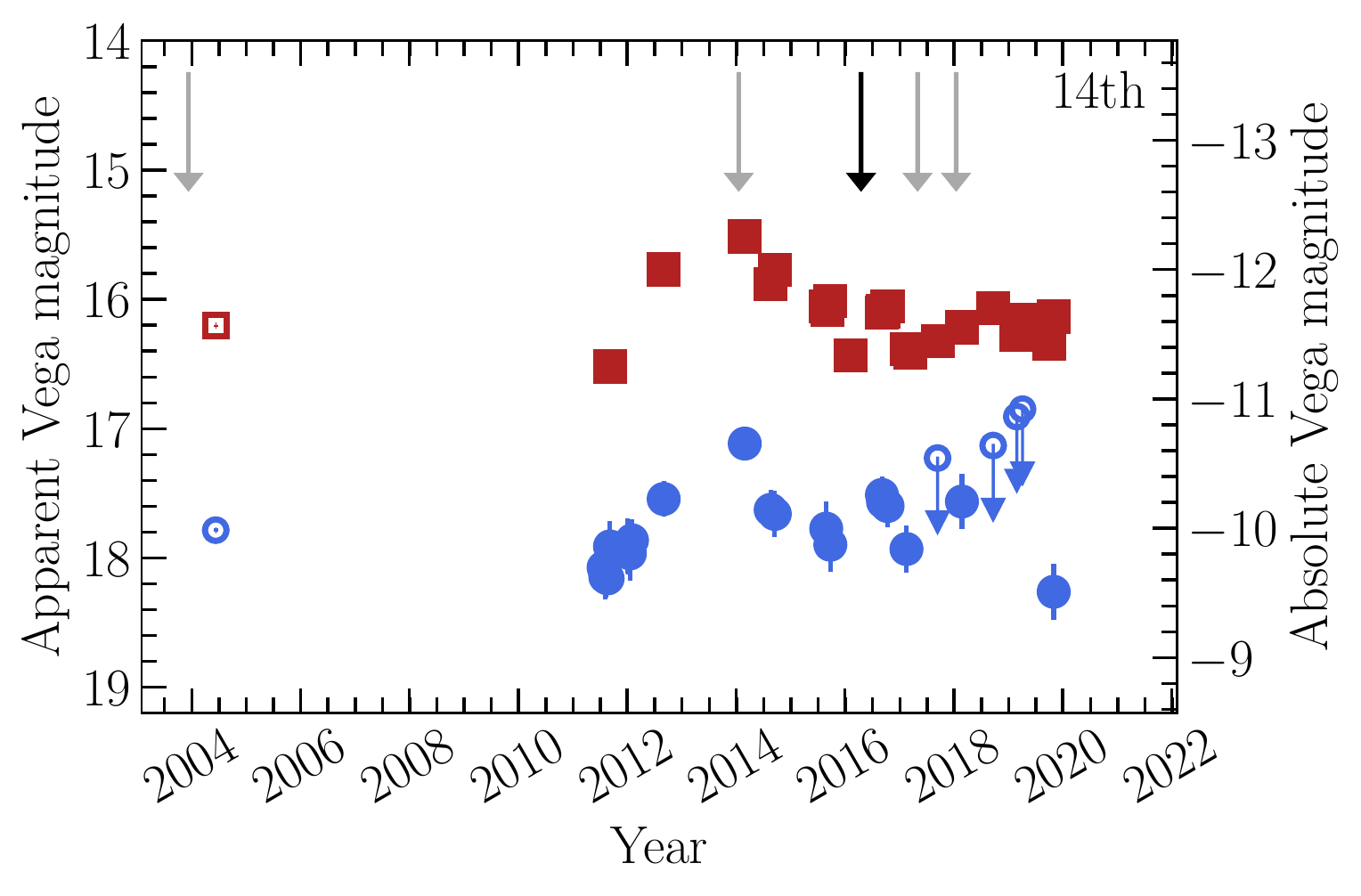}\hfill
\includegraphics[scale=.375]{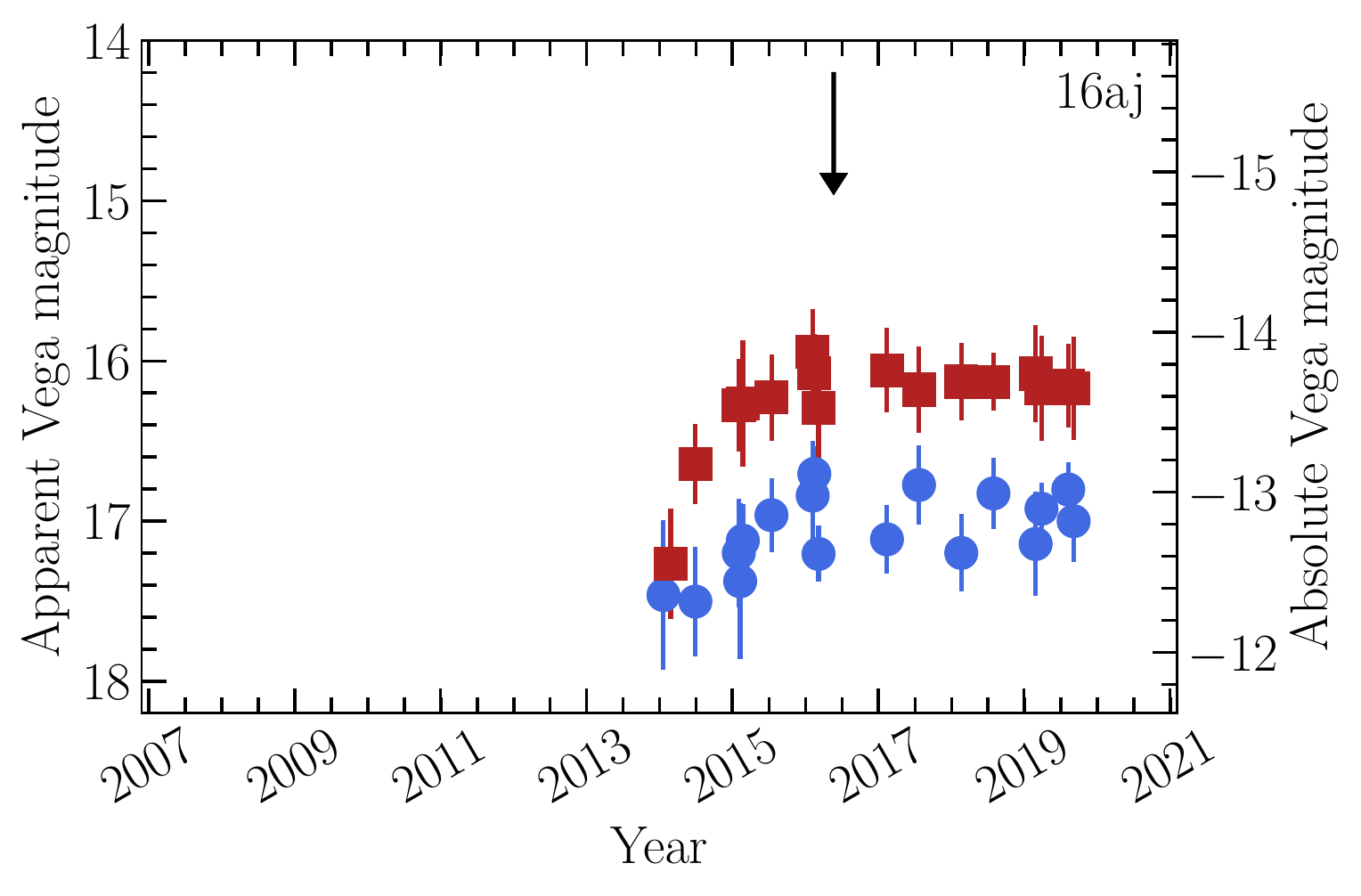}
\caption{
\Spitzer\/ IRAC light curves of six irregular variables. Plotting symbols as in Figure~1. 
\label{fig:irregular_array}
}
\end{figure*}

Several classes of stars may produce irregular, or non-periodic, variability in the IR\null. These include LBVs, which may undergo repeated outbursts or eruptive mass-loss events capable of forming copious dust, though the mechanism driving such outbursts is not well understood. In \citet{Jencson2019b}, we presented discoveries of two such sources, SPIRITS\,17pc and SPIRITS\,17qm, which underwent multiple, extremely red outbursts over the course of several years, and for which we identified luminous counterparts in archival \textit{HST\/} imaging, likely to be LBVs.  Dust-forming, colliding-wind WC binaries, like the SPIRITS sources presented in \citet{Lau2021}, are in fact periodic, but may be classified as irregular variables under our definition presented in \S\ref{sec:classes} if their orbital periods are longer than the available \textit{Spitzer\/} coverage. In this section, we discuss the six SPIRITS variables that we classify as irregular, and for which we have \HST\/ imaging. Their \Spitzer\/ light curves are shown in Figure~\ref{fig:irregular_array}. Table~\ref{table:irreg_var} gives details of the \Spitzer\/ photometry.

\begin{deluxetable*}{lcccccccc}
\tablewidth{0 pt}
\tablecaption{Properties of Irregular Variables\label{table:irreg_var}}
\tablehead{
\colhead{SPIRITS} &
\colhead{$t_{\mathrm{peak}}$\tablenotemark{a}} &
\colhead{$[3.6]$\tablenotemark{b}} &
\colhead{$\Delta [3.6]$} &
\colhead{$[4.5]$\tablenotemark{b}} &
\colhead{$\Delta [4.5]$} &
\colhead{$[3.6] - [4.5]$\tablenotemark{b}} &
\colhead{$M_{[3.6]}$\tablenotemark{b}} & 
\colhead{$M_{[4.5]}$\tablenotemark{b}} \\
\colhead{Designation} &
\colhead{[MJD]} &
\colhead{[mag]} &
\colhead{[mag]} &
\colhead{[mag]} &
\colhead{[mag]} &
\colhead{[mag]} &
\colhead{[mag]} &
\colhead{[mag]} 
}
\startdata
14qb  & 58619.73 & $15.91 \pm 0.07$ & 1.4 & $14.80 \pm 0.02$ & 1.0 & $ 1.11 \pm 0.07$ & $-13.4$ & $-14.5$ \\
14th  & 56715.49 & $17.11 \pm 0.07$ & 1.2 & $15.51 \pm 0.05$ & 1.0 & $ 1.60 \pm 0.09$ & $-10.7$ & $-12.3$ \\
14akj & 54695.30 & $14.96 \pm 0.03$ & 0.7 & $13.82 \pm 0.01$ & 0.5 & $ 1.14 \pm 0.03$ & $-13.4$ & $-14.5$ \\
14atl & 58277.58 & $15.84 \pm 0.02$ & 1.3 & $15.69 \pm 0.07$ & 1.0 & $ 0.15 \pm 0.07$ & $-12.5$ & $-12.7$ \\
15ahp & 57388.85 & $15.47 \pm 0.01$ & 0.4 & $15.56 \pm 0.03$ & 0.2 & $-0.09 \pm 0.03$ & $-12.0$ & $-12.0$ \\
16aj  & 57425.97 & $16.84 \pm 0.34$ & 0.8 & $15.94 \pm 0.27$ & 1.3 & $ 0.90 \pm 0.43$ & $-13.0$ & $-13.9$ \\
\enddata
\tablenotetext{a}{Time of the observed light curve peak (Vega magnitudes) at either [3.6] or [4.5].}
\tablenotetext{b}{Measured at time $t_{\mathrm{peak}}$.}
\end{deluxetable*}

\subsection{14qb}

IR variability of SPIRITS\,14qb in the actively star-forming galaxy NGC\,4631 ($d\simeq7.3$~Mpc) was discovered at our first two epochs of SPIRITS observations. Comparison of IRAC frames obtained in 2014 March and April, with archival images from 2004, showed that 14qb had risen $\sim$1.1 and 0.8~mag at 3.6 and 4.5~\micron, respectively, as shown in the \Spitzer\/ light curve in the top left panel of Figure~\ref{fig:irregular_array}. The source was luminous ($M_{[4.5]}\simeq-14.5$ in 2014 March) and extremely red ($[3.6]-[4.5]\simeq 1.1$). At our observation a year later, in 2015 March, 14qb had faded about 0.2~mag at both wavelengths, and was still fainter 5~months later. At that point, we considered it to be a ``slow'' SPRITE transient, and triggered our \HST\/ observations; these were obtained in 2015 November. Since then, however, 14qb did not continue to fade like a transient; instead, it slowly brightened, up to our most recent, and final, \Spitzer\/ observations in 2019 October. Thus it cannot be considered to be a SPRITE, nor is it periodic, and we classify 14qb as irregular.  

\begin{figure*}[ht]
\centering
\includegraphics[width=4.5in]{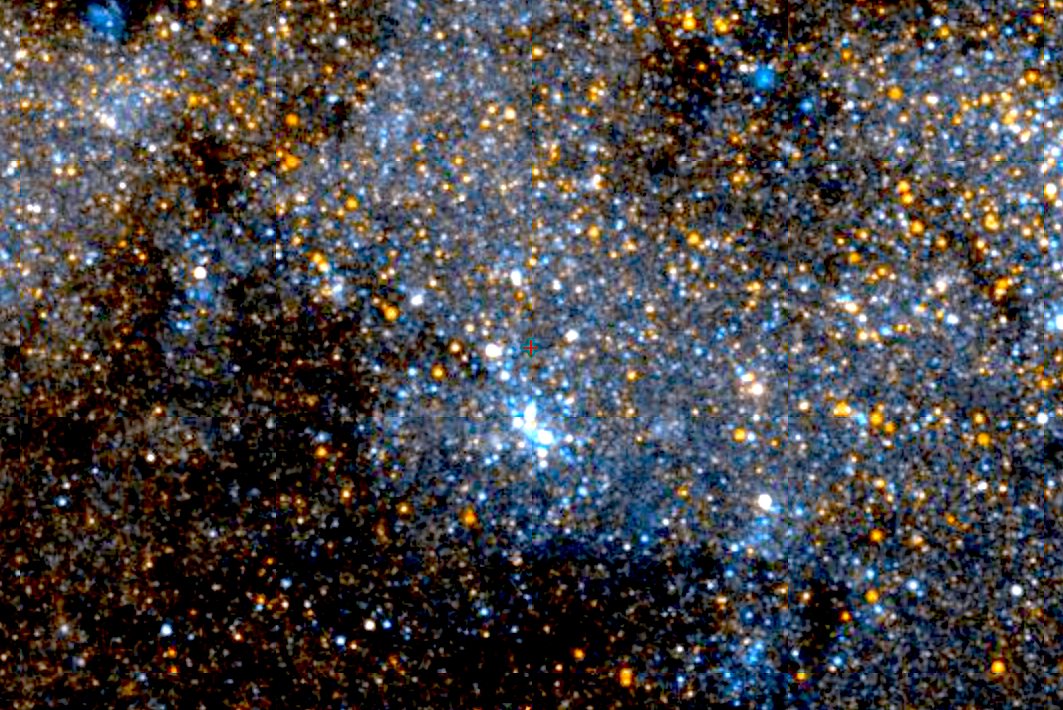}
\caption{
Color rendition of the site of the irregular variable SPIRITS~14qb in the star-forming galaxy NGC\,4631, from \HST\/ $V$ and $I$ frames in the Hubble Legacy Archive. Frame is $18''$ high ($\sim$640~pc at the distance of NGC\,4631). The site of 14qb is marked with a red cross. It lies in a region of intense star formation, with dust lanes and young associations in the vicinity. 
%{\it Bottom row:} zooms in on \HST\/ frames. The site of 14qb from astrometric registration of the \Spitzer\/ and \HST\/ frames is marked with green 3$\sigma$ error circles. Each frame is $2''$ high. The left-most frame shows $I$ in 2004. The next three frames are all from 2015, showing from left to right $I$, $J$, and $H$\null. A faint star on the southeast edge of the 3$\sigma$ circle had brightened in 2015, and is also detected at $J$ and $H$\null. It is likely an optical counterpart of 14qb.
\label{fig:14qb}
}
\end{figure*}

In addition to our triggered \HST\/ WFC3 observations in 2015, there are archival frames obtained with ACS in 2004 (GO-9765, PI R.~de~Jong). We astrometrically registered the ACS $I$-band frame with a \Spitzer\/ 4.5\,\micron\ difference image showing 14qb at maximum brightness. 14qb is located in an extremely crowded star-forming region southwest of the nucleus of NGC\,4631, with numerous young stars and dust lanes. 
Figure~\ref{fig:14qb} shows a color rendition of the location, created in the HLA from ACS frames in $V$ and $I$\null.
The vicinity of 14qb exhibits fairly high and spatially variable extinction, and the object lies near the northern edge of a rich young association. 

There are a few faint stars within a 3$\sigma$ astrometric error circle in the \HST\/ $I$ frame, none of which varied significantly from 2004 to 2015. None of these stars are conspicuous in the 2015 $J$ and $H$ frames. We conclude that there is no credible counterpart at $I$, $J$, and $H$ to this bright and very cool IR variable. Based on the limited information available for this source, we cannot conclusively determine its nature. Its high IR luminosity could be consistent with dust formation by an LBV, for example, but the star must be heavily enshrouded to explain the lack of a detectable optical or near-IR counterpart. 

%The bottom row of frames in Figure~\ref{fig:14qb} shows, from left to right, the $I$ image from 2004, and our three frames from 2015, in $I$, $J$, and $H$\null. At the southeast edge of the 3$\sigma$ error circle, marked in green, lies a faint star that has brightened by about 0.75~mag in $I$ from 2004 to 2015, consistent with the brightening seen in the IR\null. Its apparent $I$ magnitude (Vega scale) in 2015, based on a calibration using HSC stars in the field, is about 23.9, corresponding to an absolute magnitude of $-5.4$ at the distance of NGC\,4631. The star is also detected in the $J$ and $H$ frames, although it is blended with nearby objects. Given its variability and red color, this object is a candidate optical\slash NIR counterpart of 14qb. Unfortunately, there are only two epochs of \HST\/ imaging of this site.

%{\bf NOTE: we could speculate on its nature---dusty LBV, or whatever---based on the scant available information. Note that it's pretty luminous in the IR, but it's not in the optical, so it's not a standard LBV.}

\subsection{14akj and 14atl}

As noted in the last column of Table~\ref{table:observinglog}, SPIRITS 14akj and 14atl are two IR variables that serendipitously lie near the primary \HST\/ target 15nz (a periodic variable discussed above in \S\ref{sec:15nz}). These two objects belong to the actively star-forming spiral galaxy M83 ($d\simeq4.7$~Mpc). Their \Spitzer\/ light curves are shown in the top-middle and top-right panels of Figure~\ref{fig:irregular_array}. 

We classify both variables as irregular. 14akj slowly and irregularly declined in brightness over the duration of the \Spitzer\/ imaging. However, we cannot completely rule out that it might be a periodic variable with a very long period, of order 8~years. It is extremely red, with a color of $[3.6]-[4.5]\simeq1.1$; it is also luminous, having fallen from $M_{[4.5]}\simeq-14.5$ to $-14.0$ from 2008 to 2019. 
14atl differs from 14akj in several respects. It slowly rose from 2010 to mid-2018 by about one magnitude, but with two dips in brightness. Our final observations showed that another dip was underway. 14atl is relatively quite ``blue,'' with $[3.6]-[4.5]\simeq 0.15$, and it is less luminous than 14akj, with an absolute magnitude averaging about $M_{[4.5]}\simeq-12.5$ over the final few years of our monitoring. 

\begin{figure*}[ht]
\centering
\includegraphics[height=2.25in]{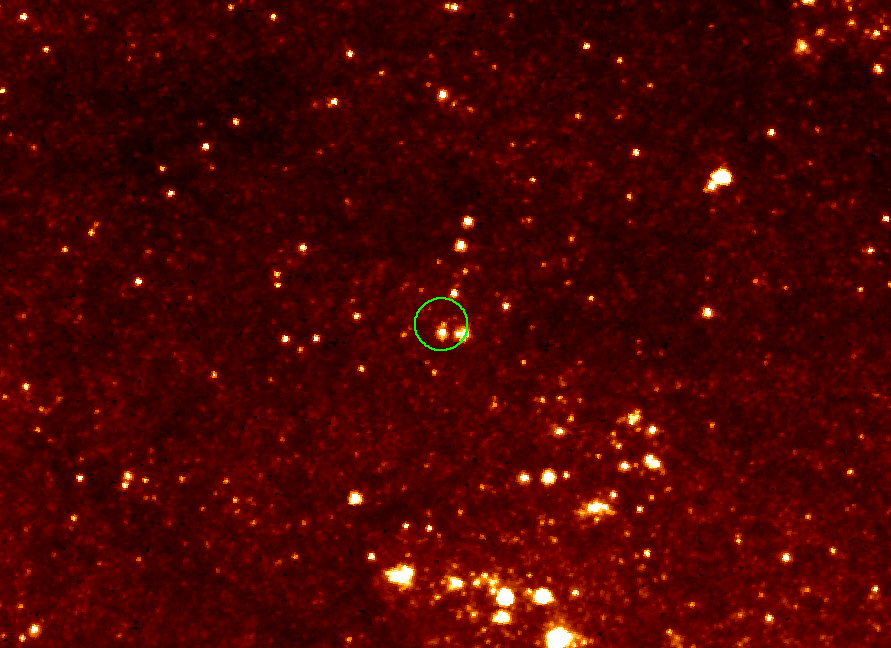}
\includegraphics[height=2.25in]{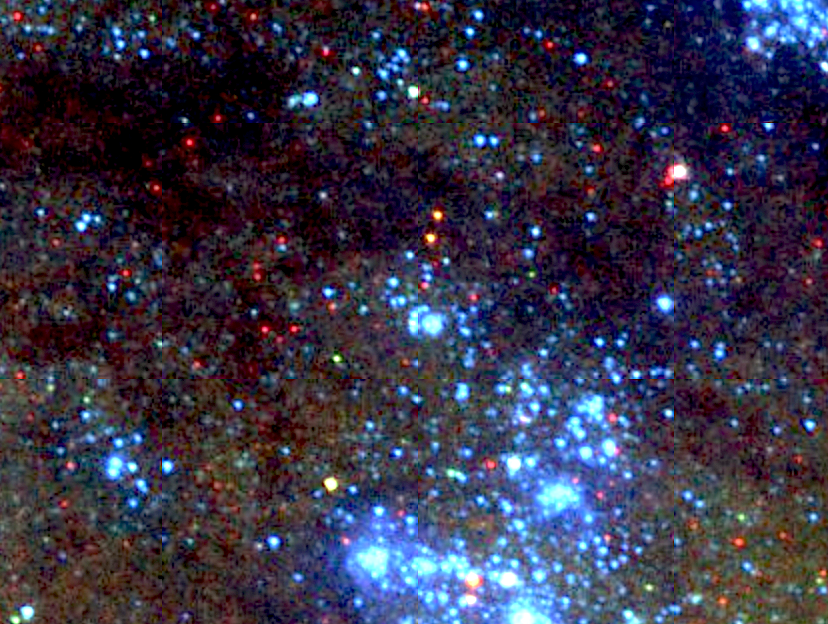}
\vskip0.024in
\includegraphics[height=2.262in]{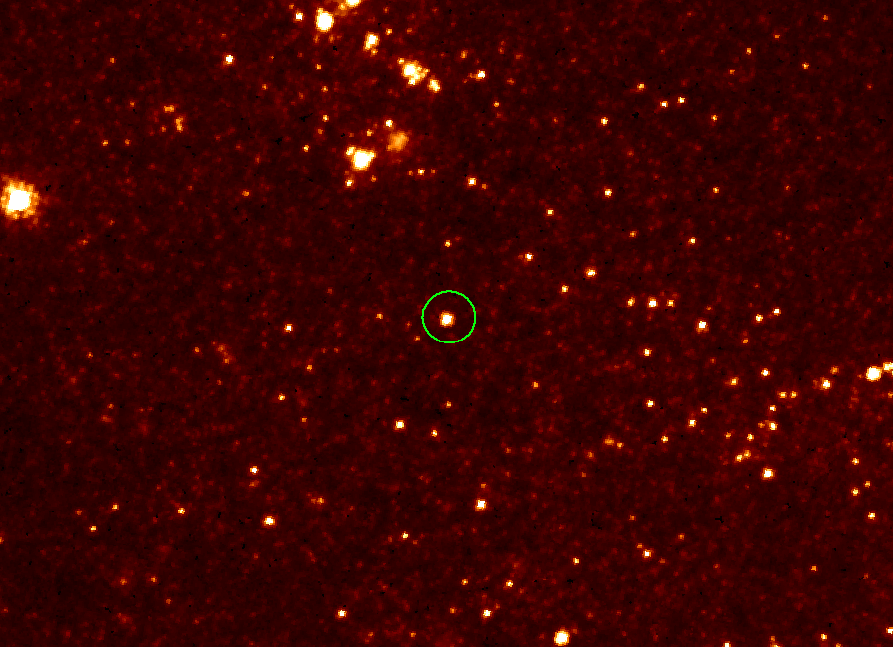}
\includegraphics[height=2.262in]{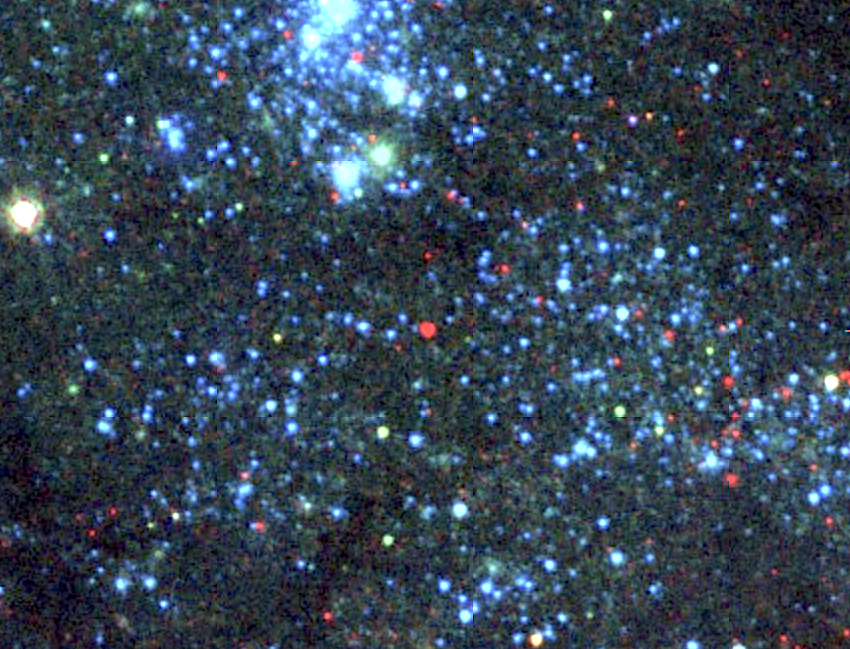}\hskip0.02in\phantom{.}
\caption{
\HST\/ images of the sites of the irregular variables SPIRITS 14akj (top) and 14atl (bottom) in star-forming regions of M83. The left-hand frames depict the $I$-band \HST\/ frames of the sites, with green circles marking the 3$\sigma$ astrometric locations of the irregular IR variables. The right-hand pictures are color renditions of the sites, from the Hubble Legacy Archive. At the location of 14akj there are no convincing optical counterparts; all of the candidate stars are blue and not conspicuously variable. However, 14atl has a bright, optically variable, and extremely red and luminous counterpart. Height of frames is $12''$ ($\sim$275~pc at the distance of M83).
\label{fig:14akjpix}
}
\end{figure*}

We astrometrically registered a \Spitzer\/ IRAC channel~2 image showing both 14akj and 14atl with an \HST/WFC3 $I$-band image obtained in 2009 (GO-11360, PI R.~O'Connell), in order to determine their locations in the \HST\/ frame. The site of 14akj lies in a very active star-forming region. Inside a 3$\sigma$ error circle there are three fairly bright stars, and several fainter ones, forming a small cluster, as shown in the top two panels of Figure~\ref{fig:14akjpix}. The top left panel shows a false-color rendition of the $I$-band image with a 3$\sigma$ error circle marking the position of 14akj. None of the three bright stars varied significantly between the two available \HST\/ $I$-band frames (the other being a 2016 image from program GO-14059, PI R.~Soria). None of these stars are prominent in the one available $J$ frame nor the two archival $H$ frames. The top right panel is a color rendition of the field, obtained from the HLA and created from WFC3 frames obtained in 2016 in $u$, $B$, and $I$\null. The image shows several rich young stellar associations in the vicinity of 14akj. All of the stars inside the error circle are seen to be quite blue. It could be that the counterpart to 14qkj was not detected in the optical or near-IR \HST\/ images, in which case it is likely to be a heavily enshrouded, and fairly massive young object. Alternatively, this source could be consistent with a WC colliding-wind binary system, where one of the bright blue stars detected with \textit{HST\/} is, in fact, the counterpart. Optical spectroscopy to search for WC features, as presented for other SPIRITS sources in \citet{Lau2021}, could test this scenario. 

The site of 14atl was also astrometrically localized in the same \HST\/ $I$ frame as described above. The bottom two pictures in Figure~\ref{fig:14akjpix} depict the $I$ frame and the 3$\sigma$ location (left panel), and the HLA color rendition (right panel). Inside the error circle is an optically bright star, and the color rendition shows that this object is extremely red. This star is very bright in a single available $J$ frame and two $H$ frames. Between 2009 and 2016 the star brightened in $I$ by about 0.5~mag, and in $H$ by 0.6~mag. These are in accordance with a similar brightening seen in the \Spitzer\/ data. There is thus little doubt that this star is the optical and near-IR counterpart of 14atl. 

Because the HSC magnitudes are averaged over widely separated epochs, we performed aperture photometry on the available individual WFC3 frames. We used the zero-points and corrections to infinite aperture from the site referenced in footnote~10. Table~\ref{table:14atl_photometry} gives our results. {The uncertainties in the magnitudes, including systematics from the camera calibrations, are generally about $\pm$0.02--0.03~mag.}

\begin{deluxetable}{lCCC}
\def\d{$\dots$}
\tablecaption{
\HST\/ WFC3 Photometry (AB Magnitudes) of Irregular Variable SPIRITS 14atl in M83\label{table:14atl_photometry}
}
\tablehead{
\colhead{Filter}
&\colhead{2009 Aug 26\tablenotemark{a}} 
&\colhead{2016 Jan 18-19\tablenotemark{b}} 
&\colhead{2016 Jan 21\tablenotemark{c}}
}
% \decimals
\startdata
F555W & 23.70 &  \d    & 23.45 \\
F606W & \d    &  22.46 & \d    \\
F625W & \d    &  22.13 & \d    \\
F814W & 20.70 &  20.20 & 20.17 \\
F110W & 19.42 &  \d    & \d    \\
F125W & \d    &  18.61 & \d    \\
F140W & \d    &  18.48 & \d    \\
F160W & 19.01 &  18.41 & \d    \\
\enddata
\tablenotetext{a}{GO-11360, PI R. O'Connell.}
\tablenotetext{b}{GO-14463, PI B. McCollum.}
\tablenotetext{c}{GO-14059, PI R Soria.}
\end{deluxetable}

\begin{figure*}
\centering
\includegraphics[width=0.49\textwidth]{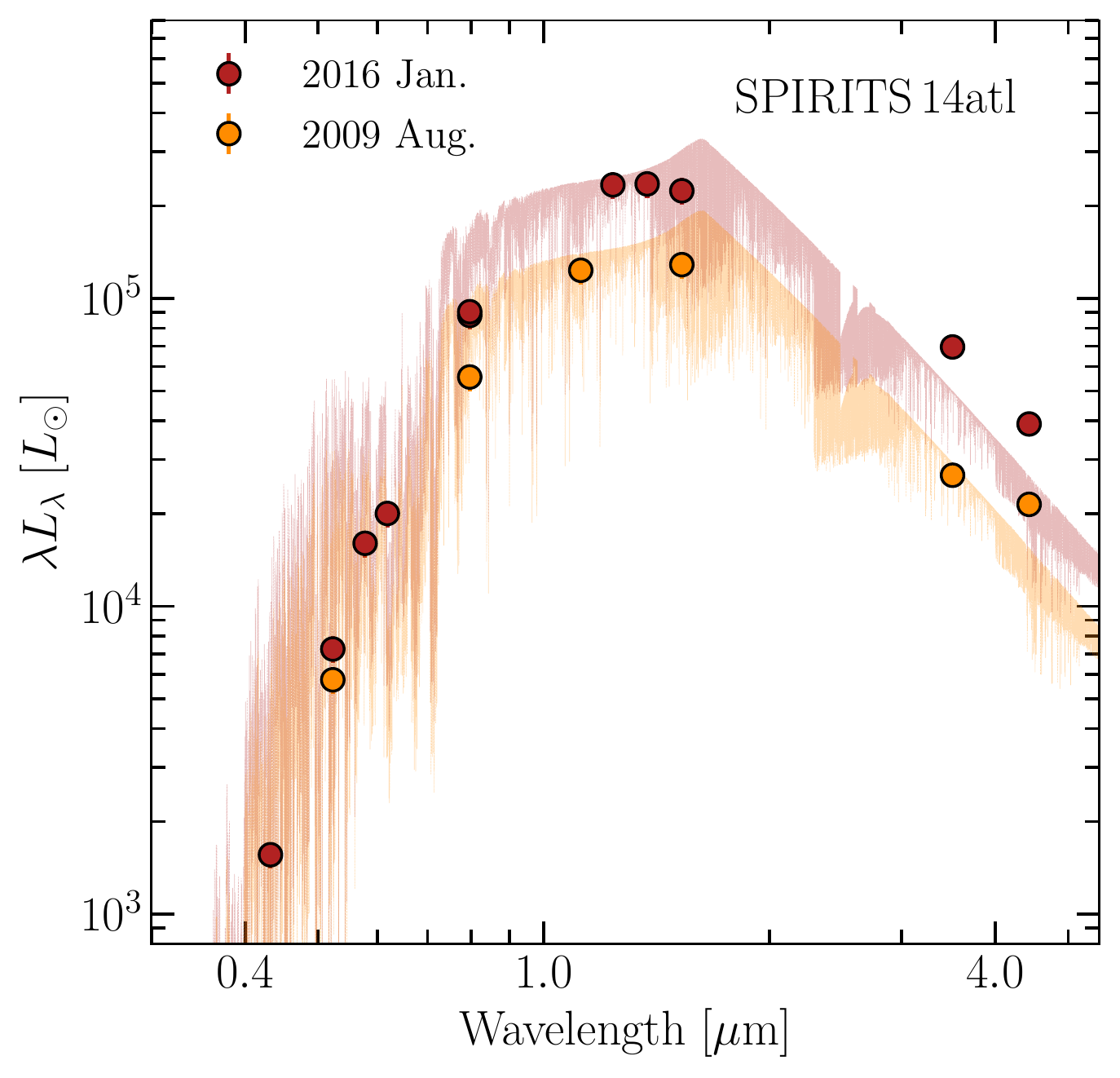}
\hfill
\includegraphics[width=0.49\textwidth]{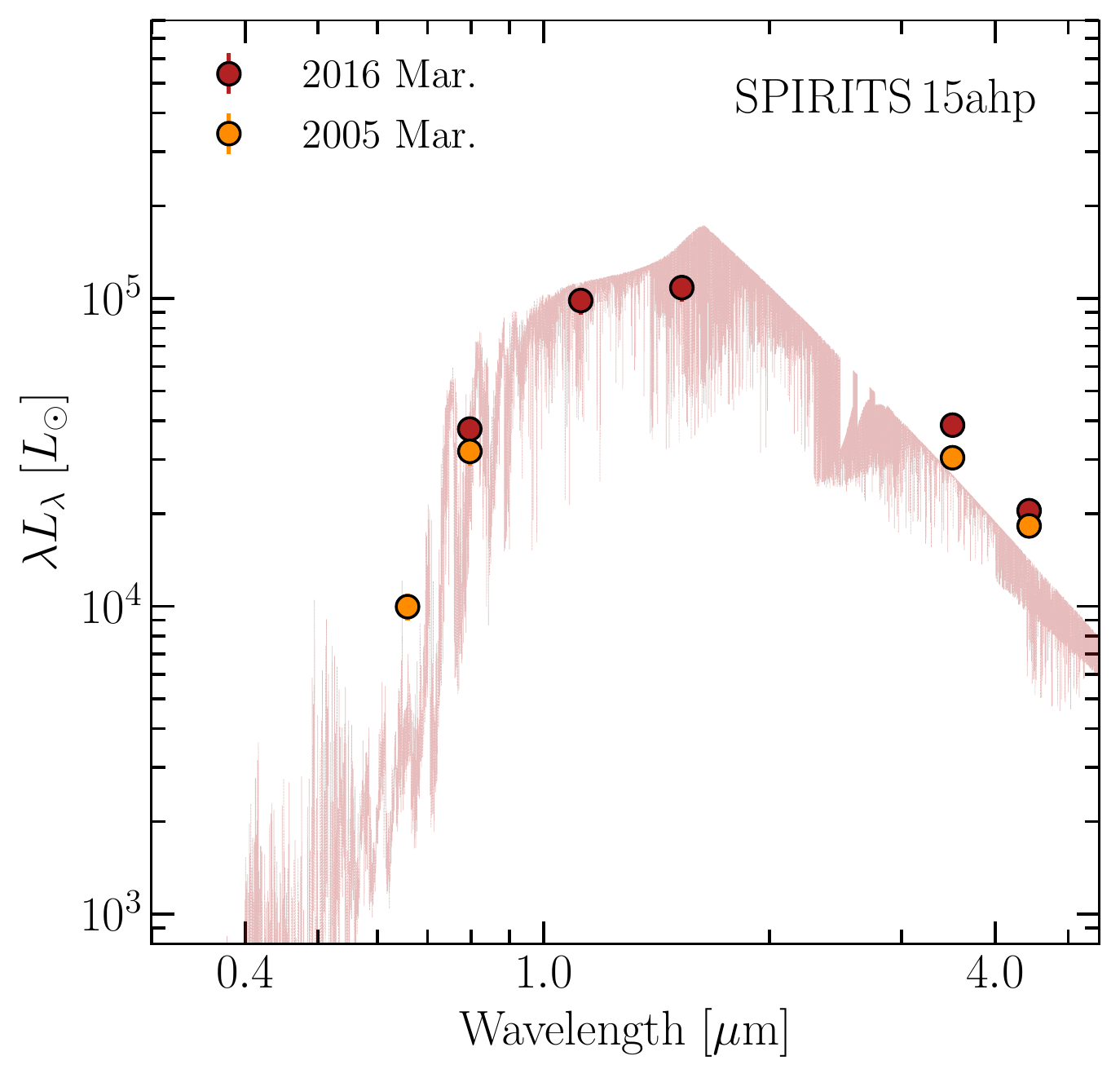}
\caption{
{\it Left:} SED of the irregular variable SPIRITS\,14atl in M83, constructed from \HST/WFC3 photometry (Table~\ref{table:14atl_photometry}) of the optical and near-IR counterpart obtained in 2009 August (orange circles) and 2016 January (red circles). The \Spitzer\/ [3.6] and [4.5] measurements are interpolated to the same two epochs. PHOENIX model stellar photospheres (nonrotating, solar-metallicity) with $T_{\mathrm{eff}} = 3500$~K, scaled to luminosities of $\log (L/L_{\odot}) = 5.24$ and $5.47$) provide reasonable approximations to both data epochs. They are superposed in the corresponding colors. {\it Right:} SED of the irregular variable SPIRITS\,15ahp in NGC\,2403, constructed from HSC magnitudes from 2005 March (orange circles) and 2016 March (red circles), along with [3.6] and [4.5] estimates from interpolations of the {\it Spitzer\/} light curves to the corresponding dates. A PHOENIX model photosphere with $T_{\mathrm{eff}} = 3300$~K, scaled to a luminosity of $\log (L/L_{\odot}) = 5.11$, is superposed, again providing a reasonable approximation to the data. 
\label{fig:14atl_15ahp_SED}
} 
\end{figure*}

We show the 2009 and 2016 SEDs of 14atl in Figure~\ref{fig:14atl_15ahp_SED}, constructed from our \textit{HST} aperture photometry of the optical/near-IR counterpart and linear interpolations of the [3.6] and [4.5] magnitudes to the corresponding epochs. The SEDs at both epochs are similar, peaking between $\approx$1--2$\,\mu$m at band luminosities of $\lambda L_{\lambda} \simeq 10^5~L_{\odot}$. In comparison with PHOENIX model photospheres (non-rotating, solar metallicity; \citealp{Kucinskas2005,Kucinskas2006}), the SEDs appear consistent with a luminous red supergiant with an effective temperature of $T_{\mathrm{eff}} \approx 3500$~K and bolometric luminosities of $\log\, L_{\mathrm{bol}}/L_{\odot} \simeq 5.24$--$5.47$. We note an excess of mid-IR flux at [3.6] and [4.5] compared to the stellar models. Interestingly, the mid-IR color is redder when the star is fainter. Together, these facts likely point to emission from warm dust that condenses in a stellar wind. The observed variability is likely associated with semiregular variations arising from pulsational instabilities, which are common in cool supergiants \citep[e.g.,][]{Yoon2010,Yang2011}. 

%According to the HSC, the magnitudes of the star (Vega scale, at the 2016 epoch [check this]) are $V=23.55$, $I=19.81$, and $H=17.23$. The star appears to be a dusty, cool, massive supergiant.

%{\bf NOTE; what more can we say about 14atl?? We should at least make a plot of, and discuss, its SED.}

%{\bf NOTE 9/18/21 There is also now an SED plot, so the text needs to be updated for this as well. And then we need to discuss its nature---dusty RSG etc}

\subsection{15ahp}

The IR variable SPIRITS 15ahp is another serendipitous target, lying by chance in the field of our \HST\/ observations of the periodic variables 15ahg, 14al, and 14dd, which were discussed above (\S\ref{sec:15ahg}). This field is in the M81 Group galaxy NGC\,2403, and the site of 15ahp lies in a star-forming spiral arm of the host galaxy. The IR light curves of 15ahp are shown in the bottom-left panel of Figure~\ref{fig:irregular_array}. This object varies in IR brightness by about 0.2~mag peak-to-peak, on a fairly short timescale of a few months. We classify it as an irregular variable. Its IR color also appears to vary. 

%[{\bf NOTE: it actually looks like there is occasionally a negative [3.6]-[4.5] color!! Can we check whether this is true???}.]
%[{\bf NOTE 2/3/2021 (Jacob): Yes, the outburst is much bright at [3.6] in the difference image uncertainty. }.]

\begin{figure*}[ht]
\centering
\includegraphics[width=4.5in]{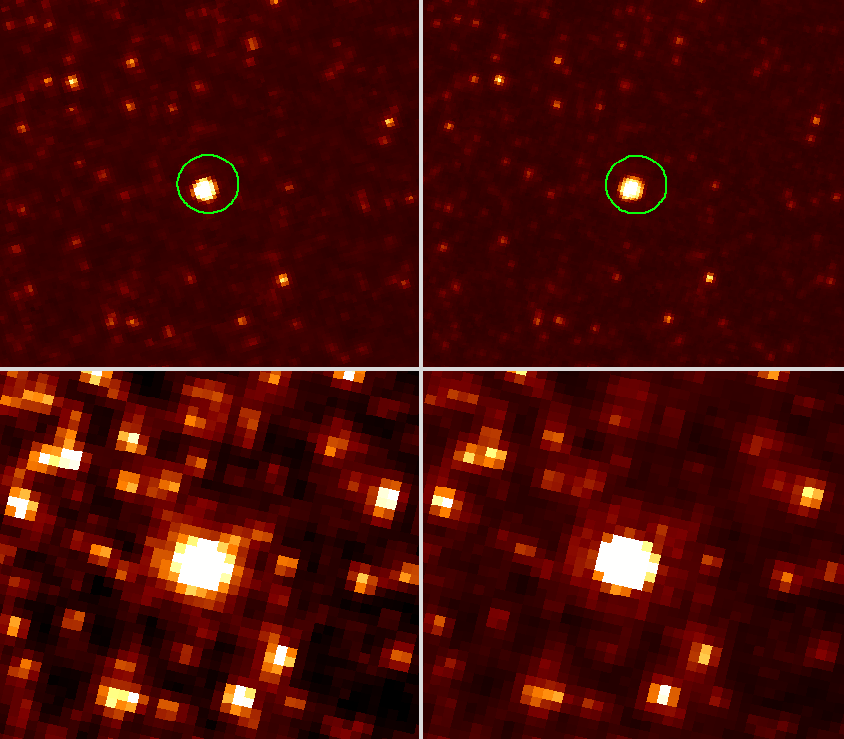}
\caption{
False-color renditions of \HST\/ images of the site of the irregular variable SPIRITS 15ahp in the nearby galaxy NGC~2403. {\it Top row:} $I$-band frames taken in 2005 (left) and 2016 (right). Green circles mark 3$\sigma$ error positions of the \Spitzer\/ variable. A conspicuous optical counterpart of 15ahp lies within the error circle, and brightened by $\sim$0.25~mag from 2005 to 2016. {\it Bottom row:} \HST\/ images of the site in 2016, taken in $J$ (left) and $H$ (right). The counterpart is very bright at $J$ and $H$\null. Height of each frame is $4\farcs8$.
\label{fig:15ahpmosaic}
}
\end{figure*}

The astrometric registration of \Spitzer\/ frames with \HST\/ images of this site was described in \S\ref{sec:15ahg}. The top two panels in Figure~\ref{fig:15ahpmosaic} show the 3$\sigma$ astrometric error circles for the location of 15ahp in two of the three available \HST\/ $I$-band images, an archival one from 2005 (GO-10402, PI R.~Chandar) and the other our frame obtained as a result of our triggered 2016 March~7 observation. A third $I$-band image from 2019 is available (GO-15645, PI D.~Sand) but not illustrated. Inside the error circle is a prominent star, which rose in $I$-band brightness by about 0.25~mag between 2005 and 2016. This is qualitatively consistent with the brightening at [3.6] and [4.5] over the same interval, as seen in Figure~\ref{fig:irregular_array}. 15ahp then faded by 0.17~mag at the 2019 \HST\/ observation. The star is very bright at $J$ and $H$, as shown in the bottom two panels in Figure~\ref{fig:15ahpmosaic}. There is thus little doubt that this object is the optical/near-IR counterpart of SPIRITS~15ahp. The HSC gives magnitudes (AB scale) of this star from the 2005 ACS frames of $\rm F658N=21.74$ and $I=20.25$, and from the 2016 WFC3 frames of $I=20.07$, $J=18.63$, and $H=18.17$.  

The multi-epoch SED, constructed in a similar manner to that of 14atl, is shown in the right panel of Figure~\ref{fig:14atl_15ahp_SED}. As with 14atl, the star appears consistent with a red supergiant, though perhaps a bit cooler and less luminous ($T_{\mathrm{eff}} \approx 3300$~K; $\log\,L_{\mathrm{bol}}/L_{\odot} \approx 5.1$), based on our comparisons with PHOENIX models. We again note a mid-IR excess compared to the photospheric models and that the mid-IR color is slightly redder when the star is fainter, likely indicating the presence of circumstellar dust. 

%10/18/21: I verified the above IJH mags---HEB

%{\bf Need to add discussion of the SED in Fig 16!}

%{\bf NOTE: here again we should probably say something about the possible nature of the source, etc. LBV or $\eta$~Car-type?? Or, collect these musings into a discussion in \S8.6} 

\subsection{14th}

SPIRITS 14th is yet another serendipitous target, which happens to lie in the field of 15wt and 14bbc in the nearby galaxy NGC~7793, which we imaged with \HST\/ as described above (\S\ref{sec:15wt}) on 2016 April~18. The bottom-middle panel in Figure~\ref{fig:irregular_array} shows the \Spitzer\/ light curves, which we classify as that of an irregular variable. 14th brightened by about 1~mag from late 2011 to a peak in early 2014, and then declined by nearly the same amount over the next two years. Since then it has remained nearly constant at [4.5] but somewhat ``noisy'' at [3.6], with our most recent and final observation showing it quite faint at [3.6]. This variable is extremely red, with a typical color of $[3.6]-[4.5]\simeq1.5$. It is not extremely luminous; the absolute magnitude at our final observation was about $M_{[4.5]}\simeq-11.5$.

% {\bf NOTE: 14th light curve plot: add a gray arrow at 2003-12-10 and at 2018-01-16 }

We registered a \Spitzer\/ frame showing 14th near maximum brightness with \HST\/ $I$-band images, as discussed for 15wt and 14bbc in \S\ref{sec:15wt}. The site is in a very rich star field. There are a few resolved stars within a 3$\sigma$ registration error circle, on top of an unresolved or partially resolved background. Comparing $I$ frames obtained in 2003, 2014, 2016 (our triggered observation), and 2017 shows no significant variations of any of these stars. Moreover, none of them are conspicuous at $J$ and $H$, nor varied between $JH$ images obtained in 2016 and 2018. We conclude that 14th is undetected in the optical and near-IR \HST\/ images, which is consistent with its extremely red IR color. Again, without an obvious counterpart detection in the optical or near-IR, it is difficult to draw strong conclusions about the nature of the source, other than that it is likely a young, fairly massive, and heavily enshrouded object. 

%\begin{figure}[ht]
%\centering
%\includegraphics[scale=.525]{Spitzer_lcs_hst_14th.pdf}
%\includegraphics[scale=.525]{Spitzer_lcs_hst_16aj.pdf}
%%\includegraphics[scale=.525]{Spitzer_lcs_hst_15ahp.pdf}
%\caption{
%\Spitzer\/ IRAC light curves of xxx eruptive variables. Plotting symbols as in Figure~1.
%\label{fig:16ajlightcurve}
%}
%\end{figure}

\subsection{16aj}

SPIRITS 16aj was announced as a possible transient by \citet{Jencson2016b}. It lies in the actively star-forming barred spiral NGC~2903, the most distant of the galaxies discussed in this paper ($d\simeq9.2$~Mpc). The \Spitzer\/ light curves of 16aj are shown in the bottom-right panel of Figure~\ref{fig:irregular_array}. This object rose in brightness by about 1~mag over the first two years of SPIRITS monitoring (early 2014 to early 2016), leading us to consider it a slow SPRITE; at that point we triggered \HST\/ imaging of the site, which was obtained on 2016 May~23. However, since early 2016 the object has remained at a nearly constant magnitude, although with possible short-term variations of a few tenths of a magnitude. Thus we now classify 16aj as an irregular variable, rather than a true transient. It is fairly red ($[3.6]-[4.5]\simeq0.9$) and of relatively high luminosity ($M_{[4.5]}\simeq-13.9$).

We astrometrically registered \Spitzer\/ frames, including a difference image showing 16aj, with an \HST\/ ACS $I$-band image taken in 2004 (GO-9788, PI L.~Ho) and with our own $I$ frame obtained in 2016. At the site there are a few resolved faint stars, lying on a sheet of unresolved starlight. We see no $I$-band variations between 2004 and 2016, and the $J$ and $H$ frames we obtained on the same date in 2016 reveal no extremely red stars.  The site lies on the edge of a rich young association and \ion{H}{2} region, and there are numerous dark dust lanes in the vicinity. We conclude that 16aj lacks a detectable optical or near-IR counterpart. Similarly to 14qb, 14th, we conclude only that the star is likely to be relatively young, massive, and heavily enshrouded. 

%Figure~\ref{fig:16ajpix} shows a color rendition of the location, taken from the HLA and derived from WFPC2 frames obtained in 1994 in $U$, $V$, and $I$ (GO-5211, PI J.~Trauger).

%{\bf NOTE HEB 6/4/21: since we claim no detection, I deleted the picture showing the site. Should we indicate the limiting magnitudes for the images here, and for the other undetected eruptives? Also, my notes indicate that there was an attempt at a spectrum with MOSFIRE. Should we mention this?  Along with any other ground-based attempted follow-ups of any of the targets in this paper??}

%\begin{figure*}[ht]
%\centering
%\includegraphics[height=3in]{16aj_hla.png}
%\caption{
%\HST\/ WFPC2 color image of the site of the eruptive variable SPIRITS 16aj in an active and dusty star-forming region in NGC\,2903. %A red cross marks the location of the variable, based on an astrometric registration of \Spitzer\/ frames with \HST\/ images. There %is no credible optical counterpart of the IR variable. Image obtained from the HLA; frame is $11\farcs7$ high.
%\label{fig:16ajpix}
%}
%\end{figure*}

%\subsection{Irregular Variables: Discussion}

%{\bf Here we have a general discussion of the irregular variables, relation to LBVs, etc etc. Or just put this in the next section}

%They may fall into two groups: those with optical counterparts (14qb [altho not very bright], 14atl, 15ahp)   and those undetected at $IJH$ (14akj, 14th, 16aj).

\section{Summary}

SPIRITS was the first large-scale monitoring survey of nearby galaxies, using the warm \Spitzer\/ telescope to search for luminous variable stars and transients at IR wavelengths. In the work described here, we employed new and archival optical and near-IR \HST\/ images to study the sites of 21 SPIRITS variables. The selected targets were of special interest because they were undetected or very faint in ground-based optical surveys. Our aims were to search for progenitors, attempt to detect the sources during outburst using deep \HST\/ imaging, and characterize their environments.

We classify the SPIRITS variables into three groups based on their photometric behavior: SPRITEs and transients; periodic variables; and irregular variables. Our main results from the \HST\/ imaging are as follows.

\begin{enumerate}

\item ``SPRITEs'' are a new class of intermediate-luminosity IR transients that lack counterparts in deep ground-based optical imaging. They are defined as objects having absolute magnitudes at maximum in the range $-14 < M_{[4.5]} < -11$. We investigated \HST\/ images of three SPRITEs, two of them with fast outburst timescales of a few days to a few weeks, and one with a slow timescale of nearly three years. Like most SPRITEs, two of the three occurred in dusty, star-forming regions, consistent with an origin from massive stars. No progenitors were found in deep pre-outburst archival \HST\/ images for the two objects in young regions. We did detect one of them---the slowly evolving 17fe---during outburst at $J$ and $H$\null. This allowed us to construct an SED, indicating that 17fe during eruption was a dusty object with a temperature of about 1050~K\null. Unusually, the third SPRITE candidate, the very fast 14axa, occurred in the old bulge of M81 rather than in a star-forming region. It appears to have been a dusty classical nova. Some, or many, of the fast SPRITEs at the low end of the luminosity range are likely to be classical novae, instead of arising from massive stars. Most SPRITEs, however, are events of an uncertain nature---possibly a mixture of massive stellar mergers, dust-obscured core-collapse supernovae, and eruptions of massive stars related to those of luminous blue variables. 

\item Variable stars are much more conspicuous among the brightest members of stellar populations in the IR than they are in the optical, particularly in late-type galaxies. More than half of our SPIRITS targets, although initially considered to be transients, proved to be periodic variables. Their pulsation periods are long, ranging from $\sim$670 to over 2100~days. These objects are likely to be highly evolved, dusty AGB stars, similar to OH/IR objects known in the Galaxy and Magellanic Clouds. The pulsators found by SPIRITS are strongly associated with star-forming regions in nearby spirals and irregular galaxies, and they likely arise from intermediate-mass stars ($\sim$5--$10\,M_\odot$). Out of the 12 periodic variables for which we have \HST\/ images, only five were warm enough to be detected with \HST\/ at $I$, $J$, and/or $H$, along with one uncertain case.

%detected:
%15ahg
%15wt
%14bbc
%16ea

%not detected:
%15nz
%15qo
%15aag
%14al
%14dd
%15afp

%uncertain:
%15mr
%15mt

\item Six SPIRITS variables did not fit our definitions for transients or (likely) periodic variables, and we classify them as irregular. Two of these sources, 14atl and 15ahp, had relatively blue IR colors ($-0.1 \lesssim [3.6] - [4.5] \lesssim0.2$ ) and had bright, red counterparts in \textit{HST\/} imaging. Their optical--near-IR SEDs are consistent with those of luminous, pulsating red supergiants. The remaining four irregular variables are much redder at IR wavelengths ($[3.6] - [4.5] \gtrsim 0.9$) and, not surprisingly, we did not identify convincing optical or near-IR counterparts in deep \HST\/ images. These objects may be consistent with eruptive, dust-forming mass-loss events, like those of LBVs, but their massive stellar counterparts are likely heavily enshrouded. For the case of 14akj, however, we noted a possible association with several luminous blue stars. 14akj could therefore instead correspond to a dust-forming colliding-wind Wolf-Rayet (WC type) binary. Optical spectroscopy to search for prominent WC emission features would test this possibility.

\item None of the SPIRITS transients observed by \HST\/ appear to be flares associated with YSOs. Although some lie in the dust lanes of the
host galaxies, none of the investigated transients are directly associated with star-forming complexes.     
However, future IR observations with facilities such as {\it JWST\/} and the Nancy Roman Space Telescope may reveal YSO transients in star-forming regions.

\end{enumerate}

The SPIRITS project was the first systematic time-domain reconnaissance of stellar variability among IR-luminous stars in nearby galaxies. It revealed several new classes of extremely cool and dusty transients and pulsating or irregular variables. An understanding of the nature of these diverse objects will require more intensive time coverage of their variations and outbursts than was possible with \Spitzer. Infrared spectroscopy would be a key new element in such investigations. Considerable progress will be possible with the powerful IR capabilities of {\it JWST\/} and the Roman Space Telescope.

%{\bf Close with more discussion; future work; role of JWST and Roman; etc etc. Mention the need for more intensive coverage of the light curves.}

%{\bf

%NOTE: discussion to be written when rest of paper is completed. Main points include:

%\begin{enumerate}

%\item There will probably be a table here summarizing the findings, e.g., a column for I J H with a check mark indicating whether or not each one was detected at that wavelength...
    
%\item Variable stars are more conspicuous among the brightest members of stellar populations in the IR than optical. Primarily due to dusty, pulsating, luminous AGB stars.

%\item The IR variables found by Spitzer lie preferentially in star-forming regions \& spiral arms, showing that they arise from massive stars.

%\item IR variables fall into the categories of SPRITE transients, periodic variables, irregular variables.

%\item The ``bluer'' IR variables tend to have optical/NIR counterparts. The redder ones generally lack such counterparts.

%\item Nature of SPRITEs remains unclear. Some fainter members of the class are likely to be dust-forming classical novae.

%\item Irregular variables may fall into 2 classes: those without optical/NIR counterparts, and those with prominent counterparts. 

%\item Information available on the IR variables to date is fragmentary (and \Spitzer\/ mission is about to end). Spectra would be crucial in elucidating their nature. Say something about JWST here?

%\end{enumerate}

%}

\acknowledgments

Support for {\it HST\/} Program numbers GO-13935, GO-14258, and AR-15005 was provided by NASA through grants from the Space Telescope Science Institute, which is operated by the Association of Universities for Research in Astronomy, Incorporated, under NASA contract NAS5-26555.

Support for this work was also provided by NASA through awards issued by JPL/Caltech.

P.A.W. acknowledges a research grant from the South African National Research Foundation and is grateful to John Menzies (SAAO) for the use of his period-finding software.

J.B. is supported by NSF grant AST-1910393.

R.D.G. was supported in part by the United States Air Force. 

{Some of the data presented in this paper were obtained from the Mikulski
Archive for Space Telescopes (MAST) at the Space Telescope Science Institute.} Our work is also
based in part on data obtained from the Hubble Legacy Archive, which is a collaboration between the Space Telescope Science Institute (STScI/NASA), the Space Telescope European Coordinating Facility (ST-ECF/ESAC/ESA), and the Canadian Astronomy Data Centre (CADC/NRC/CSA).

This research has made use of the NASA/IPAC Infrared Science Archive, which is operated by the Jet Propulsion Laboratory, California Institute of Technology, under contract with the National Aeronautics and Space Administration.

% \clearpage 

% {\it Facilities:} 

\facilities{HST (ACS, WFPC2, WFC3), IRSA, Spitzer} 

\clearpage

%\onecolumngrid

%\appendix

%\section{Light Curves \label{sec:appendix} }

%{\bf Jacob is preparing a table of light-curve data.}

%\clearpage 

%\twocolumngrid


\begin{thebibliography}{}

\frenchspacing

\bibitem[Adams et al.(2016)]{Adams2016} Adams, S.~M., Kochanek, C.~S., Prieto, J.~L., et al.\ 2016, \mnras, 460, 1645 

\bibitem[Bally \& Zinnecker(2005)]{BallyZinnecker2005} Bally, J., \& Zinnecker, H.\ 2005, \aj, 129, 2281

\bibitem[Bally et al.(2020)]{Bally2020} Bally, J., Ginsburg, A., Forbrich, J., et al.\ 2020, \apj, 889, 178

\bibitem[Blagorodnova et al.(2017)]{Blagorodnova2017} {Blagorodnova}, N., {Kotak}, R., {Polshaw}, J., et al. 2017, \apj, 834, 107

\bibitem[Blagorodnova et al.(2020)]{Blagorodnova2020} {Blagorodnova}, N., {Karambelkar}, V., {Adams}, S.~M., et al. 2020, \mnras, 496, 5503

\bibitem[Bond(2011)]{Bond2011} Bond, H.~E.\ 2011, \apj, 737, 17 

\bibitem[Bond(2018)]{Bond2018}
Bond, H.~E.\ 2018, Research Notes of the AAS, 2, 122 

\bibitem[Bond et al.(2009)]{Bond2009}
Bond, H.~E., Bedin, L.~R., Bonanos, A.~Z., et al.\ 2009, \apjl, 695, L154 

\bibitem[Cai et al.(2018)]{Cai2018}
Cai, Y.-Z., Pastorello, A., Fraser, M., et al.\ 2018, \mnras, 480, 3424 

\bibitem[Cai et al.(2019)]{Cai2019} Cai, Y.-Z., Pastorello, A., Fraser, M., et al.\ 2019, \aap, 632, L6

\bibitem[Cai et al.(2021)]{Cai2021} Cai, Y.-Z., Pastorello, A., Fraser, M., et al.\ 2021, \aap, 654, A157

\bibitem[Cao et al.(2016)]{Cao2016}
Cao, Y., Nugent, P.~E. \& Kasliwal, M.~M. 2016, \pasp, 128, 114502

\bibitem[Caratti o Garatti et al.(2017)]{CarattioGaratti2017} Caratti o Garatti, A., Stecklum, B., Garcia Lopez, R., et al.\ 2017, Nature Physics, 13, 276

\bibitem[De et al.(2021)]{De2021}
{De}, K., {Kasliwal}, M.~M., {Hankins}, M.~J., et al. 2021, \apj, 912, 19

\bibitem[Doherty et al.(2015)]{Doherty2015} Doherty, C.~L., Gil-Pons, P., Siess, L., et al.\ 2015, \mnras, 446, 2599

\bibitem[Doherty et al.(2017)]{Doherty2017} Doherty, C.~L., Gil-Pons, P., Siess, L., et al.\ 2017, \pasa, 34, e056

\bibitem[Epchtein \& Nguyen-Quang-Rieu(1982)]{Epchtein1982} Epchtein, N., \& Nguyen-Quang-Rieu\ 1982, \aap, 107, 229

\bibitem[Evans et al.(1997)]{Evans1997} Evans, A., Geballe, T.~R., Rawlings, J.~M.~C., et al.\ 1997, \mnras, 292, 192

\bibitem[Evans \& Gehrz(2012)]{Evans2012} Evans, A., \& Gehrz, R.~D.\ 2012, Bulletin of the Astronomical Society of India, 40, 213

\bibitem[Evans et al.(2012)]{Evans2012} Evans, A., Gehrz, R.~D., Helton, L.~A., et al.\ 2012, \mnras, 424, L69 

\bibitem[Evans et al.(2005)]{Evans2005} Evans, A., Tyne, V.~H., Smith, O., et al.\ 2005, \mnras, 360, 1483

\bibitem[Fazio et al.(2004)]{Fazio2004} Fazio, G.~G., Hora, J.~L., Allen, L.~E., et al.\ 2004, \apjs, 154, 10

\bibitem[Gehrz(1988)]{Gehrz1988} Gehrz, R.~D.\ 1988, \araa, 26, 377

\bibitem[Gehrz(1999)]{Gehrz1999} Gehrz, R.~D.\ 1999, \physrep, 311, 405

\bibitem[Gehrz et al.(1980a)]{Gehrz1980a} Gehrz, R.~D., Grasdalen, G.~L., Hackwell, J.~A., et al.\ 1980a, \apj, 237, 855

\bibitem[Gehrz et al.(1995a)]{Gehrz1995a} Gehrz, R.~D., Greenhouse, M.~A., Hayward, T.~L., et al.\ 1995a, \apjl, 448, L119

\bibitem[Gehrz et al.(1980b)]{Gehrz1980b} Gehrz, R.~D., Hackwell, J.~A., Grasdalen, G.~I., et al.\ 1980b, \apj, 239, 570

\bibitem[Gehrz et al.(1995b)]{GehrzJones1995b} Gehrz, R.~D., Jones, T.~J., Matthews, K., et al.\ 1995b, \aj, 110, 325

\bibitem[Goldman et al.(2019)]{Goldman2019} Goldman, S.~R., Boyer, M.~L., McQuinn, K.~B.~W., et al.\ 2019, \apj, 877, 49

\bibitem[Goldman et al.(2018)]{Goldman2018} Goldman, S.~R., van Loon, J.~T., G{\'o}mez, J.~F., et al.\ 2018, \mnras, 473, 3835

\bibitem[Goldman et al.(2017)]{Goldman2017} Goldman, S.~R., van Loon, J.~T., Zijlstra, A.~A., et al.\ 2017, \mnras, 465, 403

\bibitem[Greenhouse et al.(1988)]{Greenhouse1988} Greenhouse, M.~A., Grasdalen, G.~L., Hayward, T.~L., et al.\ 1988, \aj, 95, 172

\bibitem[Greenhouse et al.(1990)]{Greenhouse1990} Greenhouse, M.~A., Grasdalen, G.~L., Woodward, C.~E., et al.\ 1990, \apj, 352, 307

\bibitem[Hartmann \& Kenyon(1996)]{HartmannKenyon1996} Hartmann, L., \& Kenyon, S.~J.\ 1996, \araa, 34, 207 

\bibitem[Hornoch \& Kucakova(2014)]{Hornoch2014} Hornoch, K., \& Kucakova, H.\ 2014, The Astronomer's Telegram, 6176

\bibitem[Hornoch et al.(2014)]{HornochStoev2014} Hornoch, K., Stoev, H., Tudor, V., Vaduvescu, O., \& Frigo, M.\ 2014, The Astronomer's Telegram, 6188

\bibitem[Howitt et al.(2020)]{Howitt2020} Howitt, G., Stevenson, S., Vigna-G{\'o}mez, A., et al.\ 2020, \mnras, 492, 3229

\bibitem[Hunter et al.(2017)]{Hunter2017} Hunter, T.~R., Brogan, C.~L., MacLeod, G., et al.\ 2017, \apjl, 837, L29

\bibitem[Jang \& Lee(2017)]{Jang2017} Jang, I.~S., \& Lee, M.~G.\ 2017, \apj, 836, 74

\bibitem[Jencson(2020)]{Jencson2020} Jencson, J. E. 2020, PhD. Thesis, California Institute of Technology (J20)

\bibitem[Jencson et al.(2019a)]{Jencson2019a} Jencson, J.~E., Adams, S.~M., Bond, H.~E., et al.\ 2019, \apjl, 880, L20

\bibitem[Jencson et al.(2018)]{Jencson2018} Jencson, J.~E., Kasliwal, M.~M., Adams, S.~M., et al.\ 2018, \apj, 863, 20 

\bibitem[Jencson et al.(2019b)]{Jencson2019b} Jencson, J.~E., Kasliwal, M.~M., Adams, S.~M., et al.\ 2019, \apj, 886, 40

\bibitem[Jencson et al.(2017)]{Jencson2017} Jencson, J.~E., Kasliwal, M.~M., Johansson, J., et al.\ 2017, \apj, 837, 167

\bibitem[Jencson et al.(2015)]{Jencson2015} Jencson, J.~E., Kasliwal, M.~M., Tinyanont, S., et al.\ 2015, The Astronomer's Telegram 7929, 1

\bibitem[Jencson et al.(2016a)]{Jencson2016a} Jencson, J.~E., Kasliwal, M.~M., Tinyanont, S., et al.\ 2016a, The Astronomer's Telegram 8688, 1

\bibitem[Jencson et al.(2016b)]{Jencson2016b} Jencson, J.~E., Kasliwal, M.~M., Tinyanont, S., et al.\ 2016b, The Astronomer's Telegram 8940, 1

\bibitem[Jones et al.(1982)]{Jones1982} Jones, T.~J., Hyland, A.~R., Caswell, J.~L., et al.\ 1982, \apj, 253, 208

\bibitem[Kami{\'n}ski et al.(2015)]{Kaminski2015} {Kami{\'n}ski}, T., {Mason}, E., {Tylenda}, R., et al. 2015, \aap, 580, 34

\bibitem[Karambelkar et al.(2019)]{Karambelkar2019} Karambelkar, V.~R., Adams, S.~M., Whitelock, P.~A., et al.\ 2019, \apj, 877, 110 (K19)

\bibitem[Kasliwal et al.(2017)]{Kasliwal2017}
Kasliwal, M.~M., Bally, J., Masci, F., et al.\ 2017, \apj, 839, 88 (K17)

\bibitem[Kochanek(2011)]{Kochanek2011} Kochanek, C.~S.\ 2011, \apj, 741, 37 

\bibitem[Ku{\v{c}}inskas et al.(2005)]{Kucinskas2005} Ku{\v{c}}inskas, A., {Hauschildt}, P.~H., {Ludwig}, H. -G., et al.\ 2005, \aap, 442, 281

\bibitem[Ku{\v{c}}inskas et al.(2006)]{Kucinskas2006} {Ku{\v{c}}inskas}, A., {Hauschildt}, P.~H., {Brott}, I., et al.\ 2006, \aap, 452, 1021

\bibitem[Lau et al.(2021)]{Lau2021} {Lau}, R.~M., {Hankins}, M.~J., {Kasliwal}, M.~M., et al.\ 2021, \apj, 909, 113

\bibitem[Masci et al.(2017)]{Masci2017} {Masci}, F.~J., {Laher}, R.~R., {Rebbapragada}, U.~D., et al.\ 2017, \pasp, 129, 014002

\bibitem[Menzies et al.(2019)]{Menzies2019} Menzies, J.~W., Whitelock, P.~A., Feast, M.~W., et al.\ 2019, \mnras, 483, 5150

\bibitem[Metzger \& Pejcha(2017)]{Metzger2017} Metzger, B.~D., \& Pejcha, O.\ 2017, \mnras, 471, 3200

\bibitem[Mukai(2015)]{Mukai2015} Mukai, K.\ 2015, Acta Polytechnica CTU Proceedings, 2, 246

\bibitem[Ney \& Hatfield(1978)]{Ney1978} Ney, E.~P., \& Hatfield, B.~F.\ 1978, \apjl, 219, L111

\bibitem[Oskinova et al.(2018)]{Oskinova2018} Oskinova, L.~M., Bulik, T., \& G{\'o}mez-Mor{\'a}n, A.~N.\ 2018, \aap, 613, L10

\bibitem[Pastorello et al.(2019)]{Pastorello2019} Pastorello, A., Mason, E., Taubenberger, S., et al.\ 2019, \aap, 630, A75 

\bibitem[Pejcha \& Prieto(2015)]{Pejcha2015} Pejcha, O., \& Prieto, J.~L.\ 2015, \apj, 799, 215

\bibitem[Pejcha et al.(2016a)]{Pejcha2016a} {Pejcha}, O., {Metzger}, B.~D. \& {Tomida}, K. 2016a, \mnras, 455, 4351 
\bibitem[Pejcha et al.(2016b)]{Pejcha2016b} {Pejcha}, O., {Metzger}, B.~D. \& {Tomida}, K. 2016b, \mnras, 461, 2527

\bibitem[Prieto(2008)]{Prieto2008} Prieto, J.~L.\ 2008, ATel, 1550

\bibitem[Prieto et al.(2008)]{Prietoetal2008}
Prieto, J.~L., Kistler, M.~D., Thompson, T.~A., et al.\ 2008, \apjl, 681, L9 

\bibitem[Rebull et al.(2014)]{Rebull2014} Rebull, L.~M., Cody, A.~M., Covey, K.~R., et al.\ 2014, \aj, 148, 92

\bibitem[Reipurth \& Mikkola(2015)]{Reipurth2015} Reipurth, B., \& Mikkola, S.\ 2015, \aj, 149, 145

\bibitem[Riebel et al.(2015)]{Riebel2015} Riebel, D., Boyer, M.~L., Srinivasan, S., et al.\ 2015, \apj, 807, 1

\bibitem[Schaefer(2010)]{Schaefer2010} Schaefer, B.~E.\ 2010, \apjs, 187, 275 

\bibitem[Shafter(2017)]{Shafter2017} Shafter, A.~W.\ 2017, \apj, 834, 196 

\bibitem[Siess (2007)]{Siess2007}Siess, L. 2007, A\&A, 476, 893

\bibitem[Sirianni et al.(2005)]{Sirianni2005} Sirianni, M., Jee, M.~J., Ben{\'\i}tez, N., et al.\ 2005, \pasp, 117, 1049

\bibitem[Smith et al.(2020)]{Smith2020} {Smith}, K.~W., {Smartt}, S.~J., {Young}, D.~R., et al.\ 2020, \pasp, 132, 085002

\bibitem[Soszy{\'n}ski et al.(2009)]{Soszynski2009} Soszy{\'n}ski, I., Udalski, A., Szyma{\'n}ski, M.~K., et al.\ 2009, \actaa, 59, 239

\bibitem[Sparks et al.(2008)]{Sparks2008}
Sparks, W.~B., Bond, H.~E., Cracraft, M., et al.\ 2008, \aj, 135, 605 

\bibitem[Szalai et al.(2019)]{Szalai2019} Szalai, T., Zs{\'\i}ros, S., Fox, O.~D., et al.\ 2019, \apjs, 241, 38

\bibitem[Szczygie{\l} et al.(2012)]{Szczygiel2012}
Szczygie{\l}, D.~M., Prieto, J.~L., Kochanek, C.~S., et al.\ 2012, \apj, 750, 77 

\bibitem[Tinyanont et al.(2016)]{Tinyanont2016} Tinyanont, S., Kasliwal, M.~M., Fox, O.~D., et al.\ 2016, \apj, 833, 231

\bibitem[Tonry et al.(2018)]{Tonry2018}
Tonry, J.~L., Denneau, L., Heinze, A.~N., et al.\ 2018, \pasp, 130, 064505

\bibitem[Tully et al.(2013)]{Tully2013} Tully, R.~B., Courtois, H.~M., Dolphin, A.~E., et al.\ 2013, \aj, 146, 86

\bibitem[Tully et al.(2016)]{Tully2016} Tully, R.~B., Courtois, H.~M., \& Sorce, J.~G.\ 2016, \aj, 152, 50

\bibitem[Tylenda et al.(2011)]{Tylenda2011}
Tylenda, R., Hajduk, M., Kami{\'n}ski, T., et al.\ 2011, \aap, 528, A114 

\bibitem[VanderPlas \& Ivezi{\'c}(2015)]{Vanderplas2015} VanderPlas, J.~T., \& Ivezi{\'c}, {\v{Z}}.\ 2015, \apj, 812, 18

\bibitem[Walter et al.(2012)]{Walter2012} Walter, F.~M., Battisti, A., Towers, S.~E., Bond, H.~E., \& Stringfellow, G.~S.\ 2012, \pasp, 124, 1057 

\bibitem[Whitelock et al.(2003)]{Whitelock2003} Whitelock, P.~A., Feast, M.~W., van Loon, J.~T., et al.\ 2003, \mnras, 342, 86

\bibitem[Whitelock et al.(2017)]{Whitelock2017} Whitelock, P.~A., Kasliwal, M., \& Boyer, M.\ 2017, EPJWC, 152, 01009

\bibitem[Whitelock et al.(2018)]{Whitelock2018} Whitelock, P.~A., Menzies, J.~W., Feast, M.~W., et al.\ 2018, \mnras, 473, 173

\bibitem[Whitmore et al.(2016)]{Whitmore2016} Whitmore, B.~C., Allam, S.~S., Budav{\'a}ri, T., et al.\ 2016, \aj, 151, 134

\bibitem[Williams et al.(2012)]{Williams2012} Williams, P.~M., van der Hucht, K.~A., van Wyk, F., et al.\ 2012, \mnras, 420, 2526

\bibitem[Williams et al.(2020)]{Williams2020} Williams, S.~C., Jones, D., Pessev, P., et al.\ 2020, \aap, 637, A20

\bibitem[Woodward et al.(2021)]{Woodward2021} Woodward, C.~E., Evans, A., Banerjee, D.~P.~K., et al.\ 2021, \aj, 162, 183

\bibitem[Yang \& Jiang(2011)]{Yang2011} {Yang}, M., \& {Jiang}, B.~W.\ 2011, \apj, 727, 53

\bibitem[Yoon \& Cantiello(2010)]{Yoon2010} Yoon, S.-C., \& Cantiello, M.\ 2010, \apjl, 717, L62

\end{thebibliography}
\end{document}